\documentclass[sigconf]{acmart}

\usepackage{amsmath,amsfonts}
\usepackage[linesnumbered,ruled,vlined]{algorithm2e}

\usepackage{graphicx}
\usepackage{textcomp}
\usepackage{xcolor}
\usepackage{mdframed}
\usepackage{tabularx}
\usepackage{comment}
\usepackage{booktabs}
\usepackage{epsfig}
\usepackage{bm}
\usepackage[tight,footnotesize]{subfigure}
\usepackage{adjustbox}
\usepackage{multirow}
\usepackage{cleveref}

\usepackage{mdframed}
\mdfsetup{
  linecolor=red,  
  linewidth=2pt,  
}

\renewcommand\footnotetextcopyrightpermission[1]{} 
\acmConference[Conference acronym 'XX]{Make sure to enter the correct
  conference title from your rights confirmation emai}{June 03--05,
  2018}{Woodstock, NY}

\begin{document}

\title{LCP: Enhancing Scientific Data Management with \underline{L}ossy \underline{C}ompression for \underline{P}articles}

\author{Longtao Zhang}
\affiliation{%
  \institution{Florida State University}
  \city{Tallahassee}
  \state{FL}
  \country{USA}}
\email{lzhang11@fsu.edu}

\author{Ruoyu Li}
\affiliation{%
  \institution{Florida State University}
  \city{Tallahassee}
  \state{FL}
  \country{USA}}
\email{rl13m@fsu.edu}

\author{Congrong Ren}
\affiliation{%
  \institution{The Ohio State University}
  \city{Columbus}
  \state{OH}
  \country{USA}}
\email{ren.452@buckeyemail.osu.edu}

\author{Sheng Di}
\affiliation{%
  \institution{Argonne National Laboratory}
  \city{Lemont}
  \state{IL}
  \country{USA}}
\email{sdi1@anl.gov}

\author{Jinyang Liu}
\affiliation{%
  \institution{University of Houston}
  \city{Houston}
  \state{TX}
  \country{USA}}
\email{jliu217@central.uh.edu}

\author{Jiajun Huang}
\affiliation{%
  \institution{University of California, Riverside}
  \city{Riverside}
  \state{CA}
  \country{USA}}
\email{jhuan380@ucr.edu}

\author{Robert Underwood}
\affiliation{%
  \institution{Argonne National Laboratory}
  \city{Lemont}
  \state{IL}
  \country{USA}}
\email{runderwood@anl.gov}

\author{Pascal Grosset}
\affiliation{%
  \institution{Los Alamos National Laboratory}
  \city{Los Alamos}
  \state{NM}
  \country{USA}}
\email{pascalgrosset@lanl.gov}

\author{Dingwen Tao}
\affiliation{%
  \institution{Indiana University Bloomington}
  \city{Bloomington}
  \state{IN}
  \country{USA}}
\email{ditao@iu.edu}

\author{Xin Liang}
\affiliation{%
  \institution{University of Kentucky}
  \city{Lexington}
  \state{KY}
  \country{USA}}
\email{xliang@cs.uky.edu}

\author{Hanqi Guo}
\affiliation{%
  \institution{The Ohio State University}
  \city{Columbus}
  \state{OH}
  \country{USA}}
\email{guo.2154@osu.edu}

\author{Franck Cappello}
\affiliation{%
  \institution{Argonne National Laboratory}
  \city{Lemont}
  \state{IL}
  \country{USA}}
\email{cappello@mcs.anl.gov}

\author{Kai Zhao}
\affiliation{%
  \institution{Florida State University}
  \city{Tallahassee}
  \state{FL}
  \country{USA}}
\email{kzhao@cs.fsu.edu}

\begin{abstract}
Many scientific applications opt for particles instead of meshes as their basic primitives to model complex systems composed of billions of discrete entities. Such applications span a diverse array of scientific domains, including molecular dynamics, cosmology, computational fluid dynamics, and geology. 
The scale of the particles in those scientific applications increases substantially thanks to the ever-increasing computational power in high-performance computing (HPC) platforms. However, the actual gains from such increases are often undercut by obstacles in data management systems related to data storage, transfer, and processing. 
Lossy compression has been widely recognized as a promising solution to enhance scientific data management systems regarding such challenges, although most existing compression solutions are tailored for Cartesian grids and thus have sub-optimal results on discrete particle data. 
In this paper, we introduce LCP, an innovative lossy compressor designed for particle datasets, offering superior compression quality and higher speed than existing compression solutions. Specifically, our contribution is threefold. 
(1) We propose LCP-S, an error-bound aware block-wise spatial compressor to efficiently reduce particle data size while satisfying the pre-defined error criteria. This approach is universally applicable to particle data across various domains, eliminating the need for reliance on specific application domain characteristics.
(2) We develop LCP, a hybrid compression solution for multi-frame particle data, featuring dynamic method selection and parameter optimization. It aims to maximize compression effectiveness while preserving data quality as much as possible by utilizing both spatial and temporal domains.
(3) We evaluate our solution alongside eight state-of-the-art alternatives on eight real-world particle datasets from seven distinct domains. The results demonstrate that our solution achieves up to 104\% improvement in compression ratios and up to 593\% increase in speed compared to the second-best option, under the same error criteria.

\end{abstract}

\maketitle

\section{Introduction}
\label{sec:intro}
Scientific data management systems~\cite{hdf5, scientific_data_management_in_the_coming_decade, parallel_in_situ, querying_a_scientific_database_in_just_a_few_seconds} are facing ever-increasing challenges from the rapid evolution of computing power versus the comparatively slow expansion of data infrastructure in HPC facilities. The fast-growing computing power enables scientific applications to run on a larger scale with higher precision, which as a consequence produces more data beyond the memory, storage, and I/O capacities of the data systems on supercomputers. For example, the EXAALT project which focuses on molecular dynamics (MD), generates trajectories containing over a trillion time steps using exascale machines by leveraging parallel-in-time approaches~\cite{exaalt}. Storing all these frames in a scientific data management system (e.g., HDF5~\cite{hdf5}) would require hundreds of terabytes of disk space, transferring them between facilities may take hours or days, and post-analysis of all frames on a single node is impractical due to insufficient memory size. 

Error-bounded lossy compression has been widely considered a promising solution for scientific applications facing data challenges~\cite{sdrbench, interp, zfp, mgardx, qoisz, cuszx}. First, lossy compressors can reduce the data volume significantly (by a factor of $5\sim1000$ in most cases). By comparison, lossless compressors, including Zstd~\cite{zstd}, Gorilla~\cite{gorilla}, and Brotli~\cite{brotli}, can only reduce the data by a factor of 2 in most cases~\cite{interp}. Second, error-bounded lossy compression can limit the compression error, ensuring that the quality of the decompressed data remains acceptable for post-analysis.

\begin{figure}[ht]
\centering
\subfigure[Structured mesh]{
\raisebox{0cm}{\includegraphics[width=0.2\textwidth]{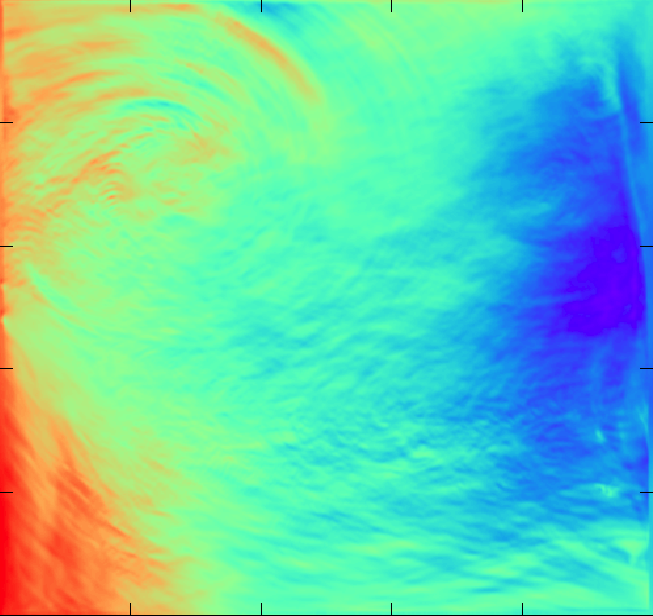}}}
\hspace{5mm}
\subfigure[Particle data]{
\raisebox{0cm}{\includegraphics[width=0.2\textwidth]{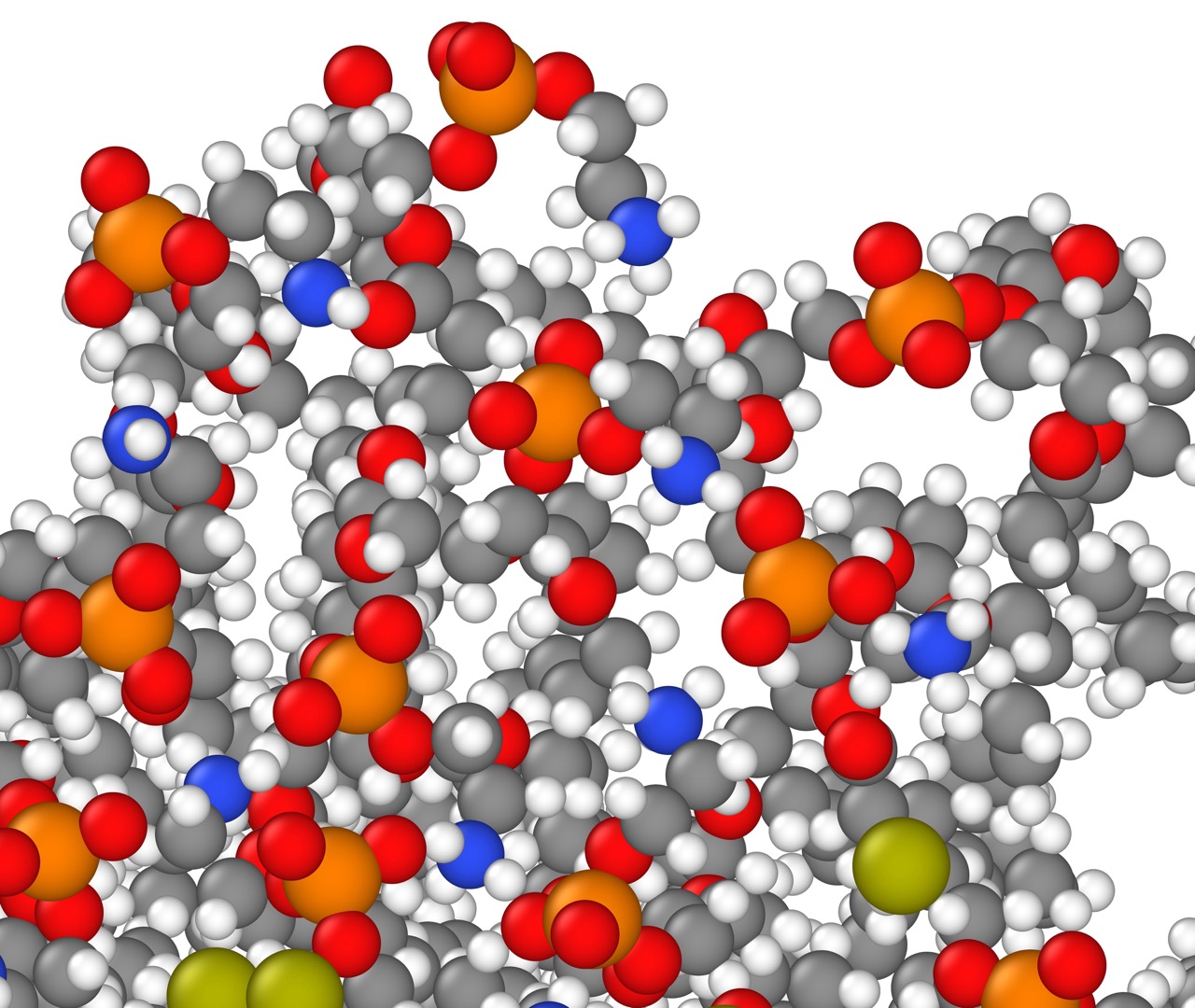}}}
\vspace{-2mm}
\caption{Visualization of structured mesh versus particles}
\label{fig:mesh_vs_particles}
\end{figure}

Scientific data can generally be classified into particle-style (e.g., locations, connectivity) and mesh-style (e.g. regular multidimensional grid in space). Existing lossy compressors, including those designed for databases (e.g., ModelarDB~\cite{modelardb, modelardb2}, SummaryStore~\cite{summarystore}, SciDB~\cite{scidb}) and those specifically designed for scientific data (e.g., SZ3~\cite{interp}, ZFP~\cite{zfp}, MGARD~\cite{mgardx}), are primarily tailored for structured mesh and suffer from low effectiveness on particle data~\cite{mdz}. However, various research fields such as material science, biology, cosmology, and computational fluid dynamics extensively utilize particle data. Scientific lossy compression techniques specific to the management of particle data remain under-explored.

In this paper, our objective is to develop an efficient scientific lossy compressor for the management of particle data. This task presents several challenges:
(1)~Most spatial compression techniques suitable for meshes are not applicable to particle data. These techniques often depend on the correlation of adjacent data inherent in structured meshes representing physical fields. In contrast, particle-style data lacks this correlation as it represents particles that are arbitrarily positioned in space 
(as depicted in~\Cref{fig:mesh_vs_particles}), leaving little structure to exploit for high compression ratios.
(2)~Applying temporal compression to multi-frame particle data is often impractical. One reason is that frames may be saved at irregular or even random intervals, such that the temporal domain may not be correlated enough to improve compression.
Second, temporal compression often requires loading consecutive frames into memory to identify global patterns or characteristics essential for compression. However, particle data frames can be exceptionally large allowing only a subset of frames to be loaded into memory.  Without large numbers of consecutive frames, the effectiveness of temporal compression diminished.
(3)~Accessing frames by batch is required by applications (discussed in~\Cref{sec: background batch compression}), which further restricts the approach. To support the retrieval of selected frames, compression is typically performed in small independent batches (each with a few frames, with no dependencies between batches). Since compression needs to be done multiple times, methods that have large amounts of metadata to store per compression introduce substantial overhead. Moreover, techniques requiring inter-batch dependencies should be excluded, otherwise decompressing a single frame will necessitate decompressing all of its preceding frames first, resulting in significant overhead on partial data retrieval. 

Taking into account all the aforementioned challenges, we introduce LCP, a novel scientific \underline{L}ossy \underline{C}ompressor for the management of \underline{P}article data. The main contributions are outlined as follows:
\begin{itemize}
    \item We propose the spatial compressor LCP-S and the temporal compressor LCP-T for particles. LCP-S is equipped with error-bound-aware quantization and spatial-block-wise coding algorithms to reach high compression effectiveness and data fidelity while guaranteeing the arbitrary error-bound defined by users before compression. Moreover, LCP-S is universally applicable to any particle data, in contrast to existing methods that are constrained to domain-specific data~\cite{mdz, tng}. 
    \item We propose LCP, our dynamic hybrid compression solution for multi-frame particle data. LCP is built on a hybrid design incorporating LCP-S and LCP-T, together with dynamic method selection and parameter optimization strategies, to maximize compression effectiveness by exploiting data characteristics in both spatial and temporal domains. Additionally, LCP is enhanced with the spatial-anchor-frame based batch compression technique to support the fast partial retrieval needs of applications.
    \item We evaluate our solution LCP on eight particle datasets from seven distinct domains with eight state-of-the-art related works. Experiments demonstrate that LCP is the best lossy compressor for the management of particle data, achieving the highest compression ratios, speed, and fidelity among all the compressors.
\end{itemize}

The remainder of the paper is structured as follows. In Section \ref{sec:background} delves into the research background. In Section \ref{sec:related}, we provide an overview of related work. Section \ref{sec:problem} formulates the research problem. Our developed particle compression framework is detailed in Section \ref{sec:overview} to Section \ref{sec: design lcp}. Section \ref{sec:evaluation} presents and discusses the evaluation results. Finally, we draw conclusions in Section \ref{sec:conclusion}.

\section{Research Background}
\label{sec:background}

In this section, we explain two fundamental sets of background concepts, the structure of particle data and scientific lossy compression, which both play crucial roles in the development of our innovative particle lossy compressor for the management of scientific data.

\subsection{Uniqueness of Particle Data}
\label{sec:background-particle}

Particle data is widely utilized in various scientific disciplines from the microscopic level to the cosmic level, as listed in Table \ref{tab:apps}. Particle data has its uniqueness when compared to other scientific data, particularly in terms of its continuity and retrieval needs. The uniqueness motivates the importance of designing lossy compression solutions specifically for particle data.

\begin{table}[ht]

    \centering
    \caption{Examples of particle data in various domains}
    \vspace{-3mm}
    \begin{adjustbox}{width=\columnwidth}
    \begin{tabular}{|c|c|c|}
    \hline
\textbf{Dataset} & \textbf{Domain} &  \textbf{Total Size}  \\ \hline
         BUN-ZIPPER~\cite{bunny} & Computer Vision &  3 MB \\ \hline
         Copper~\cite{mdz} & Material Science - Electronic & 200 MB \\ \hline
         Helium~\cite{mdz} & Material Science - Nuclear & 4 GB \\ \hline
         LJ~\cite{lammps-lj} & Computational Physics & 4 GB \\ \hline
        YIIP~\cite{yiip} & Biology & 4 GB\\ \hline
        HACC~\cite{hacc-outerrim} & Cosmology &  4 TB  \\ \hline 
        WarpX~\cite{warpx} & Plasma Physics &  8 TB \\ \hline
        3DEP~\cite{3dep, 3dep-aws} & Geology & > 200 TB \\ \hline

\end{tabular}
\end{adjustbox}
\label{tab:apps}

\end{table}

\subsubsection{Particle versus mesh}

Particle data gives researchers the flexibility to model complex systems composed of billions of discrete entities. One example is the HACC cosmology package which models the evolution of galaxies and dark matter within the universe~\cite{hacc}. The particles in HACC may represent a cluster of billions of stars or a cloud of dark matter. Scientists can gain insights into galaxy formation and the expansion of the universe by tracking the movement and interactions of these particles. In contrast, mesh data typically involves structured grids that are used to model continuous fields like temperature or pressure.

Mesh data typically has high continuity and a structured storage format, making it suitable for effective lossy compression. Particle data, on the other hand, may not have high continuity in adjacent-stored data points which is essential for lossy compression, since the data points represent values of discretely distributed particles. Moreover, the way particle data is stored may not necessarily match the order in which it is spatially organized. 




\subsubsection{Particle fields}
\label{sec: background particle fields}
Simulations of particle data often involve multiple frames to capture the temporal evolution of the system. These data typically include several fields, including position (i.e. location), velocity, and others (type, mass, connectivity, etc). Among these fields, \textbf{location} is often the focus of lossy compression because it takes up the majority of the particle data size and can tolerate some level of approximation. \textbf{Velocity} changes rapidly between time frames, making it unnecessary to store for post-analysis in most cases~\cite{mdz}. \textbf{Other fields} (type, mass, connectivity, etc) typically only need to be compressed once (such attributes stay the same in time).
As a result, when designing lossy compression for the management of scientific particle data, the focus should be on the location field~\cite{hrtc, mdz}. In the following sections, we will focus on the location fields (e.g., x, y, z).

\subsubsection{Partial retrieval}
\label{sec: background batch compression}
Particle data will be processed by applications for post-hoc analysis and decision-making. Retrieving a subset of data frames instead of all frames is an essential feature for such applications:
\begin{itemize}
    \item Getting all frames for many analyses is not necessary because they rely only on a single or a subset of frames. 
    \item Post-hoc analysis is usually executed on platforms with much less storage and memory capacity than the platforms for application execution and data generation and cannot decompress all the data at once due to hardware resource limitations. 
    \item Data retrieval speed is important. Decompressing all frames will cause delays in the post-analysis and decision-making.
\end{itemize}
To satisfy partial retrieval while maintaining some temporal information for compression, a common strategy is to compress in a batch style -- each time compress a small batch (like 8 or 16) of frames with no dependencies between batches. During retrieval, only the required batch will be decompressed to save the time and resources of decompressing the whole data. 
The problem with batch decompression is that it limits the choices of compression method one could use, as techniques requiring a large number of metadata per batch should be excluded due to metadata storage overhead. 

\subsection{Scientific Lossy Compression}  
\label{sec:background-lossy}
Lossy compression is a key technique for the management of scientific data~\cite{sz-io-hdf5, zfp, interp, mdz}.
Unlike lossless compression, lossy compression can reduce data to a much smaller size at the cost of some information distortion. Most scientific applications can tolerate such distortion to a certain level because they target the physical observations of the world which has some degree of uncertainty by nature. To keep the distortion introduced by lossy compression acceptable to scientific analysis, error-bounded mechanisms schemes have been developed for scientific lossy compression. 
One commonly adopted mechanism is the point-wise absolute error bound. As defined in~\Cref{eq:abs}, it constrains the maximum difference between each original value and its corresponding decompressed value. The error-bounded feature distinguishes scientific lossy compressors from traditional lossy compression methods like JPEG~\cite{jpeg2000} for images that do not bound error.


\section{Related Work}\label{sec:related}
In general, compression techniques could be categorized into two groups: lossless and lossy. Lossless compression techniques, including Zstd~\cite{zstd}, Gorilla~\cite{gorilla}, Brotli~\cite{brotli}, AMMMO~\cite{icde_time_series_compression}, have been adopted in variety of data management cases.
Lossless compression guarantees that the data, once decompressed, is identical to the original data. The effectiveness of lossless compression mainly depends on the frequency of repeated numbers or patterns in the data. 
Because of the predominant use of floating-point arithmetic in scientific applications where the trailing mantissa bits are uncorrelated, the applicability of lossless compression is quite limited in the scientific data management domain~\cite{sdrbench}. 

To reach higher compression ratios than lossless compressors, researchers are exploring lossy compression techniques in the data management community, with various solutions including time series databases, GPS trajectory compressors, block-wise compression, array databases, autoencoder-based compressors, and graph compressors. However, most of such solutions are not suitable for scientific particles.
Time series databases, including ModelarDB~\cite{modelardb, modelardb2} and SummaryStore~\cite{summarystore}, use techniques like the PMC-mean~\cite{pmc-mean} and the linear Swing model \cite{swing}. They are not suitable for scientific data management, due to low compression ratios~\cite{mdz}, which can be attributed to the absence of quantization and entropy coding.
GPS trajectories compressors~\cite{gps1,gps2,gps3,gps4} are designed with the assumption that GPS devices move following the road network, whereas scientific particles can move in any direction unpredictably. 
Block-wise algorithms, such as those used in SystemML~\cite{systemml} and Cumulon~\cite{cumulon}, segment data into blocks based on storage sequence; however, this sequence does not align with the spatial coordinates of particles, meaning particles close in space might not be stored adjacently on disk. Another block-wise lossless compression~\cite{cla} supports only a limited range of math operations, making it unsuitable for complex scientific computations. 
Array databases like SciDB~\cite{scidb} exhibit low compression ratios on scientific datasets~\cite{mdz}. 
Autoencoder-based solutions are mainly for mesh data instead of particles and they suffer from significantly slower speed than traditional methods~\cite{aesz}. 
Graph compression methods such as super nodes~\cite{supernode} are not suitable for scientific particles due to the lack of such super node structures in the data.

Given such limitations, certain lossy compressors have been designed specifically for particles, targeting the temporal and/or spatial domain. 
Temporal techniques aim to analyze particle trajectories to find patterns essential for compression, adopted techniques like principal component analysis (PCA)~\cite{pcadct, ED}. Such compression methods, however, are impractical for many of today's large-scale simulations. As simulation sizes grow, it becomes impractical to collect many snapshots in memory together to carry out the full-trajectory analysis.  
Spatial methods focus on individual frames. One such algorithm is to sort the particles based on space-filling curves and encode adjacent index differences with variable-length encoding~\cite{cpc2000, tao2017depth}. Another design approach is to construct a tree structure (e.g., k-d tree) to split the coordinate space and store particles in each region separately~\cite{draco, zfp-particle1}. 
Besides those two types, there are domain-specific designs for particle compression equipped with both spatial and temporal algorithms. For example, MDZ~\cite{mdz} proposed a vector-based prediction method that is mostly effective on solid material MD simulations. 
However, those domain-specific methods may not be effective on particle data from other domains, as shown in our experiments. 

\section{Problem formulation}
\label{sec:problem}
The \textbf{overall objective} of this project is to maximize the compression ratio for particle datasets with an innovated compressor and optimized parameters while ensuring that pre-defined error bounds are maintained and the decompression speed matches or exceeds that of other state-of-the-art methods (SOTA), as shown in~\Cref{eq:problem}. 
The input $D$ is a particle dataset. The compression and decompression processes are denoted as functions $f$ and $g$ respectively, with $\theta$ representing the parameters of $f$. $eb$ is the pre-defined error bound.

\begin{equation}
\label{eq:problem}
\begin{split}
&f,g,\theta= \underset{f,g,\theta}{\arg\max} \frac{size(D)}{size(f(D))} \\
s.t. \ &|d_i-d_i^{'}|\leq eb, \forall d_i \in D \\
\ & speed(g) > speed(SOTA) \\
\end{split}
\end{equation}

\textbf{Targeted data}.
Compressing particle location in multiple frames is the main target of this research, as discussed in \Cref{sec: background particle fields}. The dimension of input data $D$ comprises three components: the number of frames, the number of particles per frame, and the number of coordinates (typically 3, referring to the X, Y, and Z coordinates). $D$ can be either integers or floats. The solution has the ability to compress other fields (velocity, intensity, etc.), given that users provide them as supplementary dimensions to the spatial positions.

\textbf{High compression ratio (or low bit rate)}. Compression ratio (CR) and bit rate are two commonly used metrics to quantify the effectiveness of data reduction methods. The compression ratio is defined as $\frac{size(D)}{size(f(D))}$.
Bit rate is defined as the number of bits to store each data element on average. For instance, given a dataset in FP32 format, if a compressor reduces the data size from 20 TB to 1 TB, then its compression ratio is 20, and its bit rate is $32/20 = 1.6$. In conclusion, a higher compression ratio or lower bit rate means better compression in terms of size.

\textbf{High data fidelity while respecting the error bound}. Error bound ($eb$) and PSNR are widely adopted to control and quantify the distortion of the compression. 
Error bound is selected by users in advance, and lossy compressors must ensure they comply with this bound during the compression process. One of the most commonly used bounds is the absolute error bound defined as follows.
\begin{equation}
    eb = \max_{\forall i} |d_i - d'_i|
    \label{eq:abs}
\end{equation}
PSNR is derived by the following equation to evaluate the data fidelity after decompression. Value range of $D = max(D) - min(D)$, and $mse()$ computes the mean square error of two inputs.
\begin{equation}
    PSNR = 20*log_{10}{(\frac{\text{\textit{value range of }} D}{\sqrt{mse(D, D')}})}    
\end{equation}

\textbf{High Speed}
Speed is a critical factor for scientific lossy compression, defined as follows.
\begin{equation}
    \text{\textit{(de-)compression speed}} = {\frac{size(D)}{\text{\textit{(de-)compression time in seconds}}}}    
    \label{eq:compression-speed}
\end{equation}
The speed of decompression (data retrieval) often carries greater importance than compression (storage), as data retrieval directly affects post-analysis and decision-making, while compression can be performed asynchronously without affecting other tasks.

\section{Design Overview of LCP} 
\label{sec:overview}

\begin{figure}[ht]
\centering
\raisebox{-1cm}{\includegraphics[scale=0.26]{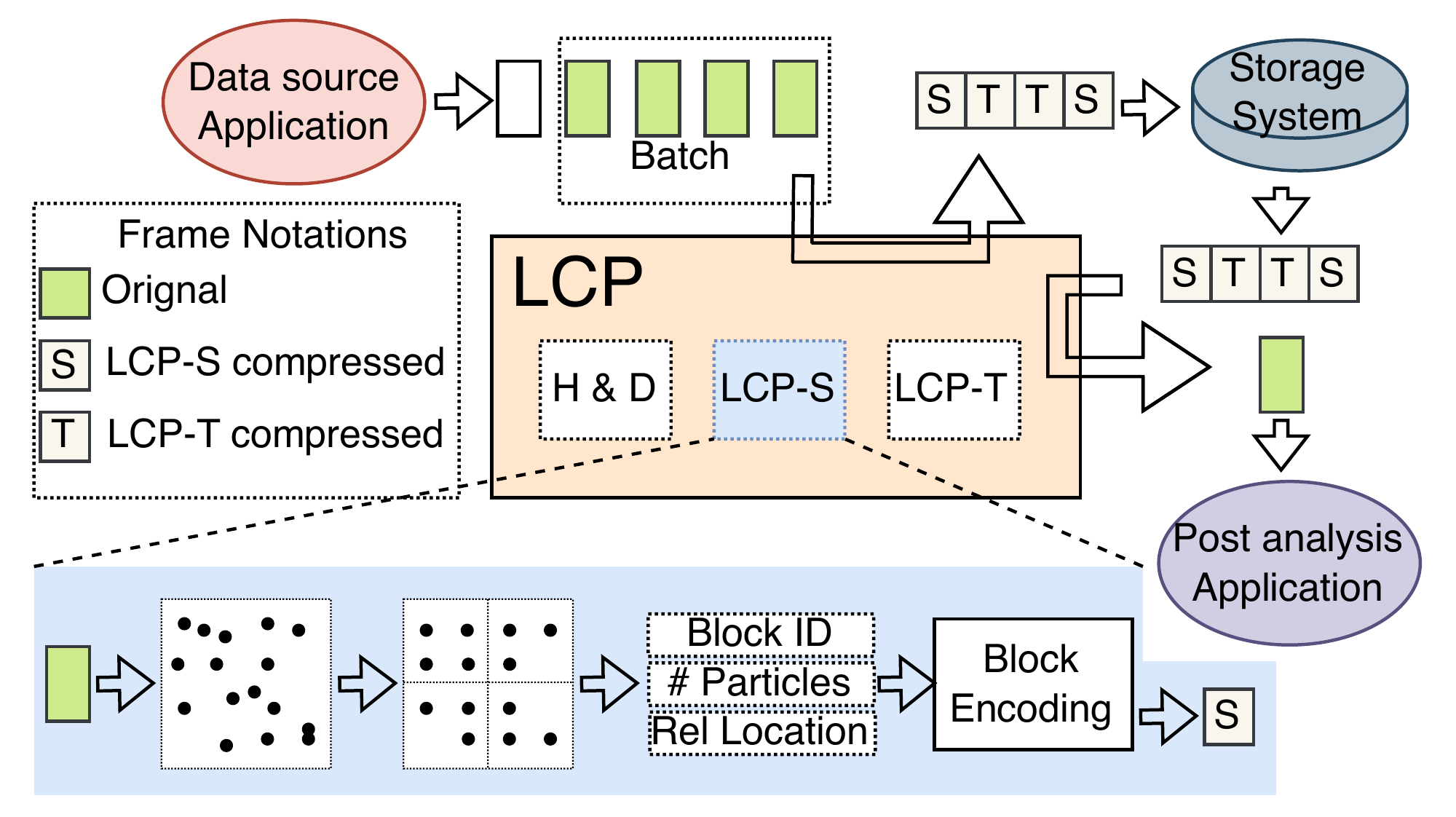}}
\vspace{-4mm}
\caption{Overall design of LCP (H\&D means hybrid selection and dynamic optimization)}
\label{fig:lcp-design}
\end{figure}

In this section, we describe the overall design of our scientific high-compression-ratio and high-speed particle compressor -- LCP. LCP is equipped with two compression algorithms (LCP-S and LCP-T), alongside hybrid algorithm selection and dynamic optimization strategies, to reach the best compression ratio and speed while keeping high data fidelity.

Figure \ref{fig:lcp-design} shows how LCP's key innovations are integrated with data storage and retrieval workflows to manage particle data. 
\begin{itemize}
    \item \textbf{Data storage}. In the storage workflow, the data source is a scientific simulation or instrument that produces a frame of particle data periodically, and the destination is a data storage or management system (e.g., Lustre~\cite{lustre}, GPFS~\cite{gpfs}, HDF5~\cite{hdf5}). LCP will compress the frames in batch, and the compressed data will be saved to the data system.
    \item \textbf{Data retrieval}. The retrieval workflow takes data from the storage system, decompresses it, and feeds it to the post-hoc analysis application.
    LCP decompresses data in batch granularity, rather than decompressing the whole dataset, to reduce the time applications spend waiting for the data, as discussed in \Cref{sec: background batch compression}. 
\end{itemize}

We build LCP upon the FZ modular framework~\cite{fz}. FZ is one of the leading compression solutions for scientific applications. It ships with various techniques and functionalities, including compression methods (e.g., interpolation~\cite{interp}, Huffman coding, Zstd~\cite{zstd}), compression integration with I/O libraries (e.g., HDF5~\cite{hdf5}) and applications (e.g.,  LAMMPS~\cite{lammps}), compression error analysis, and parameter parsing from file and command line. Most of those techniques and functionalities are provided by modules and can be integrated independently into third-party software. 
The modular design and rich API library reduce our effort in re-implementing existing techniques (e.g., lossless compression) or functionalities (e.g., compression error verification), allowing us to focus on the design and implementation of the three innovations in LCP. 


\section{LCP-S: the Error-Bound Aware Block-wise Spatial Compressor}
\label{sec: design lcp-s}
In this section, we describe LCP-S, our spatial particle compressor. Particle data typically shows significant sparsity and non-uniformity as depicted in~\Cref{fig:mesh_vs_particles}, making it challenging to achieve consistent compression results. As a result, the key idea in LCP-S is to segment the 3D space into equal-sized blocks (best block size discussed in~\Cref{sec:block_size_in_a_single_frame}) such that each block will gain more accurate compression due to closer correlation and more effective coding due to lower entropy.
In particular, LCP-S first quantizes particle coordinates to integers while respecting the pre-defined arbitrary error bound, then the coordinates are converted to relative values in each block, after that coordinates and block metadata (id and number of particles) are compressed to further reduce the size.

LCP features a distinct design compared to traditional compressors. It segments particles into groups according to their spatial coordinates, as opposed to traditional block-wise methods that divide data according to the order in which it is stored. More importantly, LCP integrates error-bound support as a fundamental aspect of its design, influencing both the quantization of coordinates and the spatial division into blocks. In addition, LCP has the ability to optimize block size and coding methods based on the data input.

Since most scientific applications, such as the HACC cosmology simulation~\cite{hacc} and WarpX plasma physics simulation~\cite{warpx}, utilize 3D Euclidean space, we primarily use 3D examples and equations to illustrate our solution in the following discussion. However, it's important to note that our solution is generalized and can be applied to data in other dimensions as well.

\subsection{Error-Bound Aware Quantization}
\label{sec: lcp-s quantization}


Quantization, by broad definition, maps a real value to a discrete value that can be represented more compactly. In the context of data compression, quantization methods typically reduce the storage volume by removing some precision from the data. 

It is vital for users to be able to set the allowed compression error a priori. Although scientific applications have the ability to tolerate or handle a certain degree of error/noise (e.g. because their data often contains noise due to measurement limitations or environmental factors, or their solvers are usually numerical-approximation-based which tolerate error~\cite{CalhounIJHPCA:LossyCompression}), different applications and use cases will have diverse tolerance on the error. 

We quantize the particles to integers with regard to the arbitrary error bound defined by users. More specifically, for each particle $d$ (coordinates as $d.x$, $d.y$, $d.z$), the quantization integer $q(d.x)$ and the reverse quantization floating point $d'$ are defined as follows.
\begin{equation}
    \begin{array}{l}
    \label{eq: quantization}
        q(d.x) = \lfloor \frac{d.x - min(D.x)}{2 \times eb} \rfloor, d \in D \\
        d.x' = ( 2 \times q(d.x) + 1 ) \times eb + min(D.x) \\
        \text{note: same for dim $y$ and $z$}
    \end{array}
\end{equation}
Our solution for quantization is different from most lossy existing compressors. First, as discussed in~\Cref{sec:background-lossy}, existing general lossy compressors primarily target on Cartesian grid -- they first estimate the data and then quantize the difference between the estimation and the original value. Since particles are arbitrarily positioned in space with much less spatial correlation than Cartesian grids, the estimation methods have limited effectiveness if applied globally. As a result, we propose to first perform quantization, and then group the quantized data into small blocks for better compression in each small region. 
Second, many existing particle compressors (e.g., Draco~\cite{draco}, GPS-based ones~\cite{gps1}) do quantization by removing the least significant bits of the data. This is often not sufficient for scientific users due to the limited choices of quantization bits and the lack of knowledge about the exact error introduced by the quantization process prior to compression. 

\subsection{Spatial-Block-Wise Coding}

After quantization, the 3D space where the particles are located will be subdivided into equal-sized blocks, as shown in~\Cref{fig:lcp-design}.
\begin{table}[ht]
\centering
\footnotesize
  \caption {Effect of blocking on entropy and autocorrelation of quantized data (lower entropy and higher correlation lead to higher compression effectiveness)} 
  \vspace{-2mm}
  \label{tab:block_vs_nonblock} 
  \begin{adjustbox}{width=\columnwidth}
  \begin{tabular}{|c|c|c|c|c|c|c|}
  \hline
    \multirow{2}{*}{Dataset}  & \multicolumn{3}{c|}{Entropy (BS: block size)} & \multicolumn{3}{c|}{Autocorrelation (BS: block size)} \\ \cline{2 - 7}
         & No block & BS = 64 & BS = 8 & No Block & BS = 64 & BS = 8 \\ \hline
Copper & 9.509 & 5.697 & 2.955 & 0.8260 & 0.9999 & 0.9999 \\ \hline     
YIIP & 12.205 & 5.999 & 2.999 & 0.3132 & 0.9997 & 0.9999 \\ \hline
BUN-ZIPPER & 6.201 & 5.948 & 2.999 & 0.5615 & 0.6234 & 0.9912 \\ \hline
\end{tabular}
\end{adjustbox}
\end{table}

This segmentation is designed to ensure that particles that are in close proximity to each other are grouped within the same block, such that we can utilize the spatial relationships and distributions of particles in each block for more efficient compression. Since each block represents a localized cluster of particles, the stronger spatial relation of particles, as demonstrated in~\Cref{tab:block_vs_nonblock} would lead to more efficient compression. Moreover, since the data has been quantized into integers and the range of integer values within a block is quite limited, the entropy within these blocks is reduced as shown in~\Cref{tab:block_vs_nonblock}, consequently making the coding more effective.

Note that our key idea is distinct from traditional block-wise compression techniques which split data into blocks by the storage order~\cite{cpc2000, tao2017depth, Sihuan-Bigdata18}. 
The storage order of particles is not the same as their spatial order in most cases~\cite{mdz} -- storage order is affected by particle type, acquisition sequence, computation methods, decomposition methods in parallel computing, etc. Such order doesn't necessarily contain a stronger local relation or lower entropy in blocks which is the key to improving the compression effectiveness of particles.


\subsubsection{Design of spatial blocks}
\label{sec: lcp-s block components}
In our design, the 3D space which contains all the particles is divided into multiple aligned fixed-sized blocks, each block containing zero or more particles.
We define block size to be proportional to $2\times eb$, such that the block id of a particle can be derived directly using its quantized position value as follows.

\begin{equation}
    \begin{array}{l}
    \label{eq: block size and id}
    \textit{block\_size} = 2\times eb \times p\\
    \textit{bid.x} = \lfloor \frac{d.x - min(D.x)}{\textit{block size}} \rfloor = \frac{q(d.x)}{p}, \text{same for dim $y$ and $z$} \\
    \textit{block\_id} = \textit{bid.x}  + \textit{bn.x} \times \textit{bid.y} + \textit{bn.x} \times \textit{bn.y} \times \textit{bid.z} \\
    \text{note: $p$ is a fixed parameter; $bn$ is block count in each dim}
    \end{array}
\end{equation}


Since we are targeting large-scale data processing scenarios, we choose to divide the space with fixed-size blocks instead of tree structures (such as k-d tree) for processing simplicity and high speed. Our solution has a $O(N)$ time complexity and is straightforward to implement in parallel. In comparison, the k-d tree has a time complexity of $O(kN)$, and parallelizing a k-d tree can be challenging especially during the construction phase. The process of building a k-d tree involves recursive partitioning of data points, which depends on the results of previous steps. As a result, our design doesn't rely on tree structures.

We keep the following three pieces of information for each non-empty block, in order to reconstruct the dataset in decompression.
\begin{itemize}

\item Block id. $block\_id$, as defined in~\Cref{eq: block size and id}, is used to locate the location of blocks in 3D space. We choose to store block id with relative coordinates instead of storing absolute coordinates because the relative coordinates yield a lower entropy which can lead to better coding effectiveness. Additionally, we only store the block that is not empty. There will be many empty blocks when the particle data is sparse in space. In such scenarios, our solution totally avoids the time and storage to handle those empty blocks.


\item Particle count. The number of points in a block is stored to indicate the length of the block. We store the particles block-by-block without gaps, thus the particle count is needed for retrieving the correct particles in a block during decompression. Note that the particle count will not be 0 as we do not store the empty block at all.

\item Relative location of particles. As discussed in the block id part, we store the block id and the relative location of particles for lower entropy. There are multiple ways to convert absolute locations to relative locations, depending on the reference or base point used. To avoid storing extra information (e.g., the median particle), we use the (0,0,0) coordinate of each block as our reference point, which means the relative location of each particle is calculated based on its difference from the origin point within the block. This approach simplifies the storage and retrieval process, as it eliminates the need for additional reference data, and simplifies the computation between relative and absolute locations.
\end{itemize}

\subsubsection{Block coding}
\label{sec: block-coding}
In this step, we lossless compress the three types of data (block id, particle count, and relative location of particles) with a chain of three coding methods. More specifically, each type of data is first delta-coded to reduce its variability and magnitude, then processed using either variable-length or fixed-length coding to efficiently translate it into a compact binary form. Finally, dictionary coding is applied to remove redundancy by exploiting repetitions in the data.

\begin{table}[ht]

\centering
\footnotesize
  \caption {Effect of fixed length coding versus Huffman coding on block id and relative position arrays} 
  \vspace{-2mm}
  \label{tab:cr_comp_coding_methods} 
  \begin{adjustbox}{width=\columnwidth}
  \begin{tabular}{|c|c|c|c|c|c|}
  \hline
    \multirow{2}{*}{Dataset} & \multirow{2}{*}{eb} & \multicolumn{2}{c|}{Block id} & \multicolumn{2}{c|}{Relative position} \\ \cline{3 - 6}
     &  & Huffman & fixed-length & Huffman & fixed-length \\ \hline
\multirow{2}{*}{Helium}  & 1E-1 & 35 B & \textbf{34 B} & 18.76 KB & \textbf{33.13 KB} \\ \cline{2 - 6}
  & 1E-2 & 35 B & \textbf{34 B} & \textbf{70.24 KB} & 102.68 KB \\ \cline{2 - 6}
  & 1E-3 & \textbf{76.46 KB} & 110.55 KB & 116.94 KB & \textbf{115.68 KB} \\ \cline{1 - 6}
\multirow{2}{*}{Copper}  & 1E-1  & 35 B & \textbf{34 B} & 2.80 KB & \textbf{2.42 KB} \\ \cline{2 - 6}  
  & 1E-2 & 35 B & \textbf{34 B} & 7.70 KB & \textbf{5.09 KB} \\ \cline{2 - 6} 
  & 1E-3 & 9.07 KB & \textbf{5.98 KB} & 776 B & \textbf{720 B} \\ \cline{1 - 6} 
\multirow{2}{*}{3DEP}  & 1E-1 & \textbf{2.45 KB} & 3.71 KB & \textbf{11.93 MB} & 16.42 MB \\ \cline{2 - 6}
  & 1E-2 & \textbf{2.90 MB} & 5.02 MB & \textbf{26.20 MB} & 26.73 MB \\ \cline{2 - 6}
  & 1E-3 & \textbf{34.39 KB} & 58.69 KB & \textbf{35.79 MB} & 50.53 MB \\ \cline{1 - 6}
\end{tabular}
\end{adjustbox}

\end{table}

For delta coding, we replace each data $D_i$ by $D_i - D_{i-1}$  which is its difference from the previous data. Since we process blocks in order, the delta coding will convert the block ids from an array with increasing values to an array with many repeated numbers, such that it makes the following variable-length/fixed-length coding and dictionary coding more effective. For particle count array and relative location array, although they are not strictly monotonic, due to the limited value range, delta coding will still result in many repeated numbers thus benefitting the encoding stage.

The next step in the encoding stage is to shrink the number of bits used to store each value by fixed-length or variable-length coding. We use Huffman coding as our variable-length coding method. While fixed-length coding stores each value with exactly the same number of bits, Huffman coding assigns shorter bits to more frequently occurring values. Thus Huffman coding may lead to shorter output, but storing its coding table may offset this advantage. In \Cref{tab:cr_comp_coding_methods}, we compare the effectiveness of Huffman coding and fixed-length coding for the block id array and relative location array. The table shows that the optimal coding method varies, reflecting the diverse distributions of different datasets and their specific constraints. As a result, we will calculate the expected coding length of both methods and select the one with a shorter length to use. 

Dictionary coding is the last step in the encoding stage. We use Zstd~\cite{zstd} as our dictionary coder to further reduce the data size by removing redundancy in the data.

\section{LCP: The Dynamic Multi-Frame Hybrid Compressor}
\label{sec: design lcp}
In this section, we propose LCP, our dynamic hybrid compression solution for multi-frame data based on our spatial compressor LCP-S and the temporal compressor LCP-T. 
LCP aims to maximize compression effectiveness while respecting the fixed-batch requirements of scientific applications, by using a combination of spatial and temporal compression methods and adjusting them dynamically.
For instance, when processing the first frame in a batch, LCP can apply either spatial or temporal compression to achieve the highest compression ratio, supported by its spatial anchor frames. Traditional video compression schemes like the group of pictures (GOP) structures~\cite{h265}, on the other hand, only compress the first frame (I-frame) spatially. 
Moreover, LCP employs a finite-state machine to select the most effective compression method for each frame with minimal overhead. In contrast, existing particle compressors such as MDZ~\cite{mdz}, could only select the best method at the batch level.

We first introduce the temporal compressor LCP-T and discuss how we hybrid it with LCP-S, and then we bring the spatial-anchor-frame-based solution for batch compression. Dynamic optimization is discussed in the end.

\subsection{LCP-T: The Temporal Compressor}
LCP-T is our temporal compressor for multi-frame particle datasets. The objective of LCP-T is to boost compression effectiveness by exploiting temporal redundancies and correlations. LCP-T compresses data by frames. For each frame, LCP-T first quantizes input data with the error-bound aware technique of LCP-S as discussed in~\Cref{sec: lcp-s quantization}. Then LCP-S predicts the current frame based on the previous frame and calculates the difference between the two frames. In the end, the difference array will be handled by Huffman coding and dictionary coding (Zstd) to reduce the data volume.

\subsection{Hybrid Compression with LCP-S \& T}
\label{sec: hybrid lcp s and t}
\begin{figure}[ht] \centering
\raisebox{-1cm}{\includegraphics[scale=0.33]{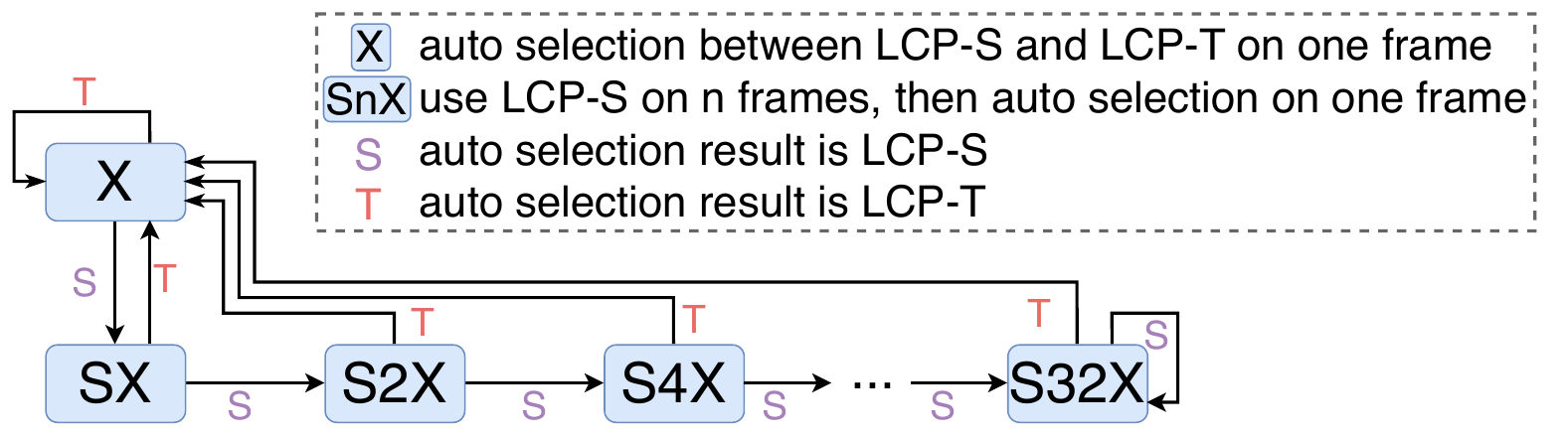}}
\vspace{-2mm}
\caption{LCP-FSM: LCP uses a finite-state machine to determine the available compression methods for each frame}
\label{fig:design_s_t}
\vspace{-2mm}
\end{figure}

LCP contains both the spatial compressor LCP-S and the temporal compressor LCP-T. To maximize compression effectiveness by considering both spatial and temporal factors, it is equipped with a selection mechanism LCP-FSM, as shown in~\Cref{fig:design_s_t}, for determining the best compression approach (LCP-T or LCP-S) for each frame.

\begin{figure}[ht] \centering
\hspace{-8mm}
\subfigure[Helium]
{
\raisebox{-1cm}{\includegraphics[scale=0.25]{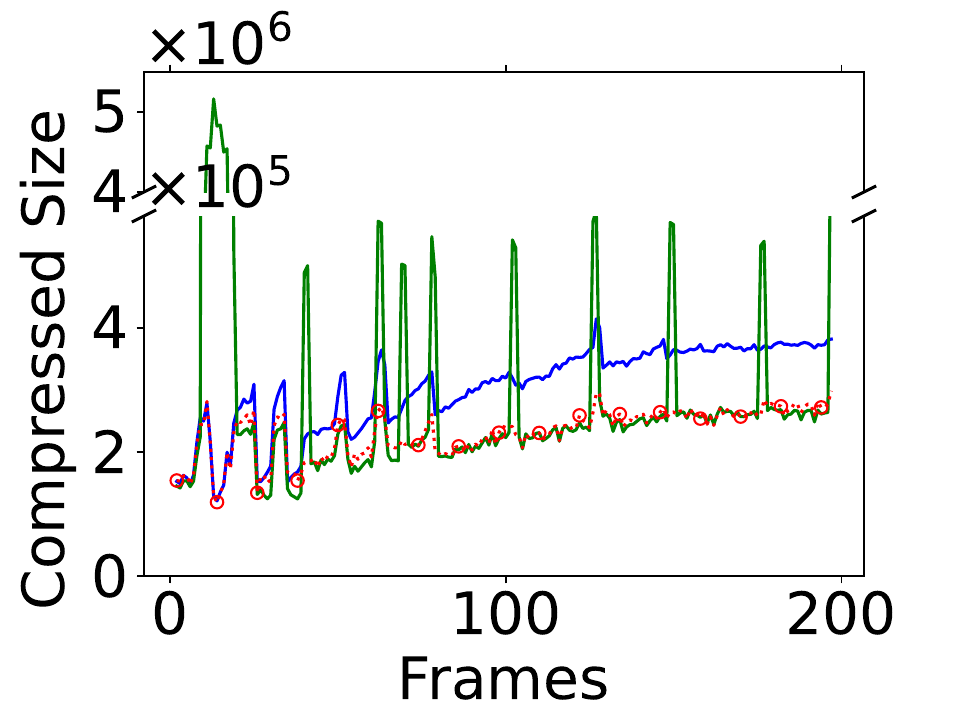}}
}
\hspace{-4mm}
\subfigure[YIIP]
{
\raisebox{-1cm}{\includegraphics[scale=0.25]{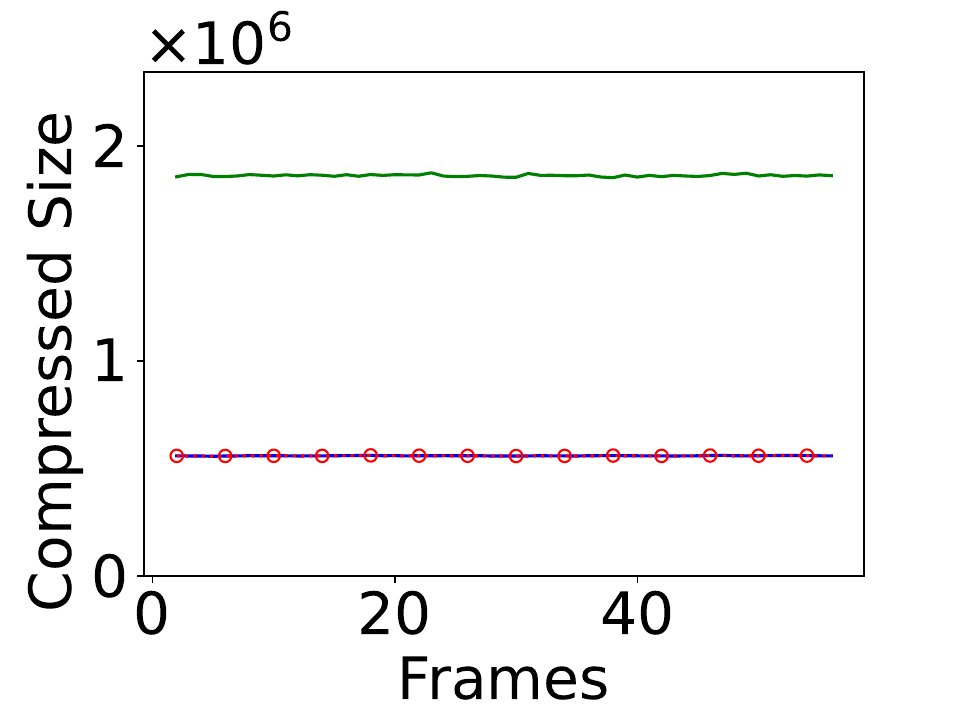}}
}
\hspace{-8mm}

\includegraphics[scale=0.25]{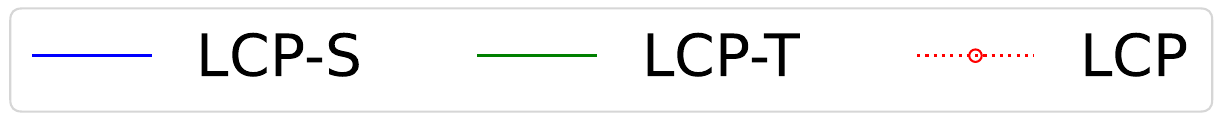}

\vspace{-2mm}
\caption{LCP-S and LCP-T exhibit varying effectiveness due to the dynamic nature of spatial and temporal relationships across different datasets and over time. Our LCP can consistently choose the best method for compression because of our spatial-anchor-frame-based design detailed in \Cref{sec: compress-in-batch} (low compressed size in figures indicates better compression)}
\label{fig:s_t_selection}
\vspace{-2mm}
\end{figure}
LCP will select the best compression approach by comparing their estimated compressed size for each frame. For LCP-S, we use the size of the most recently LCP-S compressed frame, as LCP-S tends to maintain a stable compression size over time, as shown in~\Cref{fig:s_t_selection}. On the other hand, temporal compression may have distinct results on frames of the same dataset, thus we have to test LCP-T to get the accurate compressed size. LCP-FSM significantly reduces unnecessary LCP-T executions by using a state machine to manage the necessity of testing LCP-T. For example, suppose we are in the $S2X$ state, if the LCP-S/T comparison on the third frame still chooses LCP-S, the necessity of testing LCP-T further decreases, thus the state changes to $S4X$. As a result, the overhead of method selection is under 5\% even in cases when LCP-S outperforms LCP-T in every frame. Moreover, \Cref{fig:s_t_selection} demonstrates that our method has high selection accuracy as LCP consistently chooses the optimal method for all the frames.




\subsection{Spatial-Anchor-Frame-Based Batch Compression}
\label{sec: compress-in-batch}

\begin{algorithm}[ht]
\caption{LCP Multi Frame Compression}
\label{alg: multi frame compression}
\SetAlgoLined

\SetKwArray{CompressedFrames}{comp\_frames}
\SetKwArray{CompressedBatch}{comp\_batch}
\SetKwArray{CompressedAnchorFrames}{comp\_anchor\_frames}
\SetKwData{LastAnchorFrame}{last\_anchor\_frame}
\SetKwData{Frame}{frame}
\SetKwData{PrevFrame}{frame.prev}
\SetKwData{Method}{method}
\SetKwData{Batch}{batch}
\SetKwData{Data}{data}
\SetKwData{Compressed}{compressed}

\KwIn{data - the original data}
\KwOut{compressed data}

\CompressedFrames{} $\leftarrow$ []\;
\CompressedAnchorFrames{} $\leftarrow$ []\;
\LastAnchorFrame $\leftarrow$ nil\;

\For{\Batch in \Data}{
 \CompressedBatch{} $\leftarrow$ []\;

\For{\Frame in \Batch}{
  \uIf{is \Frame first in batch}{
    \PrevFrame $\leftarrow$ \LastAnchorFrame}
   \uElse{
   \PrevFrame $\leftarrow$ previous frame
  }
  \Method $\leftarrow$ LCP-FSM(\Frame, \PrevFrame)\;
  \Compressed $\leftarrow$ false\;
  
  \uIf{\Method == "compare"}{
    estimate LCP-S and LCP-T with \Frame \\
    \Method $\leftarrow$ the one with higher effectiveness \\
    \Compressed $\leftarrow$ true\\
  }
  \uIf{\Method == "spatial"}{
    \uIf{compressed == false}{
       compress \Frame by LCP-S}
    \uIf{\Frame is first frame in batch}{
        add result to \CompressedAnchorFrames{}\\
        \LastAnchorFrame $\leftarrow$ frame\\
    }\uElse{
        add result to \CompressedBatch{}\\
    }
  }
  \uElse{ \tcp{method == "temporal"}
      \uIf{compressed == false}{
    compress \Frame by LCP-T with \PrevFrame \\
    }
    add result to
    \CompressedBatch{} \\
  }
}
add \CompressedBatch{} to \CompressedFrames{}
}
return(\CompressedFrames{}, \CompressedAnchorFrames{})

\end{algorithm}


In this section, we discuss how we support batch compression in LCP. All temporal compressors, including LCP-T, have inter-frame dependencies in compression. To support partial retrieval in this scenario, frames are usually compressed in batches, as we discussed in~\Cref{sec: background batch compression}. The issue for batch compression is that the first frame in each batch cannot be temporal compressed (otherwise it brings inter-batch dependencies) which significantly limits the compression ratios in such solutions~\cite{mdz}.

LCP is equipped with the spatial-anchor-frame-based batch compression design to facilitate temporal compression without introducing inter-batch dependencies. As presented in~\Cref{alg: multi frame compression}, if a frame is the first in its batch, and is compressed by LCP-S, we call it a spatial anchor frame and store such anchor frames in a separate array. Then to compress a frame that is the first in its batch by LCP-T, we will use its nearest anchor frame for the temporal compression. During decompression, we first locate the compressed frame from either its batch or the compressed anchor frame array. Then if the frame is compressed by LCP-S, we decompress it directly by LCP-S, otherwise we first decompress its dependent frame recursively and then decompress it by LCP-T. 

Our solution offers a significant advantage in both efficiency and effectiveness. On one hand, compared with the original batch compression design, our solution has negligible runtime overhead. For compression, we change the way to store some frames without introducing any extra computation. For decompression, in the worst-case scenario when all frames in a batch are compressed by LCP-T, the entire batch needs to be decompressed to get the last frame, while our solution only needs one more decompression for the anchor frame. On the other hand, as shown in~\Cref{fig:s_t_selection}, our solution achieves optimal compression results while adhering to the batch compression constraints.


\subsection{Dynamic Optimizations}
In this section, we discuss the two dynamic optimization strategies in LCP which determine the best block size and error-bound scale.

\subsubsection{Best block size}
\label{sec:block_size_in_a_single_frame}
\begin{figure}[ht] \centering
\hspace{-8mm}
\subfigure[USGS, eb=1e-3]
{
\raisebox{-1cm}{\includegraphics[scale=0.25]{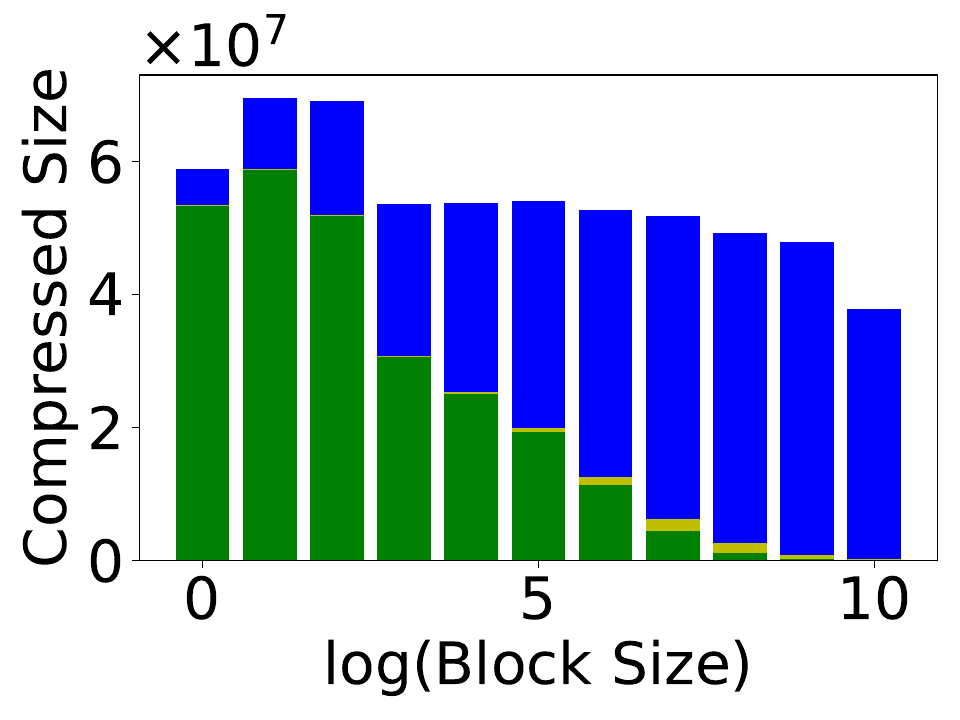}}
}
\hspace{-4mm}
\subfigure[LJ, eb=1e-3]
{
\raisebox{-1cm}{\includegraphics[scale=0.25]{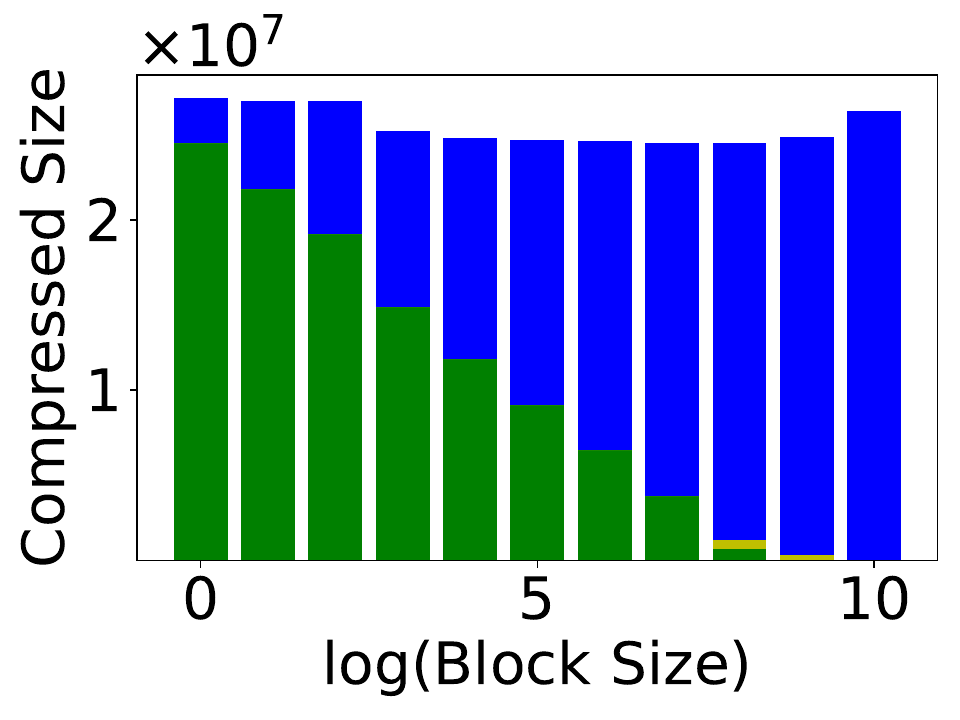}}
}
\hspace{-8mm}

\includegraphics[scale=0.25]{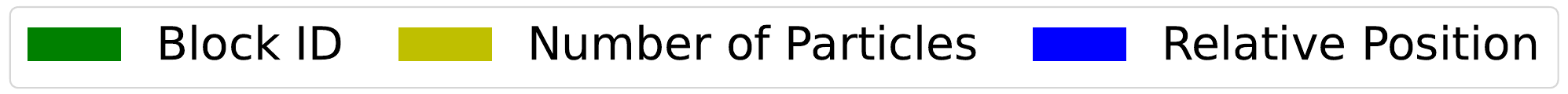}

\vspace{-2mm}
\caption{Effect of block size on LCP-S, with a breakdown view of the compressed size of the three components introduced in~\Cref{sec: lcp-s block components} (log base 2)}
\label{fig:block_size_cr}
\vspace{-2mm}
\end{figure}

\begin{figure}[ht] \centering

\includegraphics[scale=0.25]{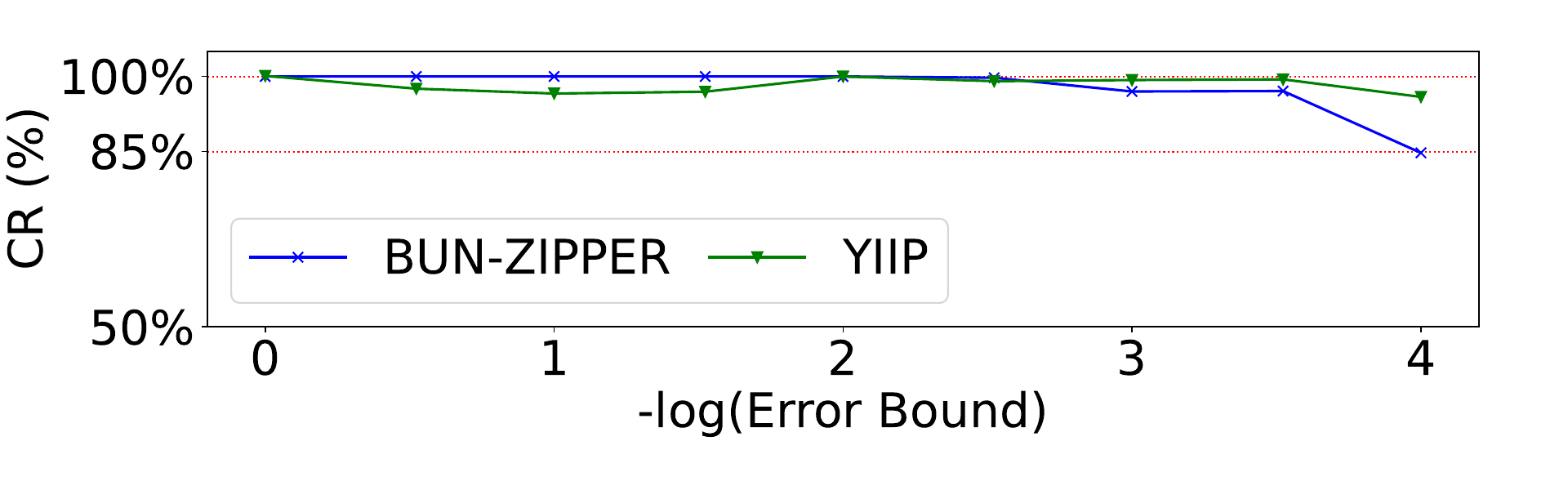} 
\caption{Our block size optimization strategy is highly effective for different inputs and error bound settings (CR(\%) = CR with optimized block size / best CR from extensive offline block size search, log base 10)}
\label{fig:block_size_est_methods}
\vspace{-2mm}

\end{figure}

The choice of block size can significantly impact the compression ratio, with the impact varying on different data input, as shown in~\Cref{fig:block_size_cr}. Therefore a solution is needed to identify the best block size for the specific data input. Traditional algorithms like binary or ternary search are unsuitable because the relationship between block size and compression ratio is not monotonous or unimodal. Instead, we first perform an offline analysis to identify the candidates of optimal block size which are $2^k, 0 \le k \le 16$. Those candidates have higher chances to become the best option than other choices based on our comprehensive testing. Then we evaluate those candidates on the sampled input to find the best choice of block size. As demonstrated in~\Cref{fig:block_size_est_methods}, our solution is highly effective, reaching at least 85\% of the compression ratio of the theoretical best block size in most cases.

\subsubsection{Best error-bound scale for anchor frames}
\label{sec:error_bound_in_multiple_frames}

\begin{figure}[ht] \centering

\hspace{-8mm}
\subfigure[Copper]
{
\raisebox{-1cm}{\includegraphics[scale=0.25]{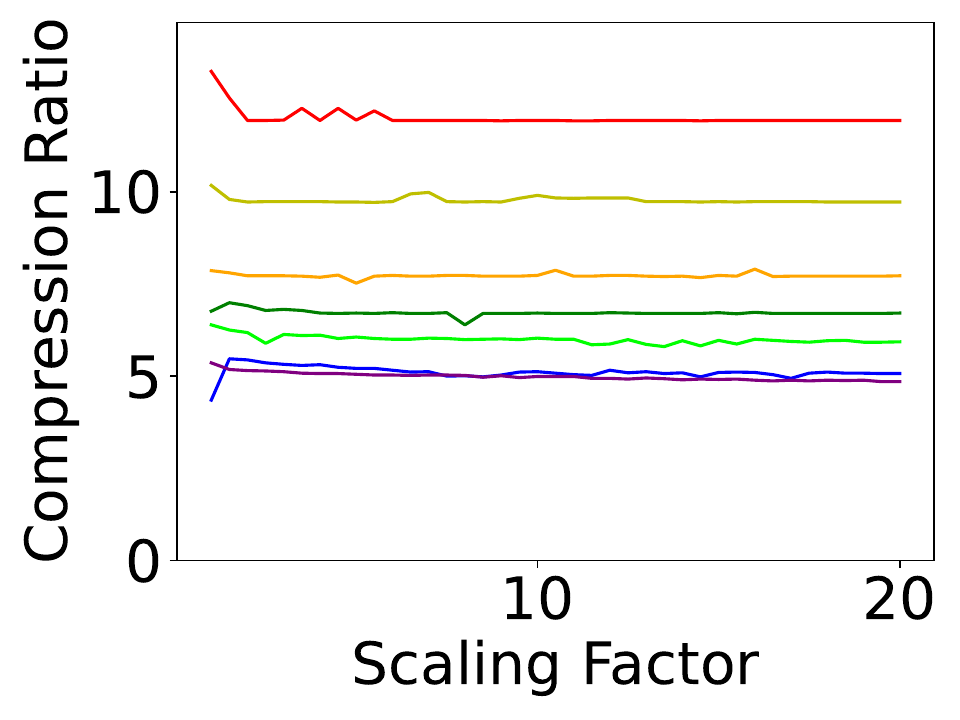}}
}
\hspace{-4mm}
\subfigure[Helium]
{
\raisebox{-1cm}{\includegraphics[scale=0.25]{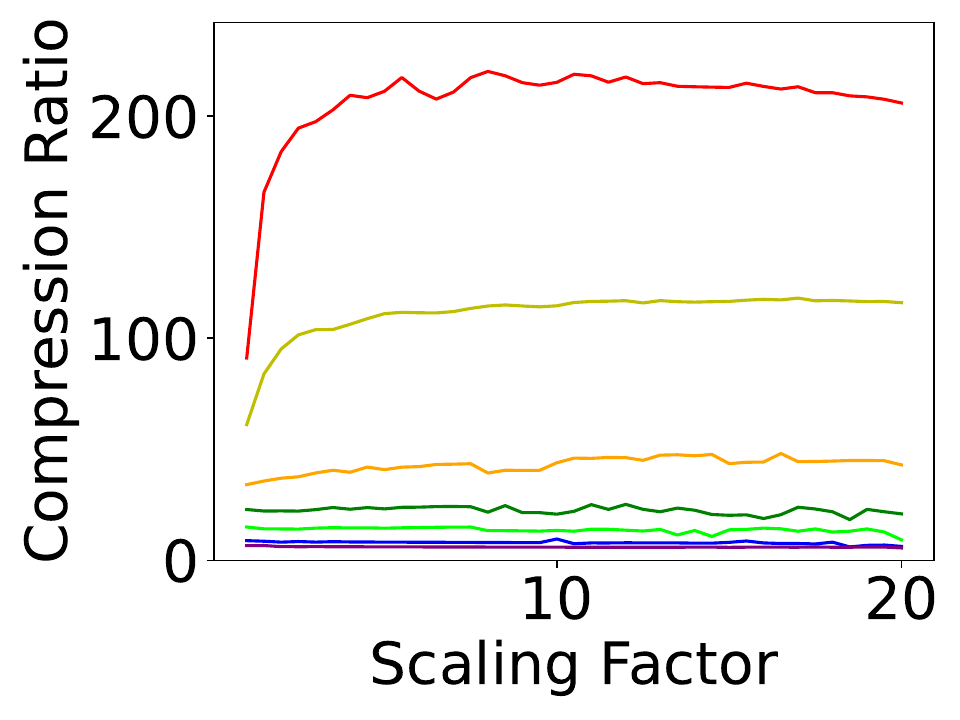}}
}
\hspace{-8mm}

\includegraphics[scale=0.25]{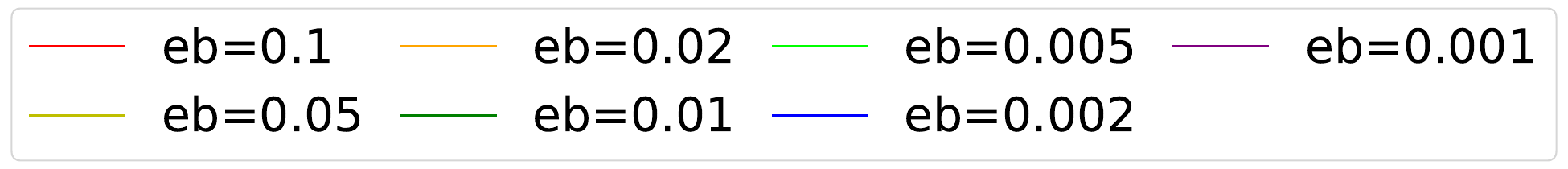}

\vspace{-2mm}
\caption{Effect of scaling error bound of anchor frames on the overall compression ratios. (error bound scaling factor = user set error bound / error bound used by LCP-S)}
\label{fig:fflag_cr}
\vspace{-2mm}

\end{figure}

LCP may increase the compression precision (lower the error bound) of anchor frames toward overall better compression ratios.
For data with high temporal correlation, most of the frames will be compressed by LCP-T, except for the anchor frames which have to be compressed by LCP-S. The temporal compression accuracy relies on frame similarities. By increasing the compression precision of anchor frames, the similarity between anchor frames and to-be-temporal-compressed frames will be less impacted by the compression error of anchor frames, such that the temporal compression will be more effective. As demonstrated in~\Cref{fig:fflag_cr}, when scaling the error bound to a smaller value, the compression ratio improves, but the rate of improvement diminishes after a certain threshold, so we choose the value of 5 as the error bound scale factor. On the other hand, this scaling method is limited to data with high temporal correlation, because for other cases, increasing the precision of many anchor frames would not offset the compressed size saving from temporal frames. As a result, we will dynamically analyze input data and apply this error-bound scaling method selectively based on the temporal correlation of frames.



\section{Experimental Evaluation}
\label{sec:evaluation}

In this section, we present the experimental settings and the evaluation results of our solution on eight particle datasets compared with eight state-of-the-art baselines. 

\subsection{Experimental Setting}

\subsubsection{Execution Environment} The experiments are performed on the Purdue Anvil supercomputer~\cite{anvil} through NSF ACCESS~\cite{access}. Each computing node in Anvil features two AMD EPYC 7763 CPUs with 64 cores at a 2.45GHz clock rate and 256 GB DDR4-3200 RAM. 

\subsubsection{Datasets} The experiments are evaluated on seven scientific particle datasets and one 3D model. The details of the datasets can be found in~\Cref{tab:apps}.
We evaluate the locations fields (x, y, z) of all the datasets, as discussed in~\Cref{sec: background particle fields}. For terabyte-scale datasets (3DEP, HACC, WarpX), we use subsets $\geq$ 100 GB for evaluation. Regarding the time domain, the Copper, Helium, LJ, and YIIP datasets include multiple frames. HACC and WarpX contain several frames, but their analyses typically treat each frame separately thus compression needs to be done per frame. The rest datasets consist of a single frame each.




\subsubsection{State-of-the-Art Lossy Compressors in Our Evaluation}
We compare our solution with eight leading lossy compressors.

\begin{itemize}
    \item SZ2 (git version f466775)~\cite{liangErrorControlledLossyCompression2018}: a generic scientific error-bounded lossy compressor framework with Lorenzo and regression predictors. 
    \item SZ3 (git version 90c66be)~\cite{interp, sz3}: the next generation compressor after SZ2. Its interpolation prediction algorithm is more effective than many compressors including SZ2 and Mgard+~\cite{mgardx} on mesh data~\cite{interp}. However, SZ3 has sub-optimal results on particle data which further indicates the importance of designing compression algorithms for particles.
    \item MDZ (git version 90c66be)~\cite{mdz}: a molecular dynamics particle compressor that mainly takes advantage of the specific features of solid material MD simulations. Note that in multi-frame compression, MDZ treats the first frame as metadata and thus does not include its size in the compressed format. In this evaluation, we count all metadata sizes as well.
    \item ZFP (v1.0.1)~\cite{zfp}: a fast scientific lossy compressor based on orthogonal transformation.
    \item SPERR (v0.8.1)~\cite{sperr}: a scientific lossy compressor based on wavelet transform, offering better compression ratios in some cases but slower speeds due to the transform’s complexity.
    \item Draco (v1.5.7)~\cite{draco}: a lossy compressor developed by Google for 3D geometric meshes and point clouds. Draco does not support arbitrary error bound -- users can only specify the number of bits before compression without knowing the corresponding error range.
    \item TMC13 (git version a3d15c5)~\cite{gpcc}: Developed by the MPEG group, TMC13 is the implementation of the MPEG geometry-based point cloud compression standard (MPEG G-PCC). TMC13 supports both intra and inter prediction.
    \item TMC2 (git version 5d9900a)~\cite{vpcc}: TMC2 is developed by the MPEG group. It follows the MPEG video-based point cloud compression standard (MPEG V-PCC). TMC2 converts particles to 2D video frames and leverages existing video codecs (like H.265~\cite{h265}) for compression. 
\end{itemize}


\subsection{Evaluation Results and Analysis}
The evaluation covers three aspects -- ablation study, compression quality \& fidelity, and speed. To be more specific, the ablation study proves the essential contributions of LCP's components. The evaluation of compression error, compression ratio, rate distortion, and visual quality shows that our solution LCP has better compression quality and fidelity than the state-of-the-art lossy compressors in most cases. The speed evaluation demonstrates that LCP has the best speed over all other compressors for efficient data retrieval. Those evaluations also cover non-scientific datasets to show the strong generalizability of LCP.

We exclude TMC2's results from this section due to several fundamental limitations that make it unsuitable for compressing scientific particles. First, TMC2 cannot guarantee the same point count between the original and reconstructed data, which is a critical requirement for scientific particle simulations and analysis. Second, its use of 16-bit integers for quantization is not versatile enough to accommodate the precision requirements of particle data (LCP supports arbitrary bounds). Third, finding optimal parameters (such as the settings to maintain the original particle count) requires substantial manual tuning effort. 
Our preliminary results on two datasets (BUN-ZIPPER and Copper) revealed that TMC2 is 200 $\sim$ 50,000 times slower than LCP with a lower compression ratio in most error bounds.
Given such limitations and drawbacks, we omit TMC2 from the evaluation sections below.


\begin{figure}[ht]

\vspace{-3mm}
    \centering
    \hspace{-8mm}
    \subfigure[Helium]
    {
        \raisebox{-1cm}{\includegraphics[scale=0.08]{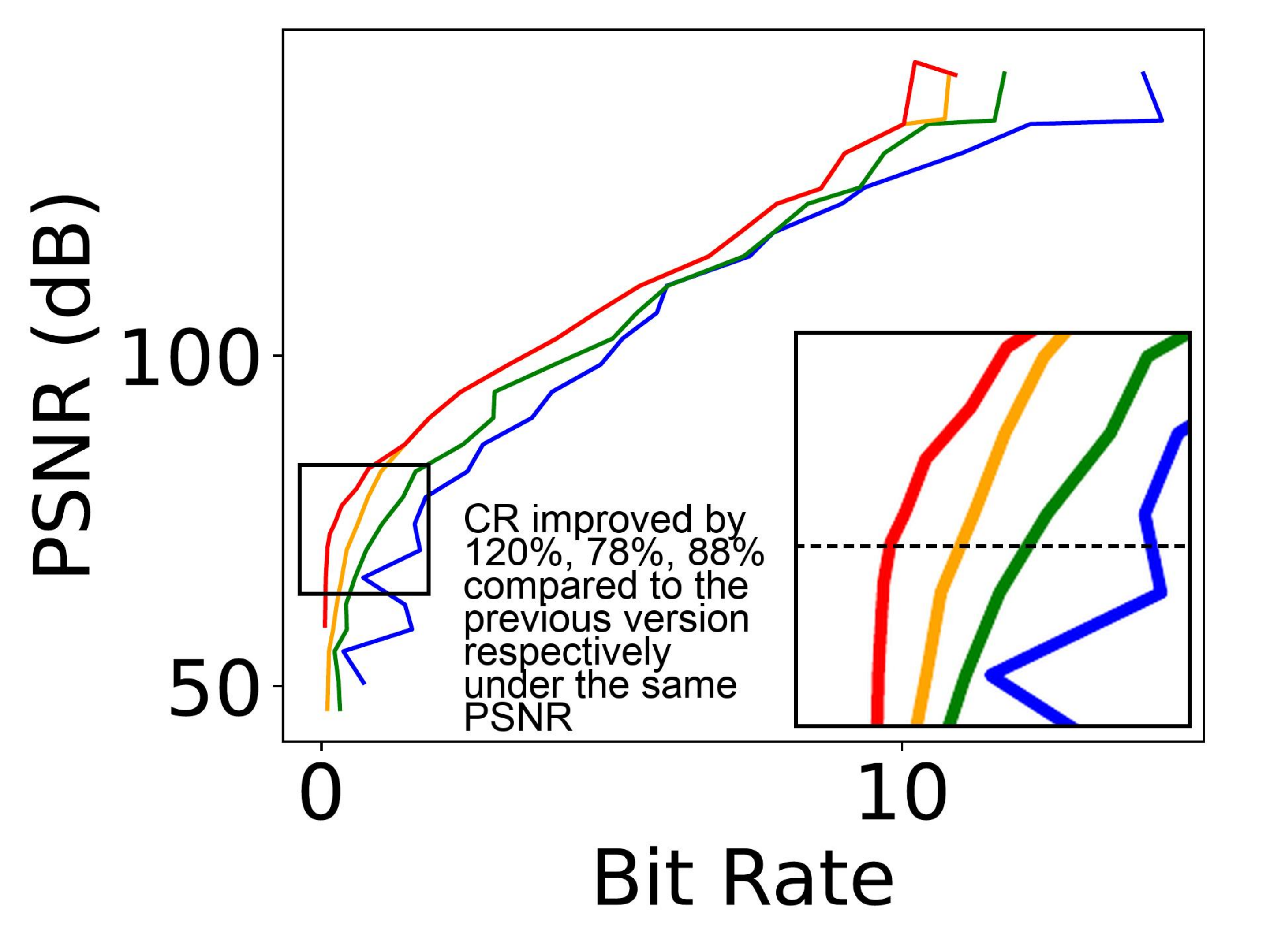}}
    }
    \hspace{-4mm}
    \subfigure[LJ]
    {
        \raisebox{-1cm}{\includegraphics[scale=0.25]{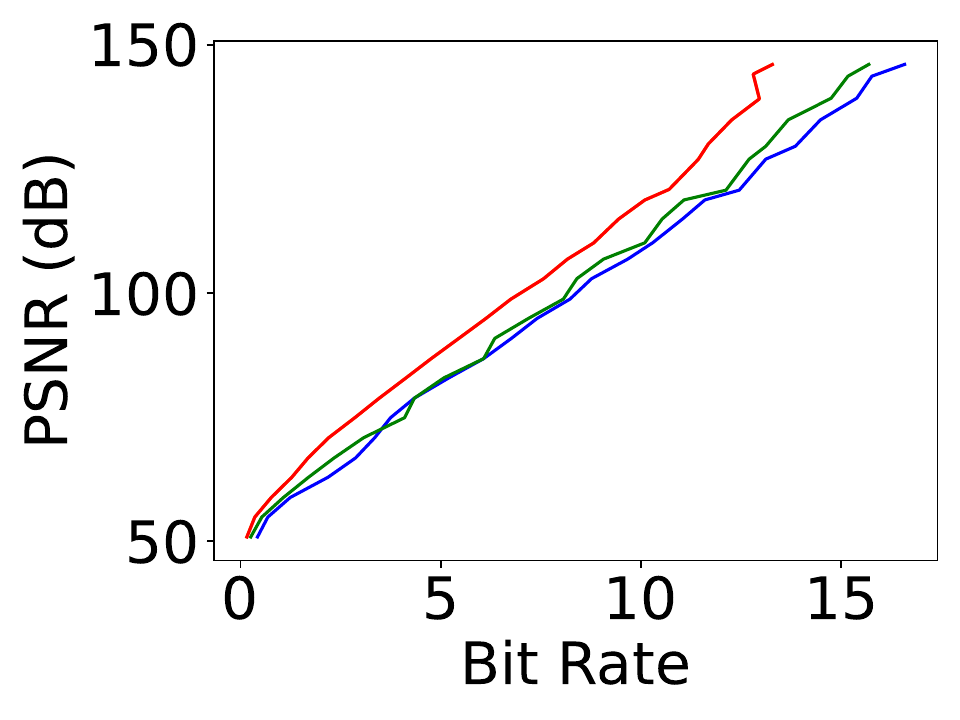}}
    }
    \hspace{-8mm}
    \vspace{-3mm}

    \hspace{-8mm}
    \subfigure[Helium (eb=5e-2)]
    {
        \raisebox{-1cm}{\includegraphics[scale=0.25]{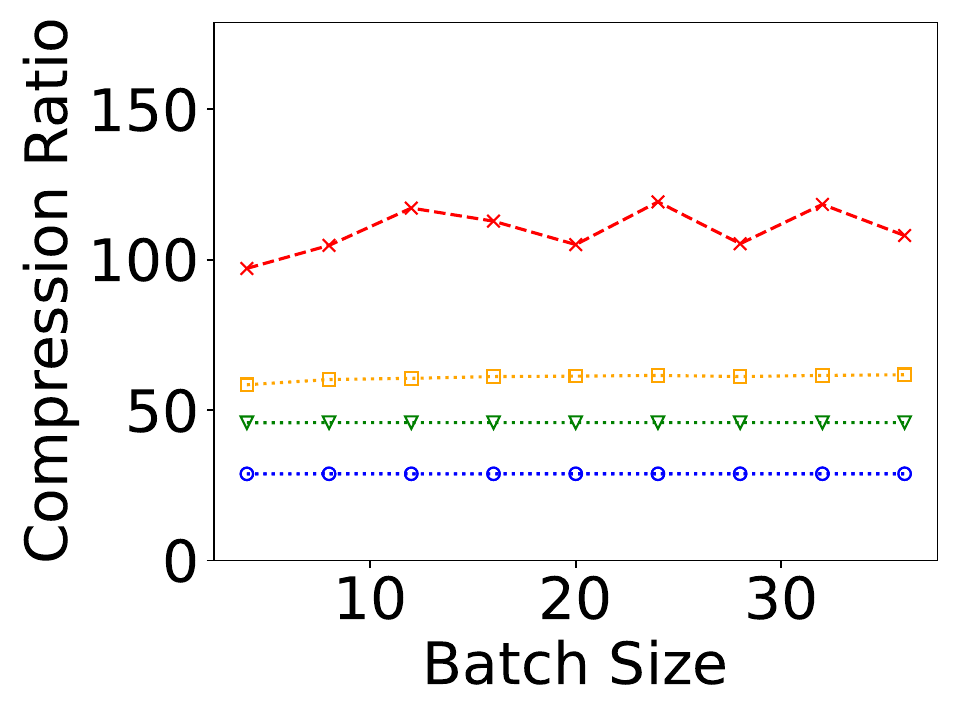}}
    }
    \hspace{-4mm}
    \subfigure[LJ (eb=1e-1)]
    {
        \raisebox{-1cm}{\includegraphics[scale=0.25]{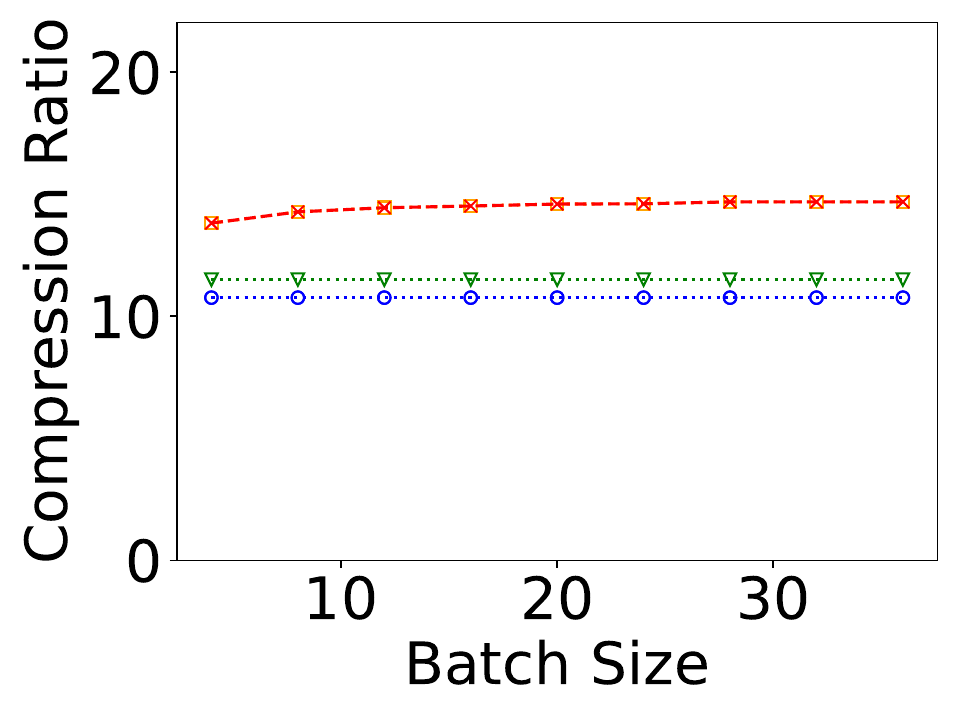}}
    }
    \hspace{-8mm}

    \vspace{3mm}
    \includegraphics[scale=0.25]{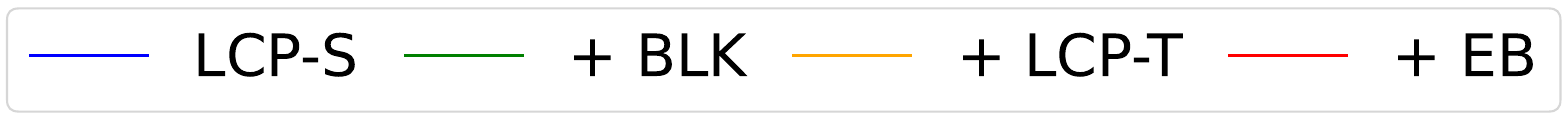}

    \vspace{-2mm}
    \caption{Ablation study demonstrates the essential contributions of each LCP component. LCP-S:~\Cref{sec: design lcp-s} spatial compressor; BLK:~\Cref{sec:block_size_in_a_single_frame} dynamic block size optimization; LCP-T:~\Cref{sec: hybrid lcp s and t} hybrid compressor with LCP-S and LCP-T; EB:~\Cref{sec:error_bound_in_multiple_frames} best error-bound scale for anchor frame}
    \label{fig:version_iteration_comparison}
    \vspace{-2mm}

\end{figure}


%





\begin{figure}[ht] \centering
\includegraphics[scale=0.25]{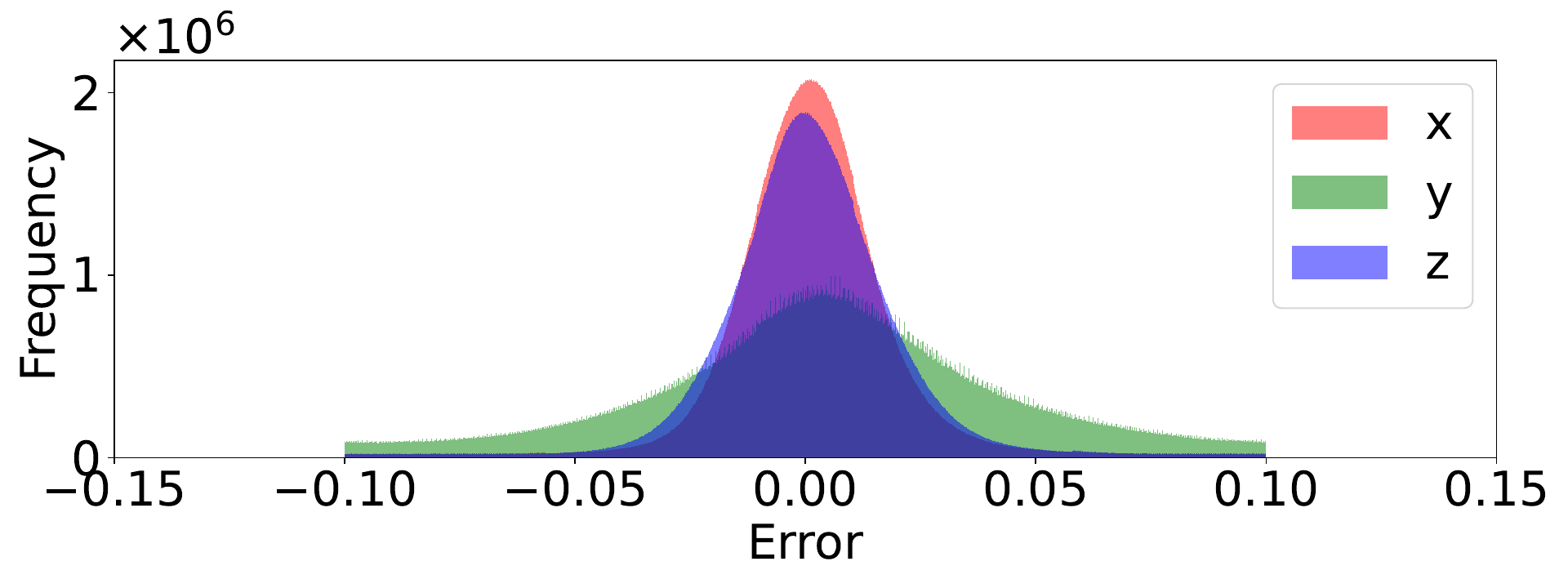}
\vspace{-2mm}
\caption{The error distribution graph shows LCP strictly keeps the actual compression error under the user pre-defined bound (Helium,  error bound =  0.1)}
\label{fig:error_dist}
\vspace{-2mm}
\end{figure}

\begin{figure}[ht] \centering

\includegraphics[scale=0.35]{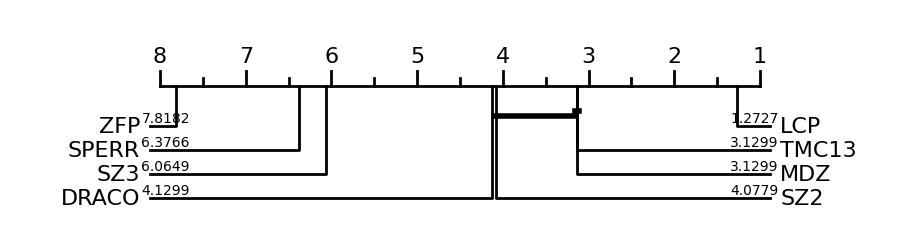}
\vspace{-2mm}
\caption{The critical difference diagram ranks compressors based on compression ratios across various settings and datasets. LCP ranks first among all baselines (a higher ranking indicates a better compression ratio)}
\label{fig:cd}
\vspace{-2mm}

\end{figure}

\begin{figure}[ht] \centering

\hspace{-8mm}
\subfigure[Helium (eb=1e-2)]
{
\raisebox{-1cm}{\includegraphics[scale=0.25]{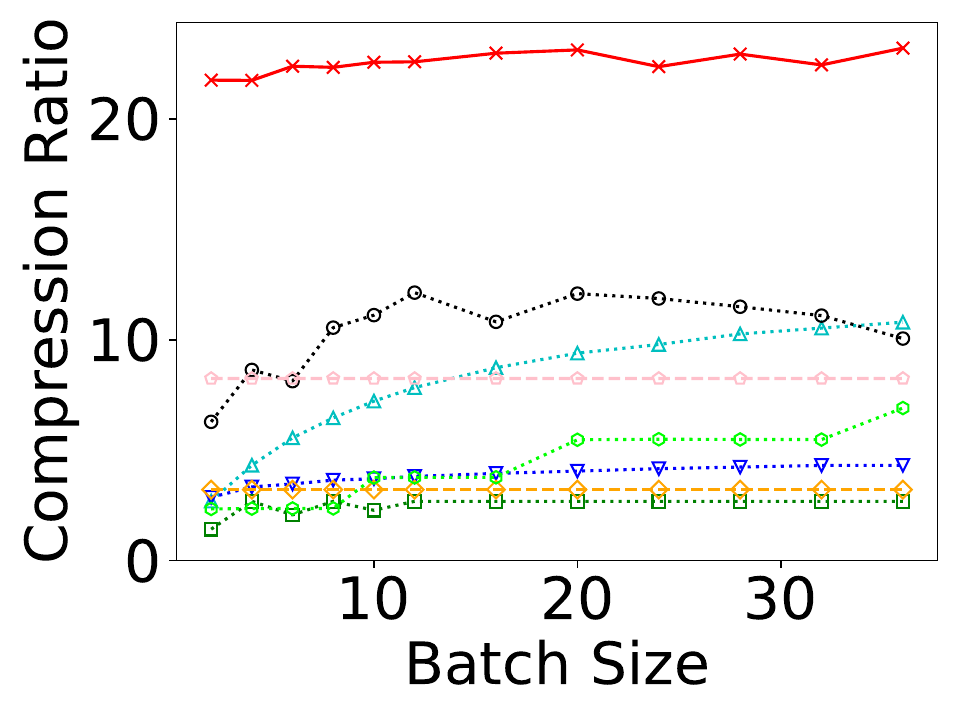}}
}
\hspace{-4mm}
\subfigure[Copper (eb=1e-3)]
{
\raisebox{-1cm}{\includegraphics[scale=0.25]{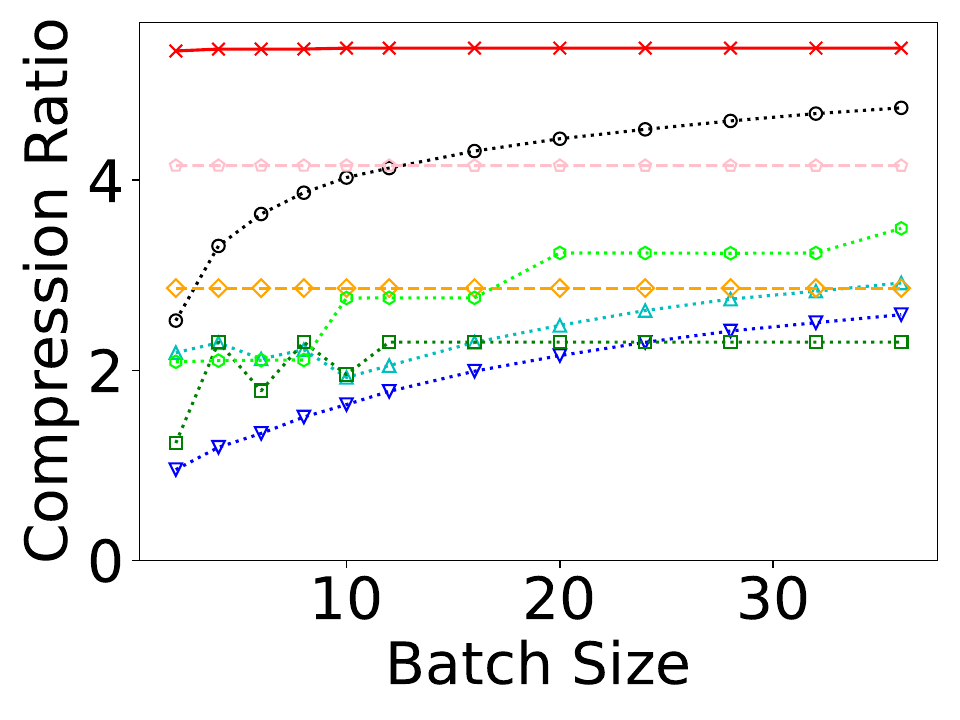}}
}
\hspace{-8mm}
\vspace{-3mm}

\hspace{-8mm}
\subfigure[LJ (eb=1e-2)]
{
\raisebox{-1cm}{\includegraphics[scale=0.25]{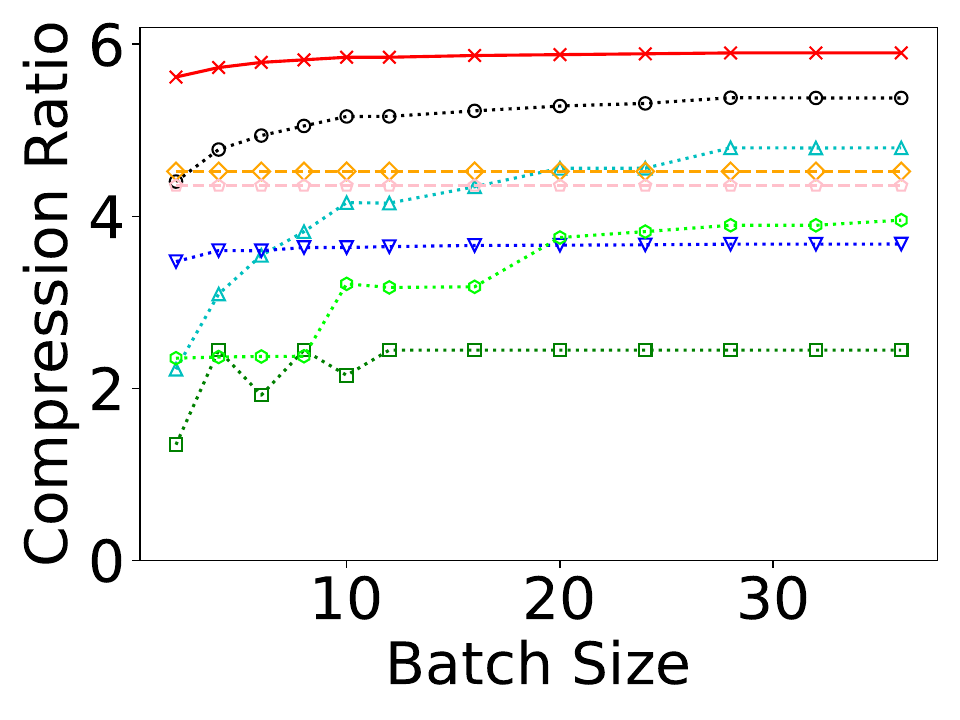}}
}
\hspace{-4mm}
\subfigure[YIIP (eb=1e-1)]
{
\raisebox{-1cm}{\includegraphics[scale=0.25]{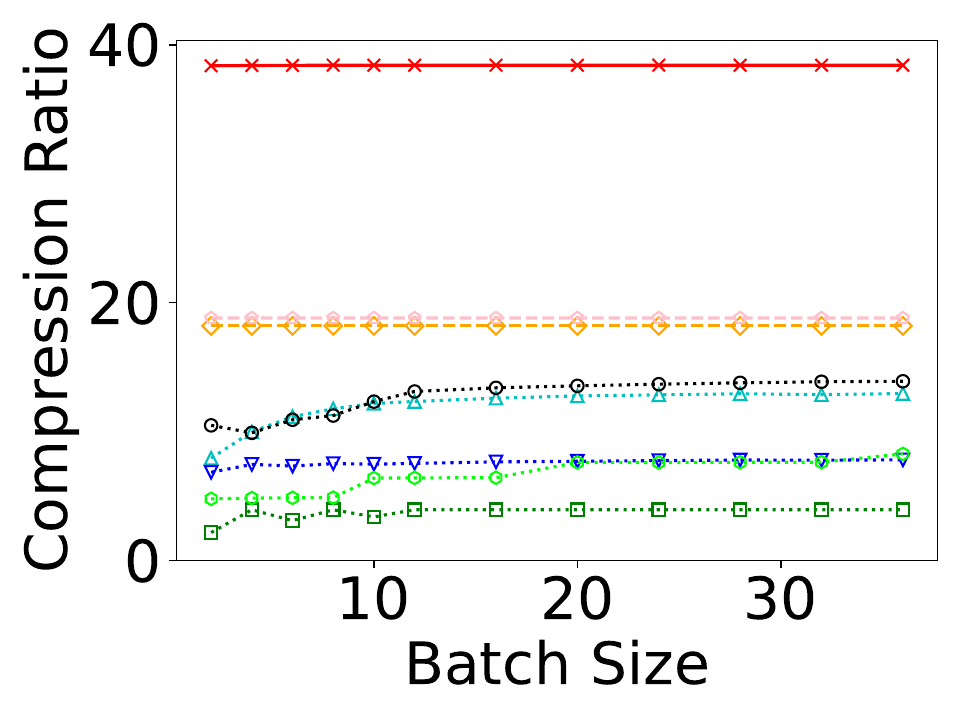}}
}
\hspace{-8mm}
\vspace{-2mm}

\vspace{3mm}
\includegraphics[scale=0.25]{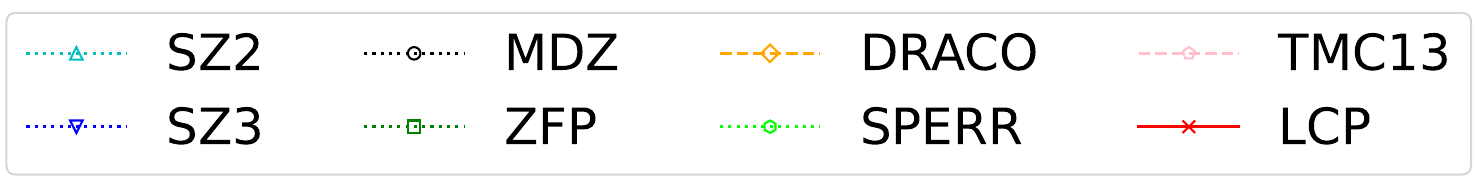}

\caption{LCP has the highest compression ratio on all multi-frame datasets and under different batch sizes and error-bound settings}
\label{fig:various_batch}
\vspace{-2mm}

\end{figure}

\begin{figure}[ht] \centering

\hspace{-8mm}
\subfigure[Helium]
{
\raisebox{-1cm}{\includegraphics[scale=0.25]{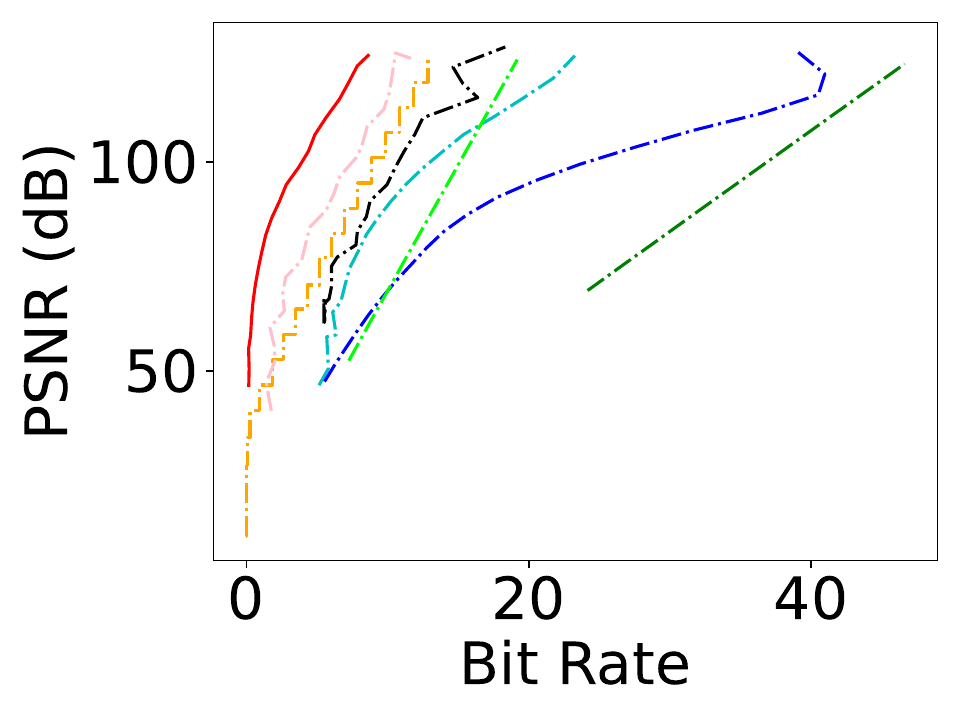}}
}
\hspace{-4mm}
\subfigure[Warpx]
{
\raisebox{-1cm}{\includegraphics[scale=0.25]{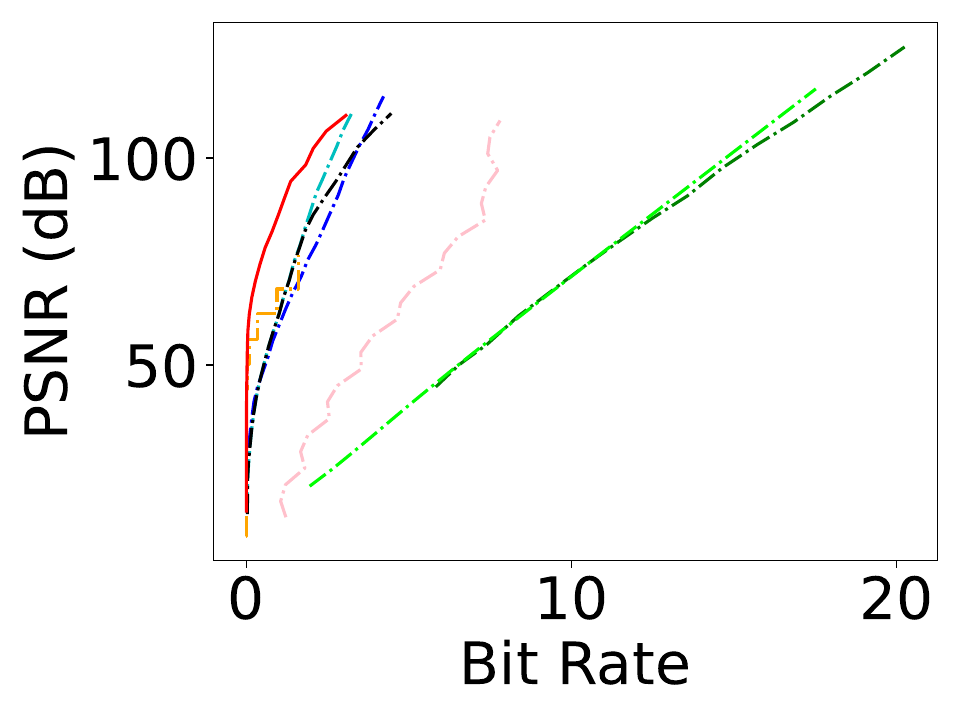}}
}
\hspace{-8mm}
\vspace{-3mm}

\hspace{-8mm}
\subfigure[Copper]
{
\raisebox{-1cm}{\includegraphics[scale=0.25]{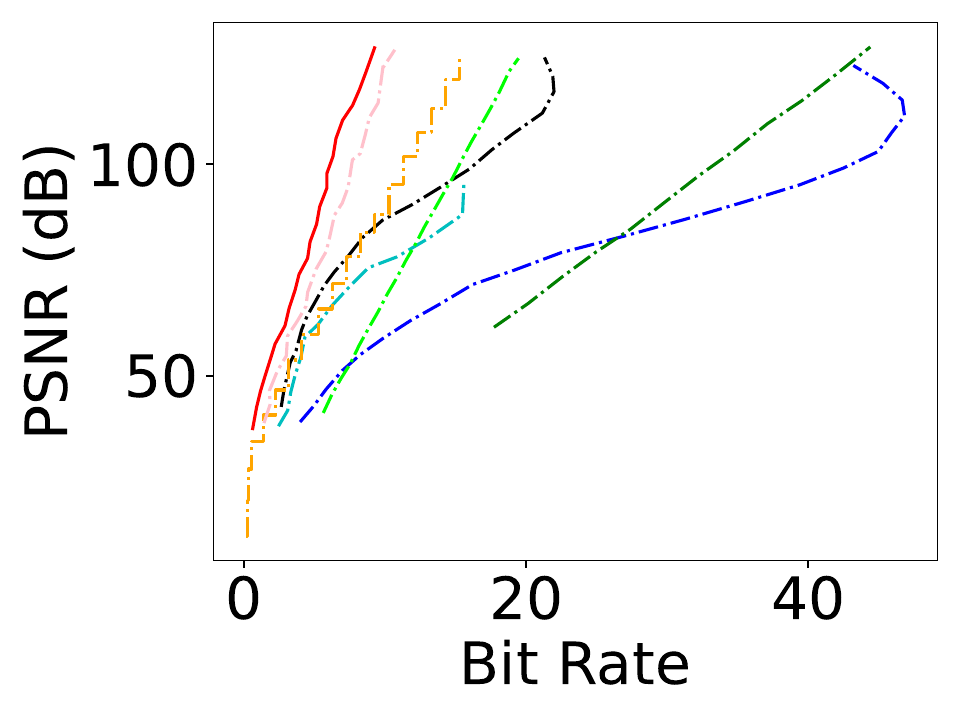}}
}
\hspace{-4mm}
\subfigure[HACC]
{
\raisebox{-1cm}{\includegraphics[scale=0.25]{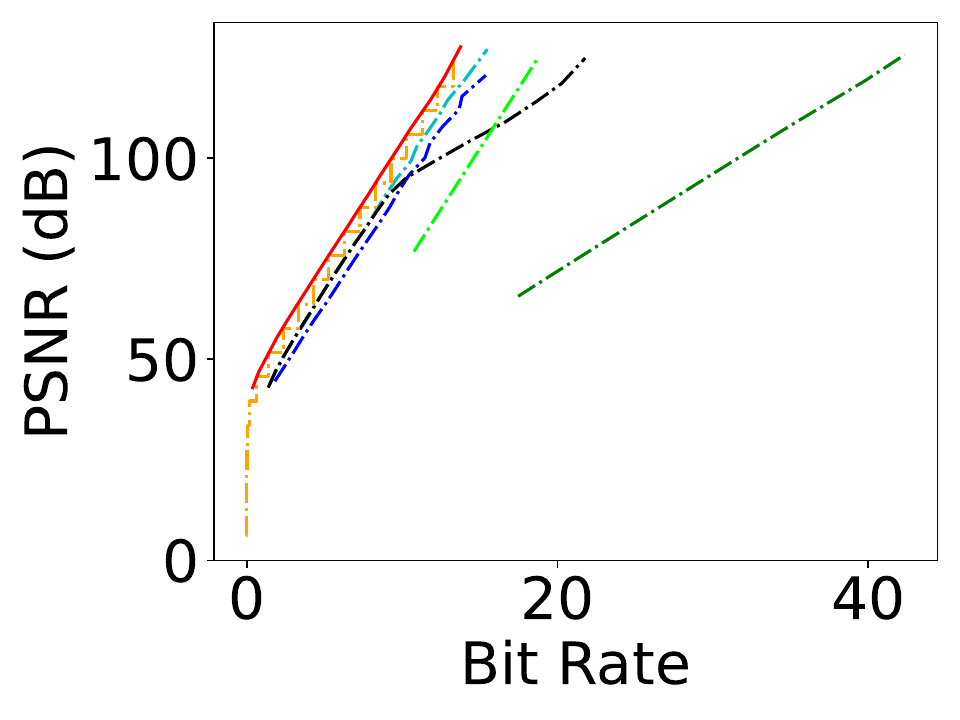}}
}
\hspace{-8mm}
\vspace{-3mm}

\hspace{-8mm}
\subfigure[LJ]
{
\raisebox{-1cm}{\includegraphics[scale=0.25]{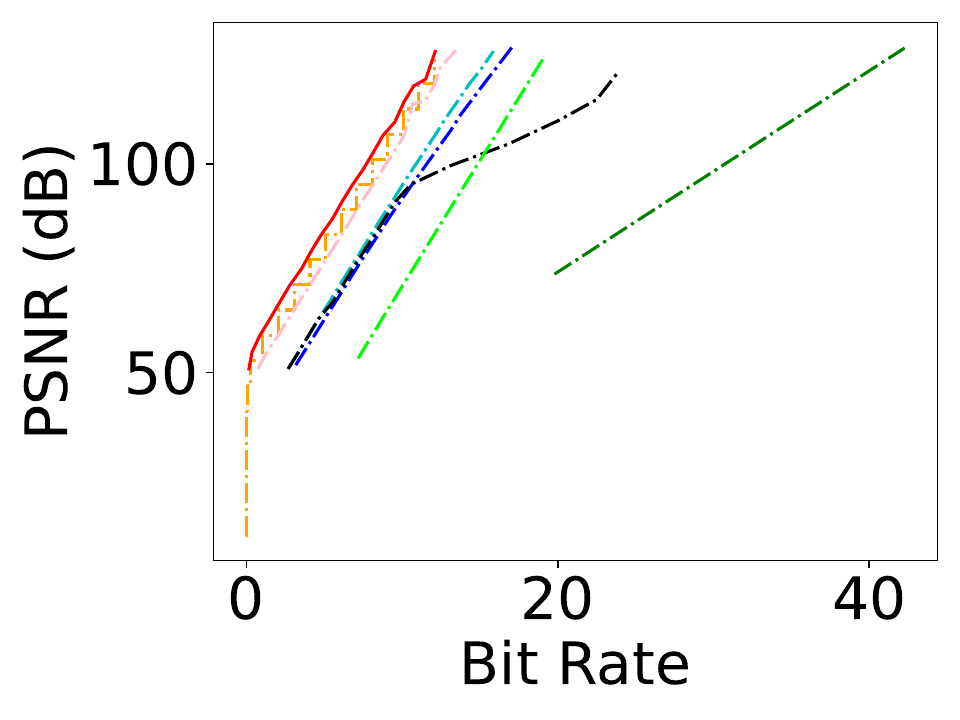}}
}
\hspace{-4mm}
\subfigure[YIIP]
{
\raisebox{-1cm}{\includegraphics[scale=0.25]{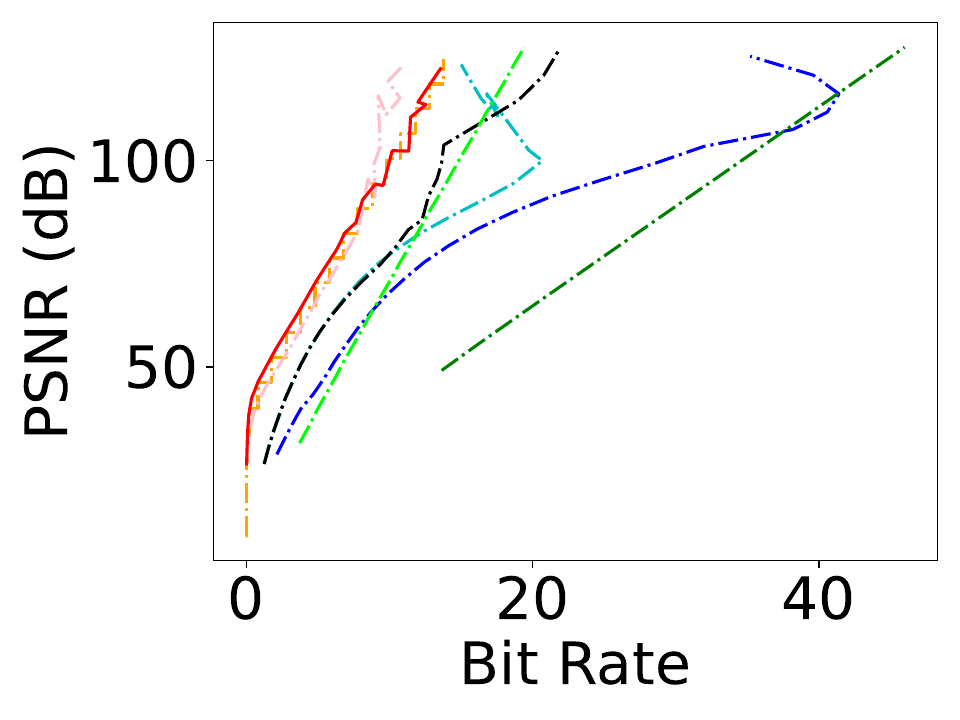}}
}
\hspace{-8mm}
\vspace{-3mm}

\hspace{-8mm}
\subfigure[BUN-ZIPPER]
{
\raisebox{-1cm}{\includegraphics[scale=0.25]{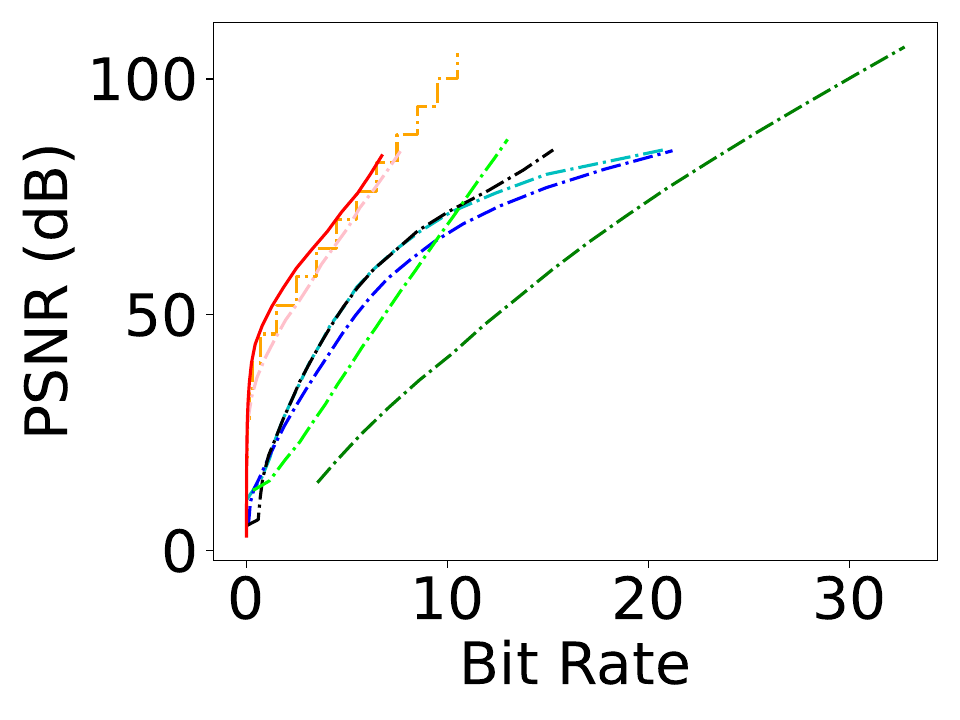}}
}
\hspace{-4mm}
\subfigure[3DEP]
{
\raisebox{-1cm}{\includegraphics[scale=0.25]{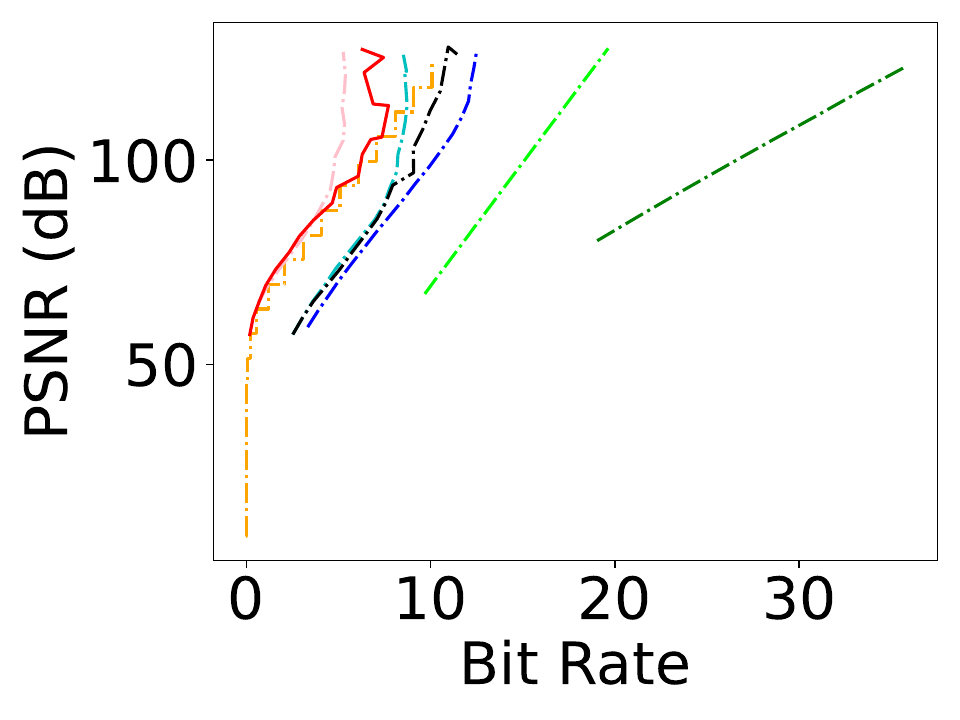}}
}
\hspace{-8mm}

\includegraphics[scale=0.25]{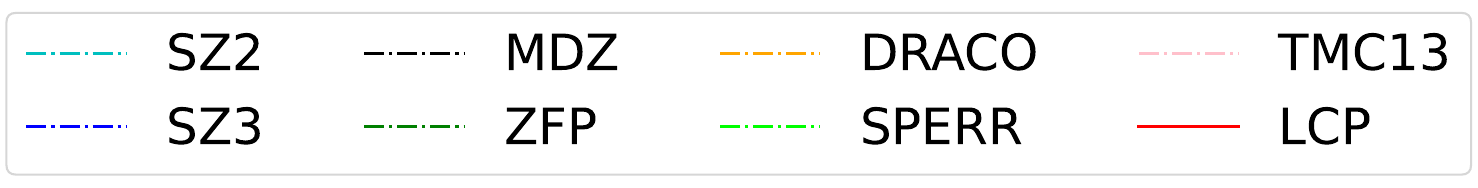}

\vspace{-2mm}
\caption{In single frame cases (spatial compression only), LCP demonstrates the best compression quality  on the rate-distortion graphs (lower bit rate and higher PSNR indicate better compression quality)}
\label{fig:br_psnr_1}
\vspace{-2mm}

\end{figure}

\begin{figure}[ht] \centering
\hspace{-8mm}
\subfigure[Helium]
{
\raisebox{-1cm}{\includegraphics[scale=0.25]{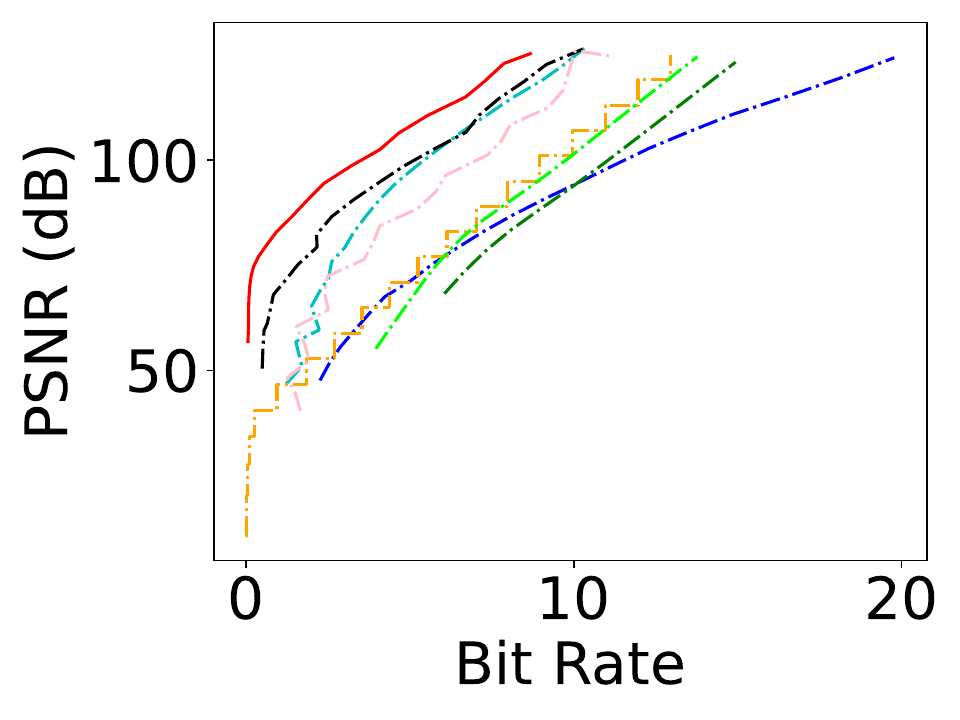}}
}
\hspace{-4mm}
\subfigure[Copper]
{
\raisebox{-1cm}{\includegraphics[scale=0.25]{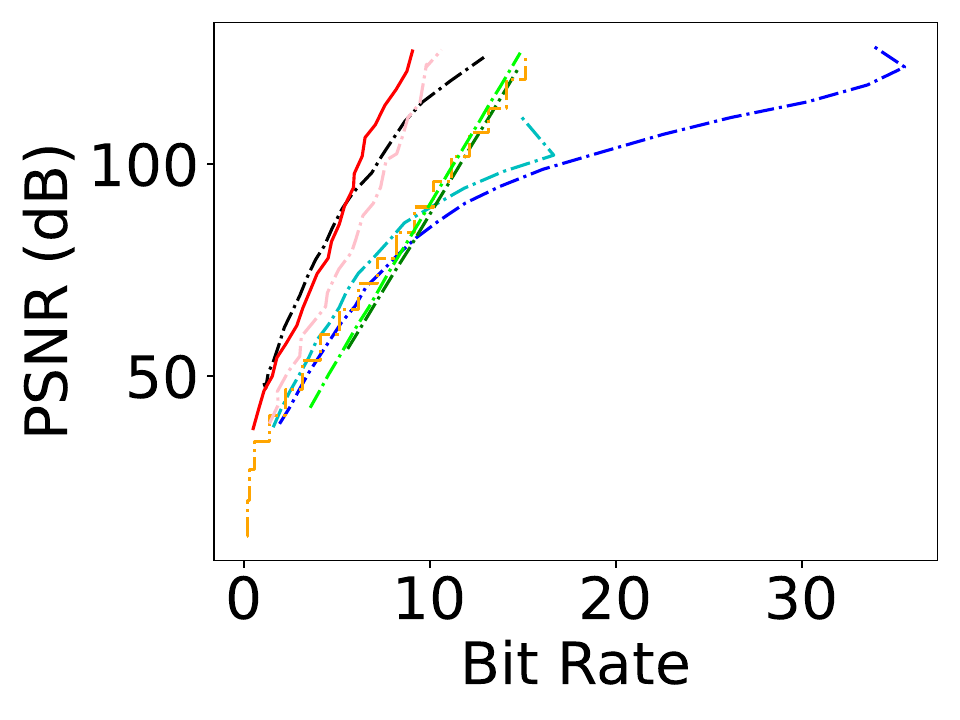}}
}
\hspace{-8mm}
\vspace{-3mm}


\hspace{-8mm}
\subfigure[LJ]
{
\raisebox{-1cm}{\includegraphics[scale=0.25]{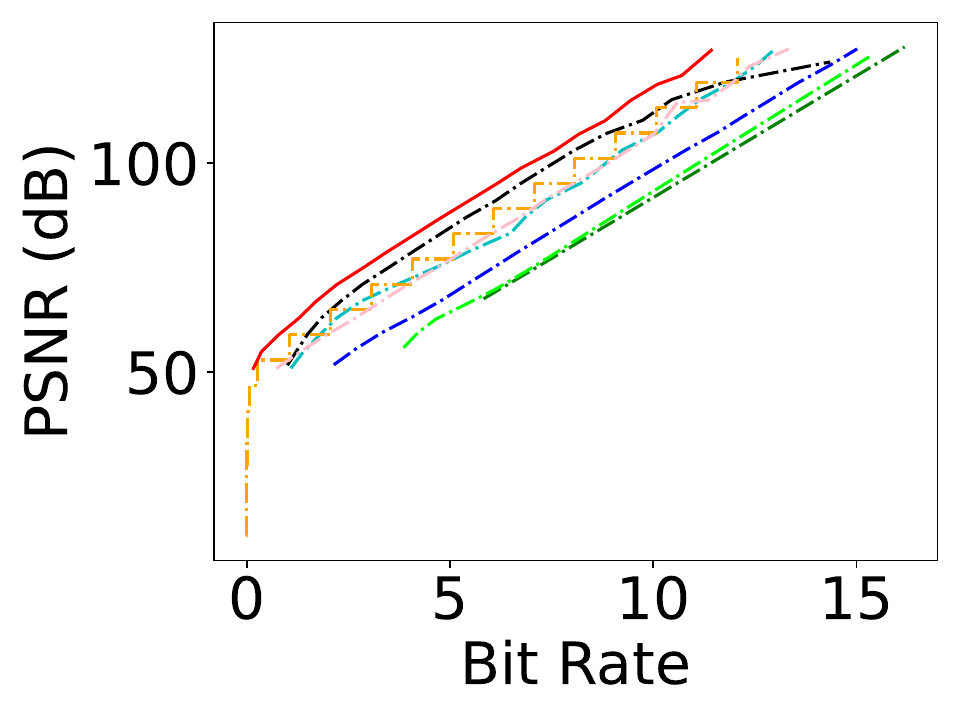}}
}
\hspace{-4mm}
\subfigure[YIIP]
{
\raisebox{-1cm}{\includegraphics[scale=0.25]{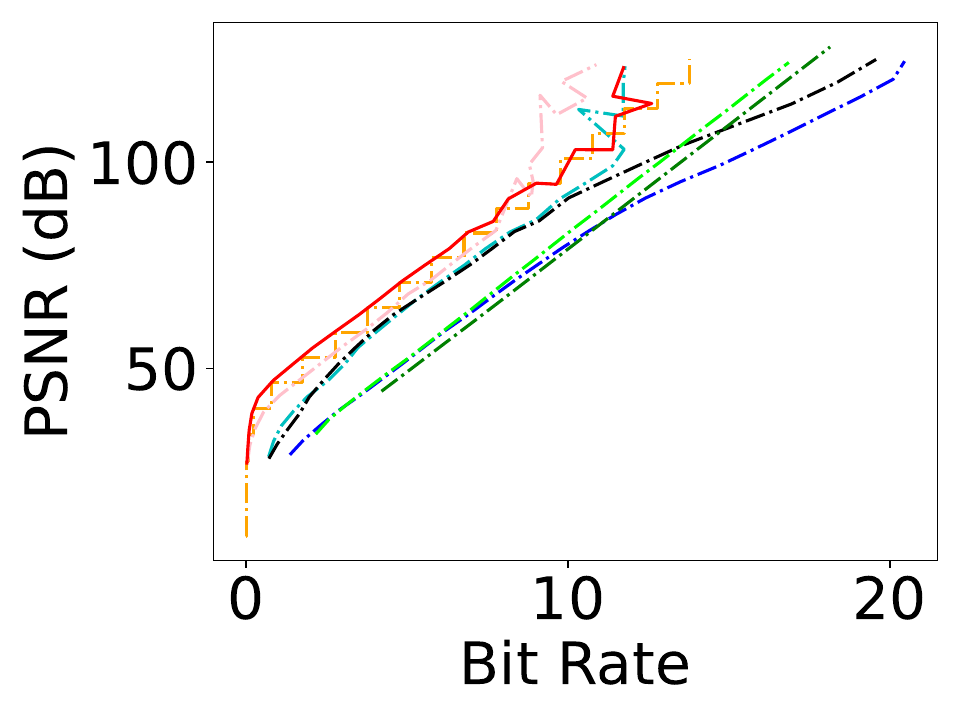}}
}
\hspace{-8mm}

\includegraphics[scale=0.25]{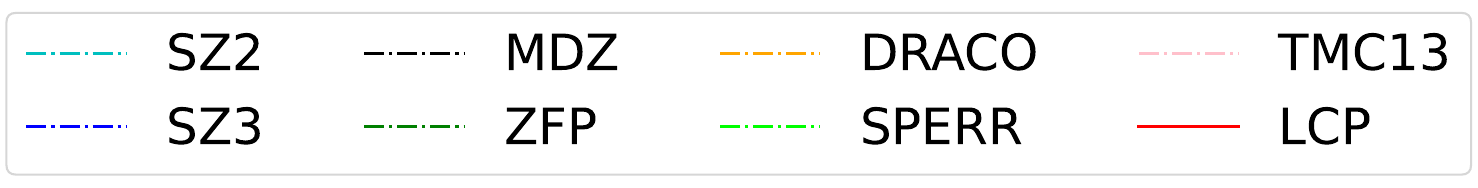}

\vspace{-2mm}
\caption{In multi-frame cases (spatial and temporal compression), LCP has the best compression quality showing on the rate-distortion graphs (batch Size = 16)}
\label{fig:br_psnr_16}
\vspace{-2mm}
\end{figure}

\begin{figure*}[ht] \centering

\vspace{-3mm}

\hspace{0mm}
\subfigure[{Original}]
{
\raisebox{-1cm}{\includegraphics[scale=0.06]{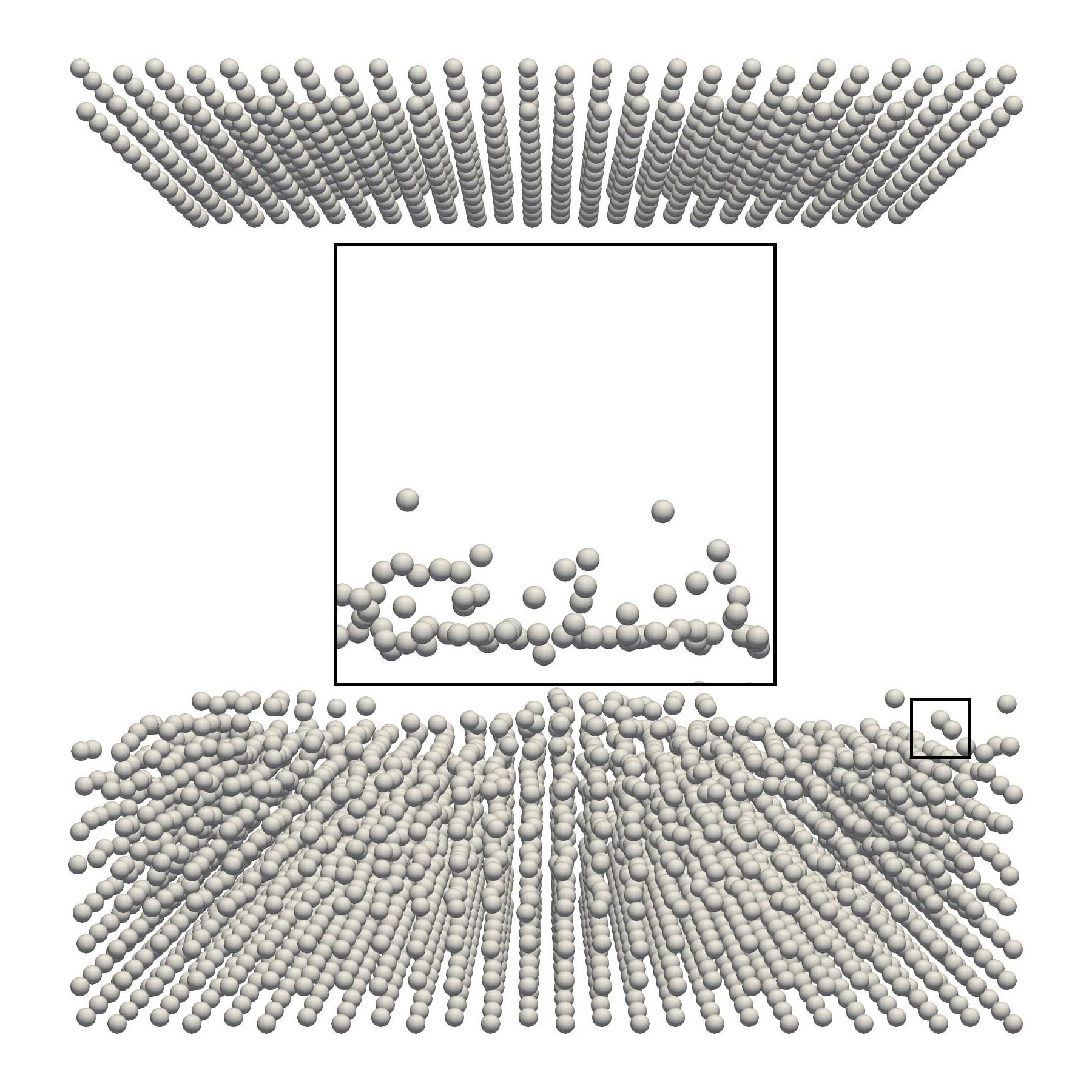}}
}
\hspace{1mm}
\subfigure[{SZ2 (CR=13.12, PSNR=38.07})] 
{
\raisebox{-1cm}{\includegraphics[scale=0.06]{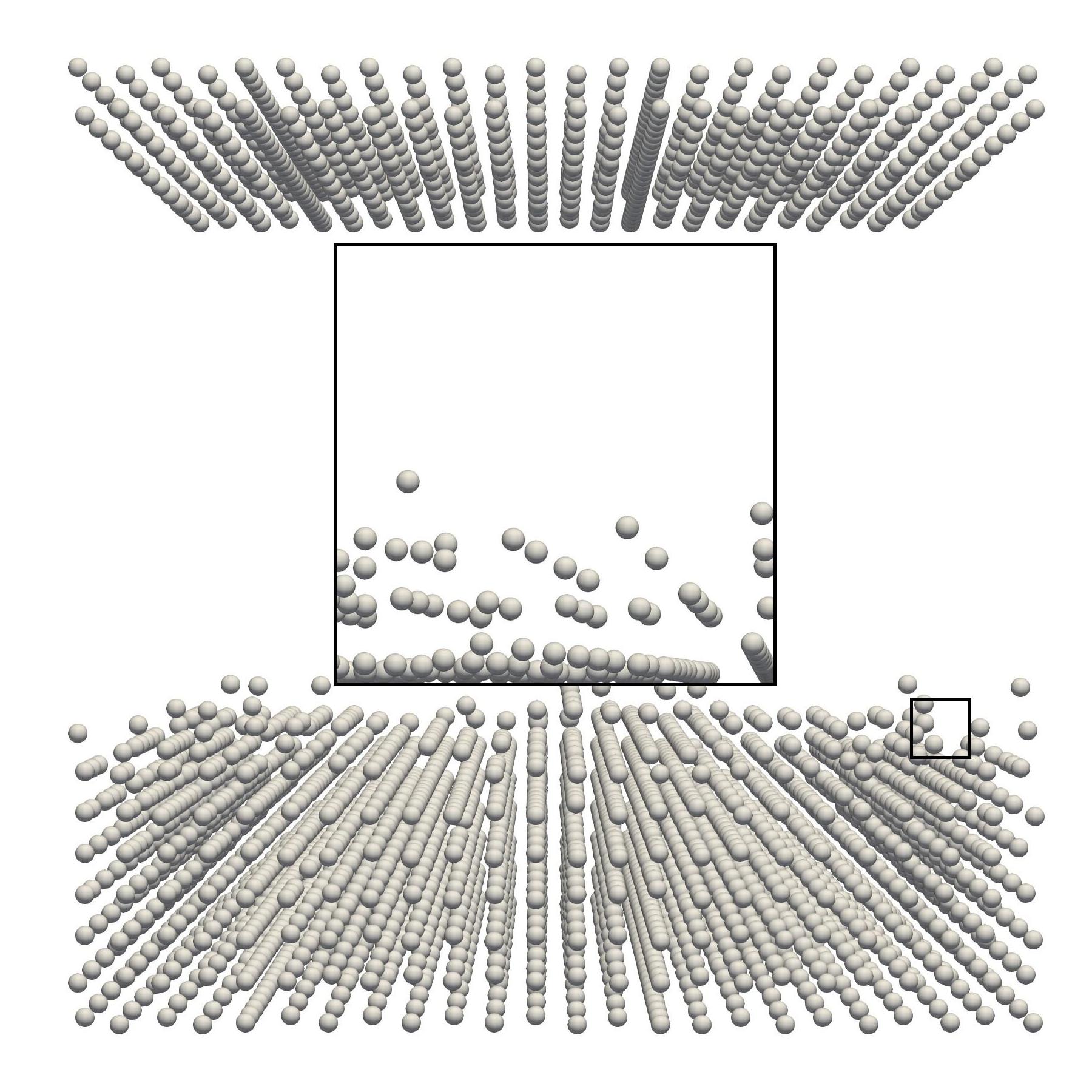}}
}
\hspace{1mm}
\subfigure[{SZ3 (CR=14.05, PSNR=36.26})] 
{
\raisebox{-1cm}{\includegraphics[scale=0.06]{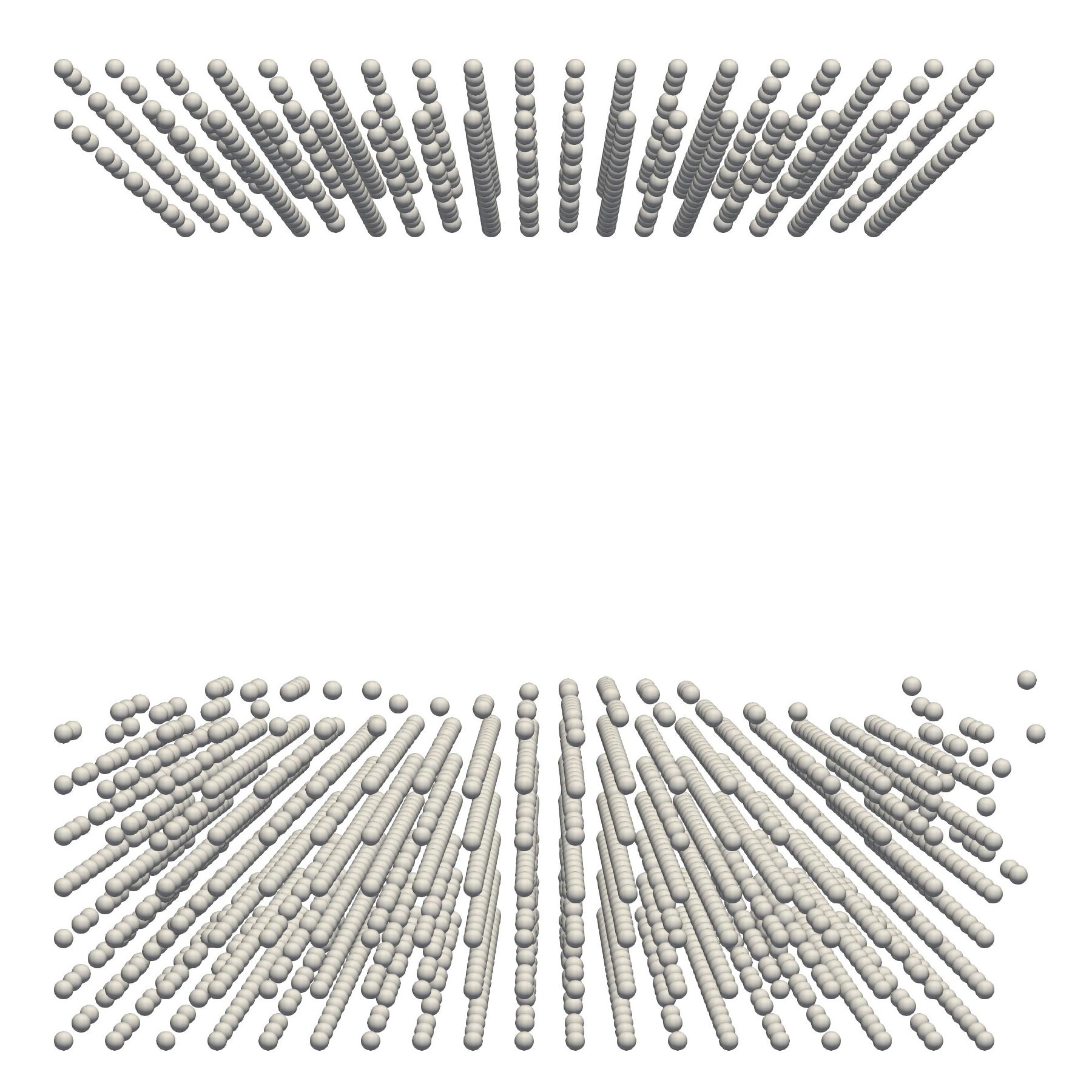}}
}
\hspace{1mm}
\subfigure[{MDZ (CR=14.42, PSNR=34.47})] 
{
\raisebox{-1cm}{\includegraphics[scale=0.06]{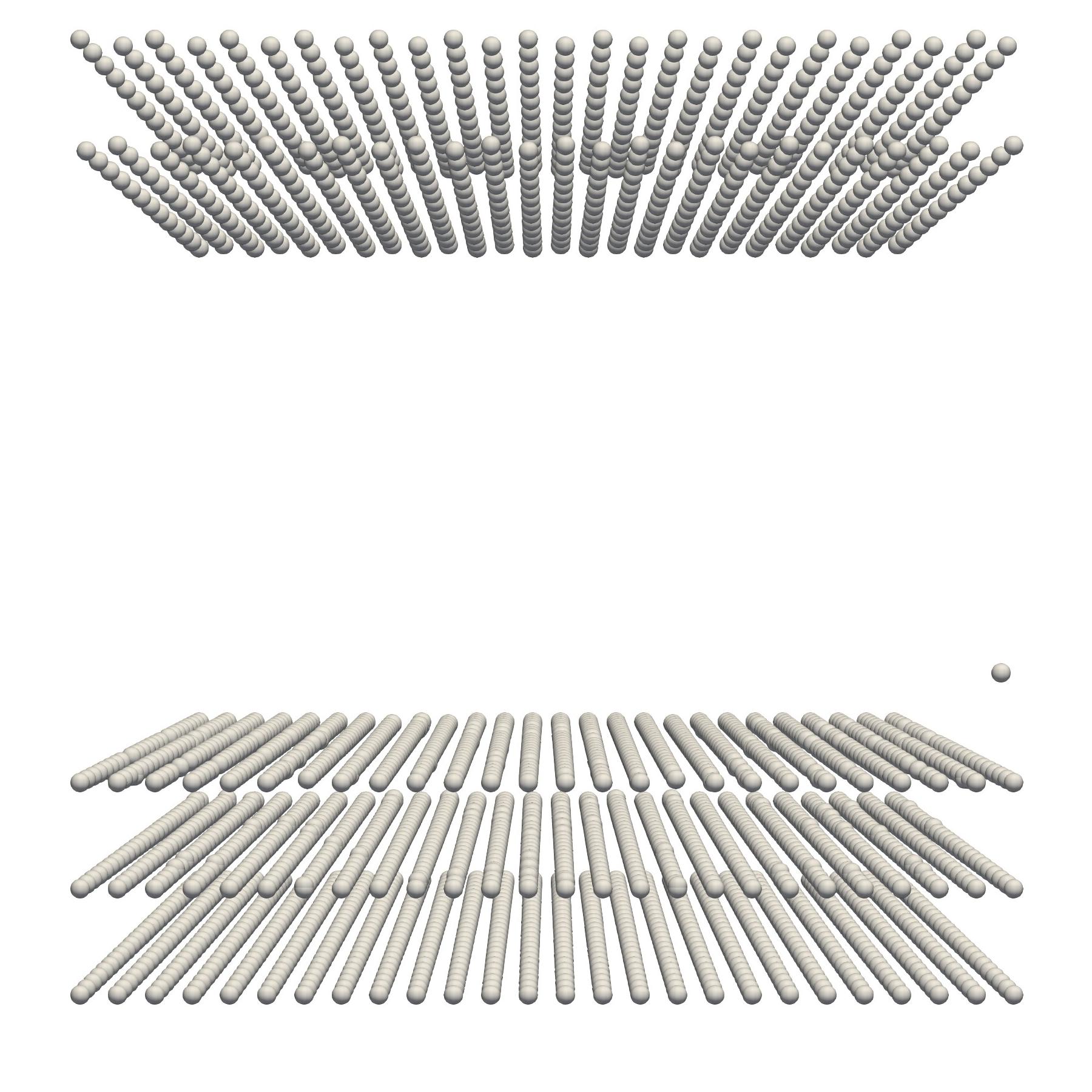}}
}
\hspace{-8mm}

\vspace{-2mm}

\hspace{0mm}
\subfigure[{TMC13 (CR=13.74, PSNR=48.52})] 
{
\raisebox{-1cm}{\includegraphics[scale=0.06]{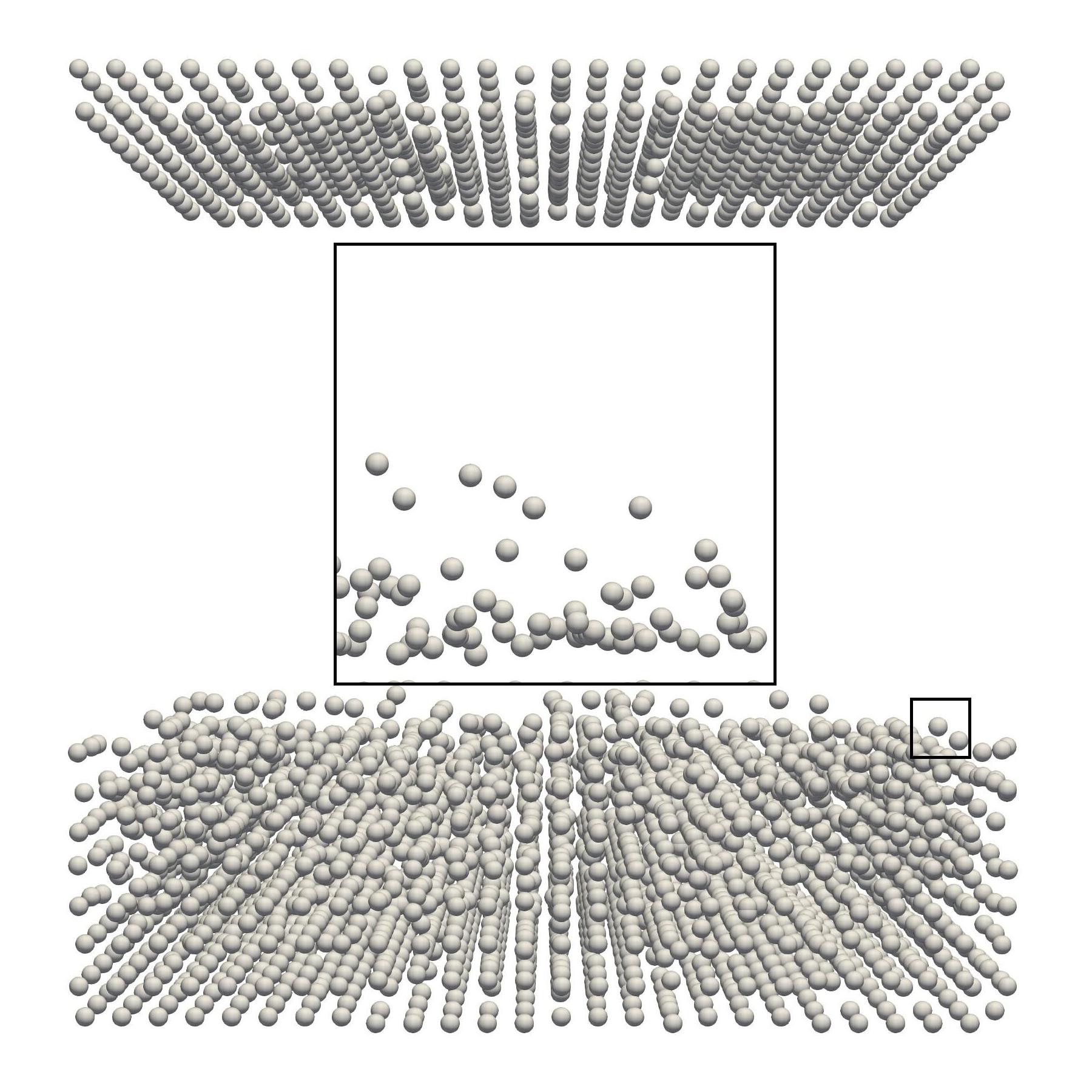}}
}
\hspace{1mm}
\subfigure[{DRACO (CR=14.23, PSNR=46.61})] 
{
\raisebox{-1cm}{\includegraphics[scale=0.06]{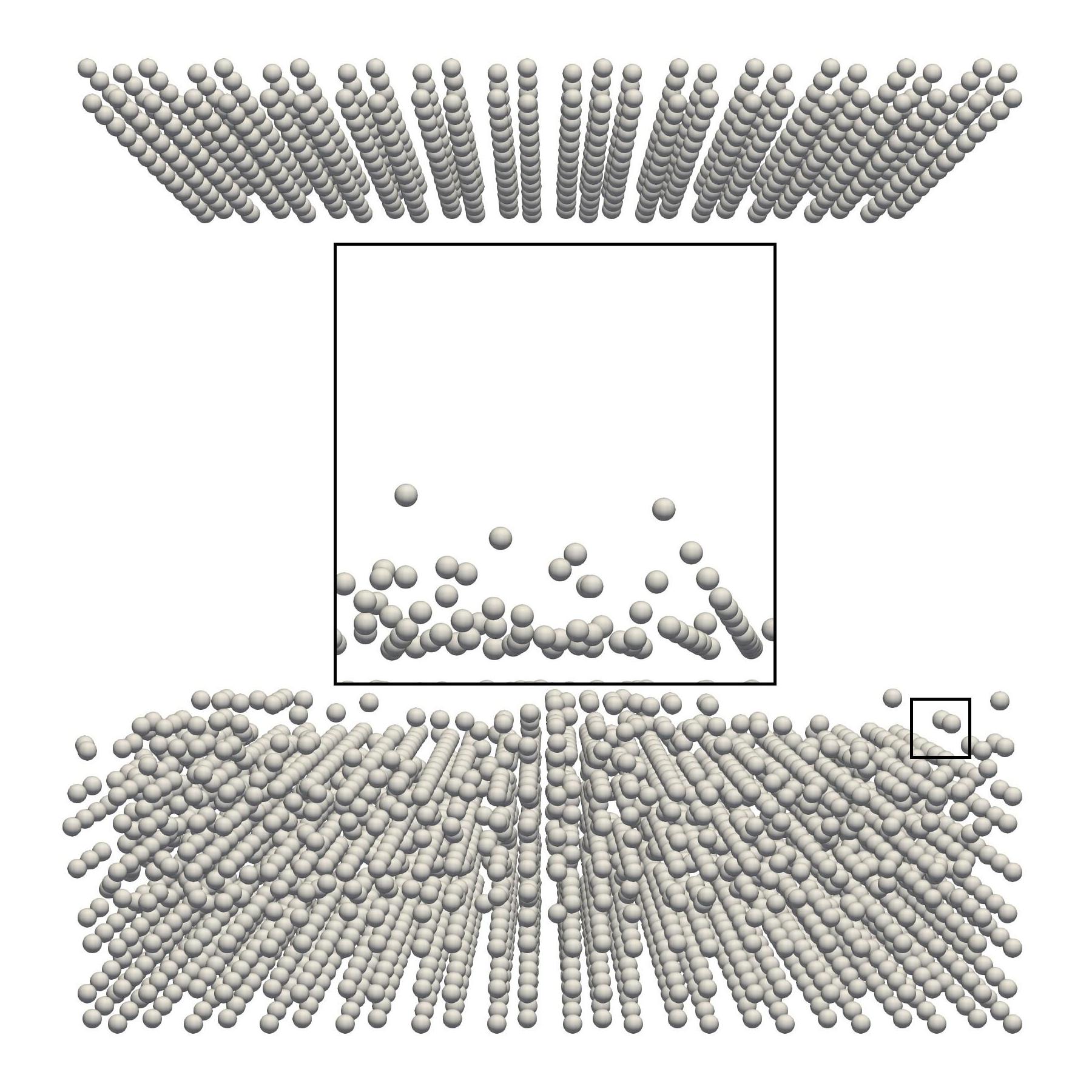}}
}
\hspace{1mm}
\subfigure[{SPERR (CR=12.62, PSNR=21.27})] 
{
\raisebox{-1cm}{\includegraphics[scale=0.06]{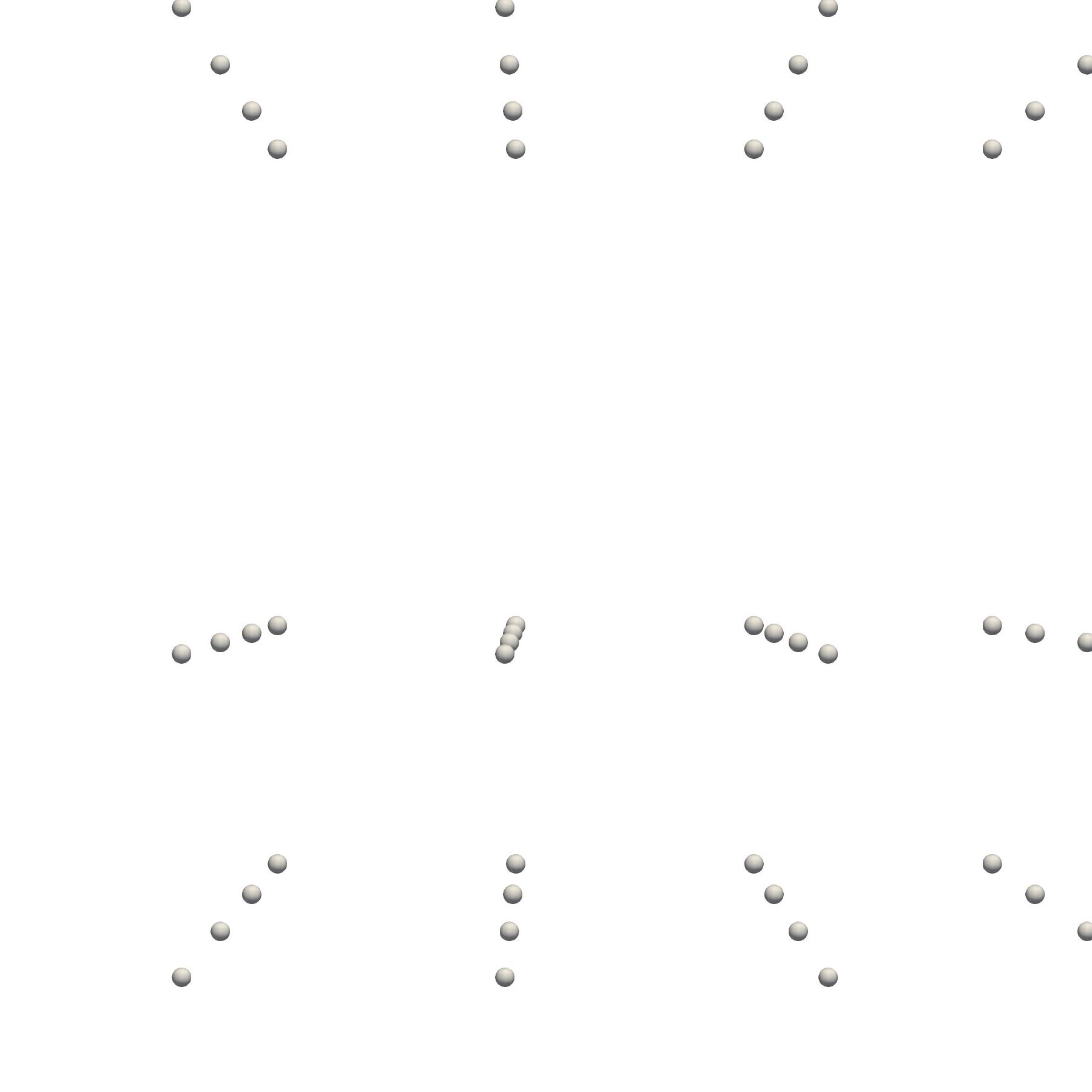}}
}
\hspace{1mm}
\subfigure[{LCP (CR=14.54, PSNR=57.49})] 
{
\raisebox{-1cm}{\includegraphics[scale=0.06]{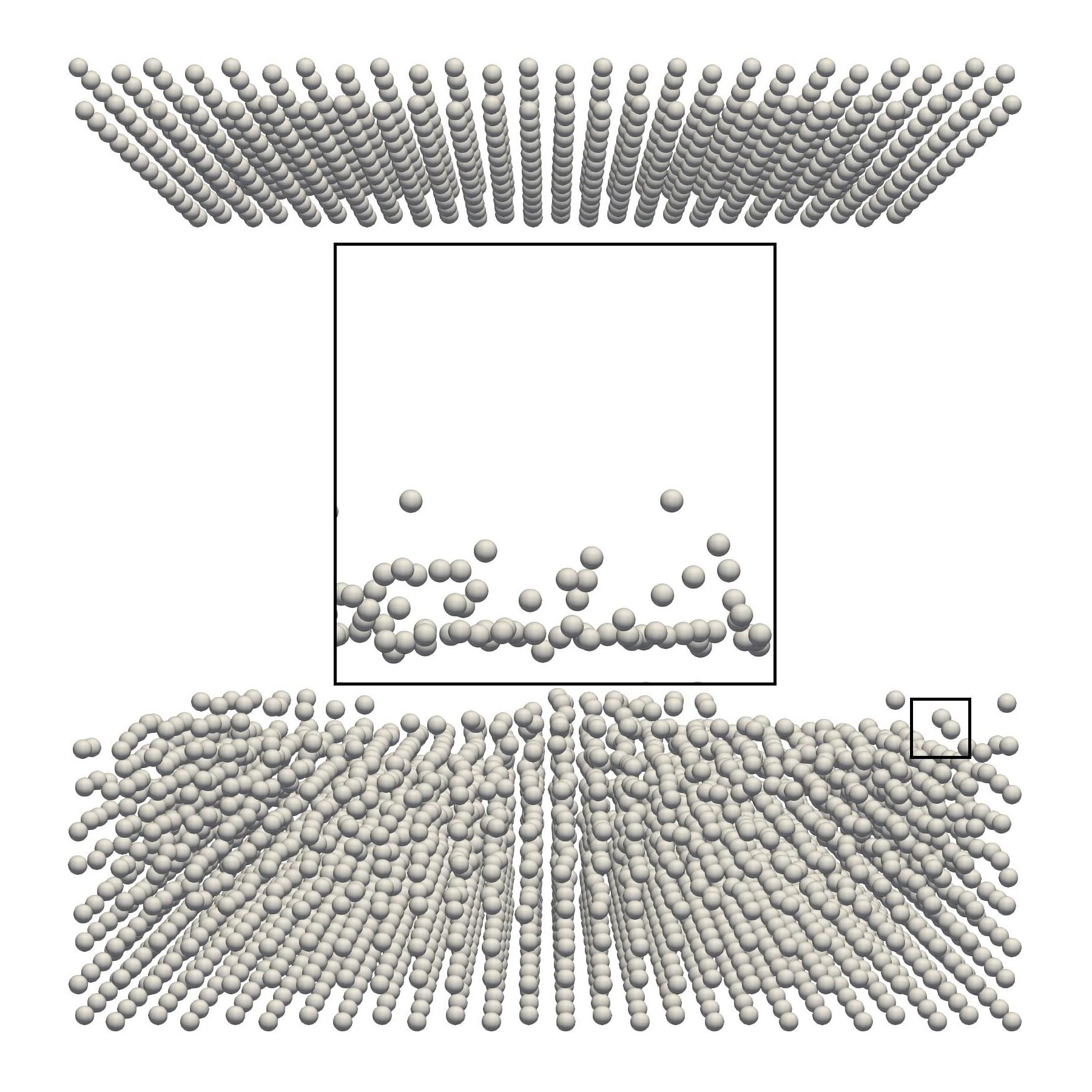}}
}
\hspace{-8mm}
\vspace{-2mm}
\caption{Visual quality comparison on Copper shows LCP has the highest data fidelity on scientific data over other compressors (ZFP is omitted as it has more distortion than SPERR. ZFP's PSNR is 6.84 when CR is 11.8)}
\label{fig:visulization_copper}
\vspace{-2mm}

\end{figure*}

\begin{figure*}[ht] \centering
\vspace{-3mm}
\hspace{0mm}
\subfigure[{Original}]
{
\raisebox{-1cm}{\includegraphics[scale=0.06]{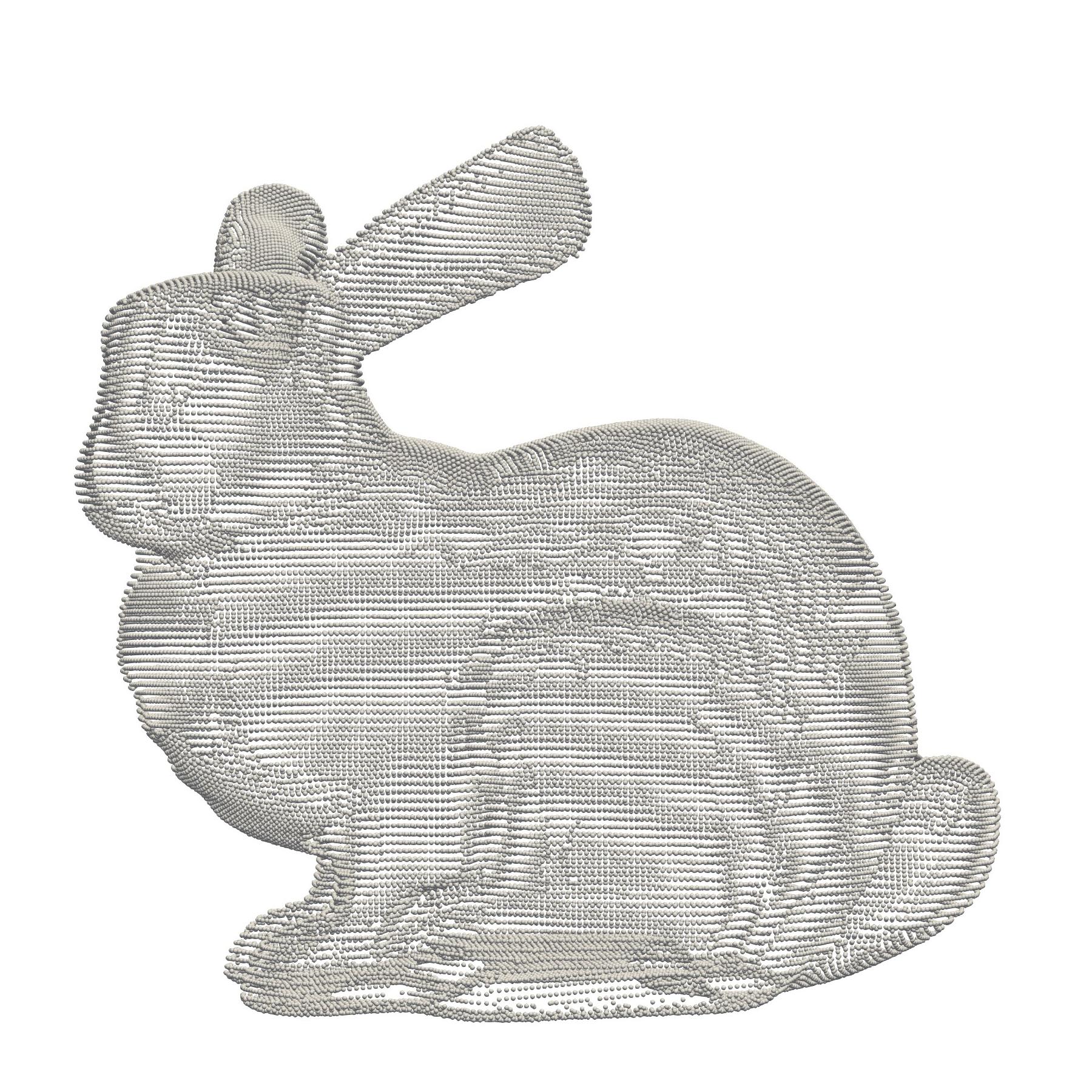}}
}
\subfigure[{SZ2 (CR=12.94, PSNR=33.77})] 
{
\raisebox{-1cm}{\includegraphics[scale=0.06]{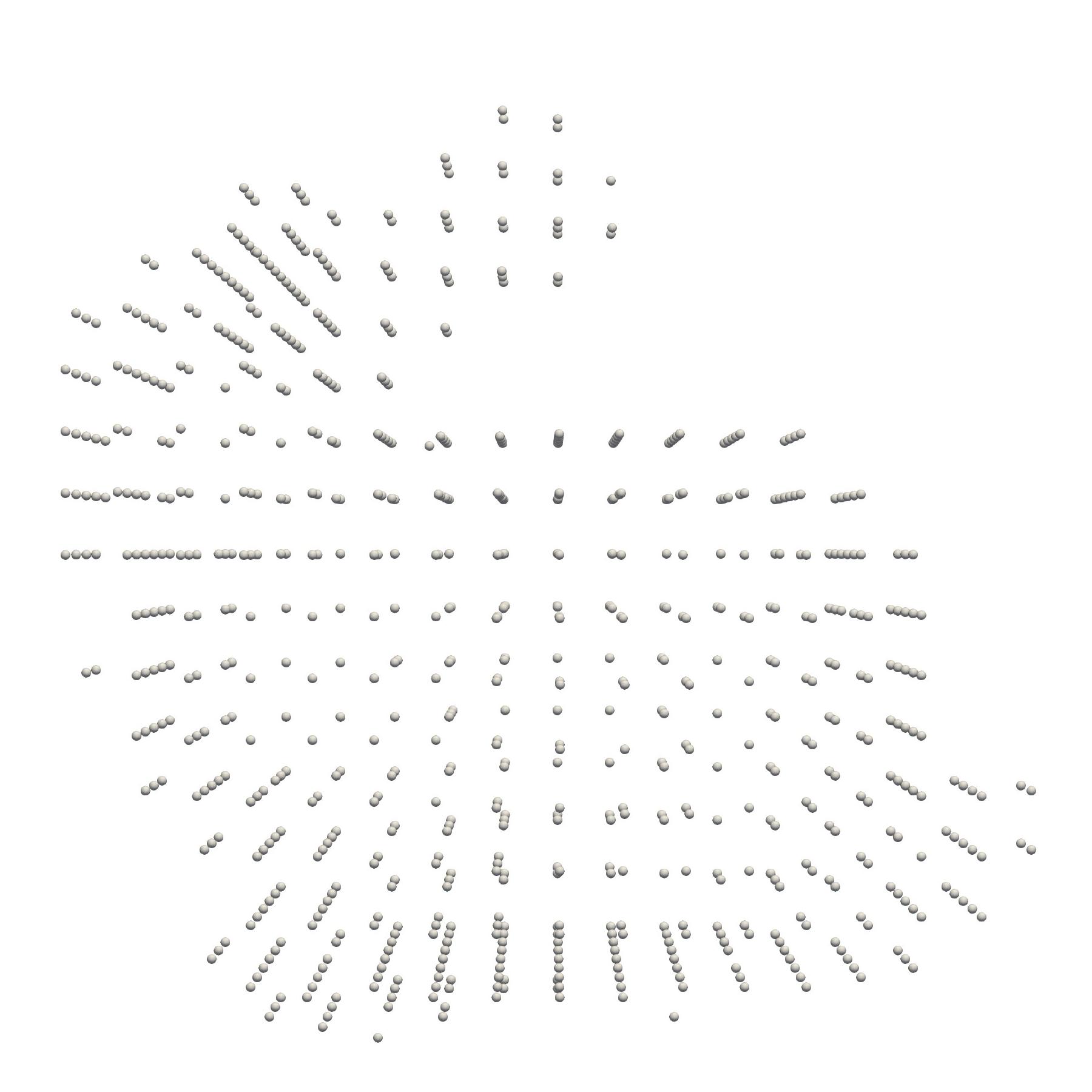}}
}
\subfigure[{SZ3 (CR=12.87, PSNR=33.69})] 
{
\raisebox{-1cm}{\includegraphics[scale=0.06]{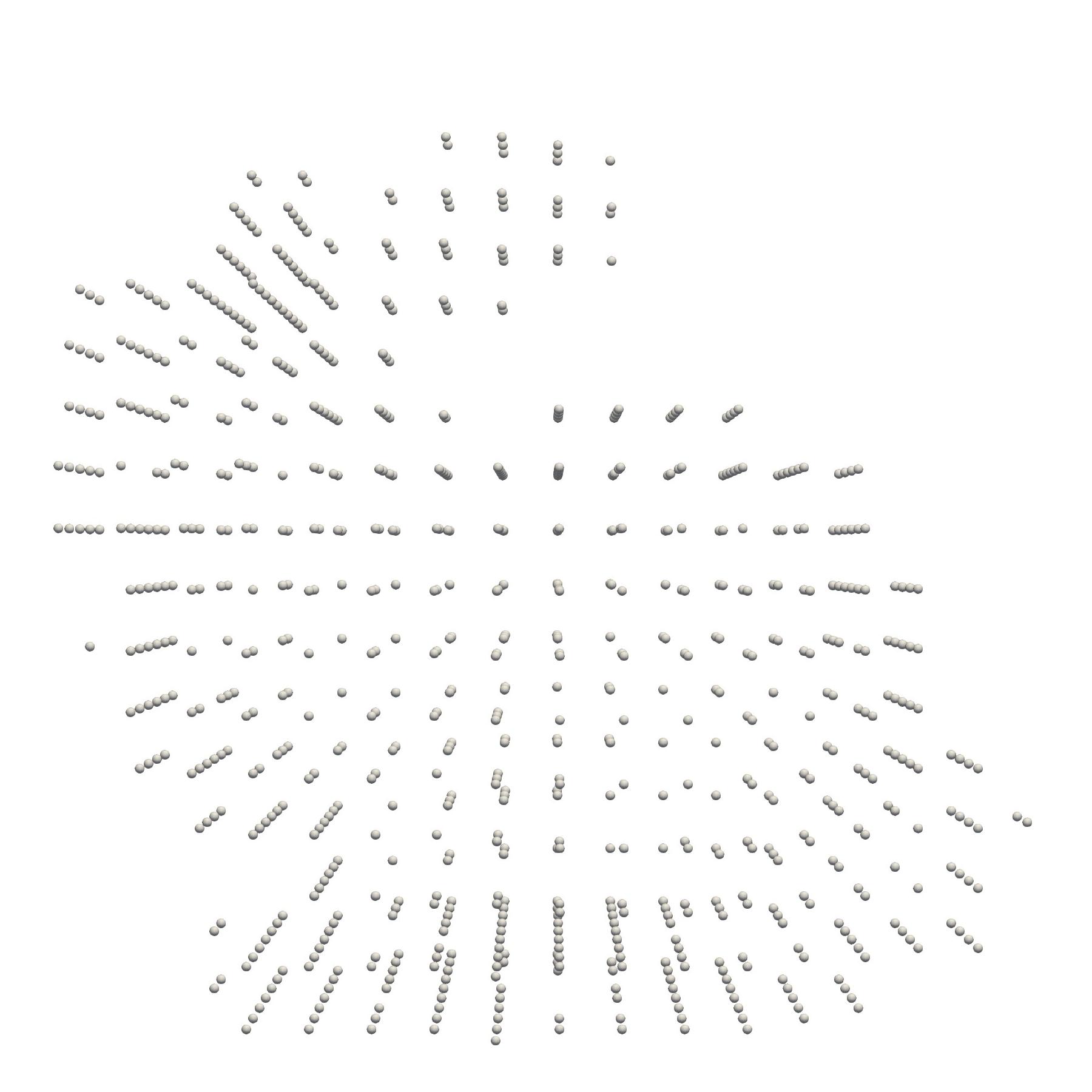}}
}
\subfigure[{MDZ (CR=12.91, PSNR=33.69})] 
{
\raisebox{-1cm}{\includegraphics[scale=0.06]{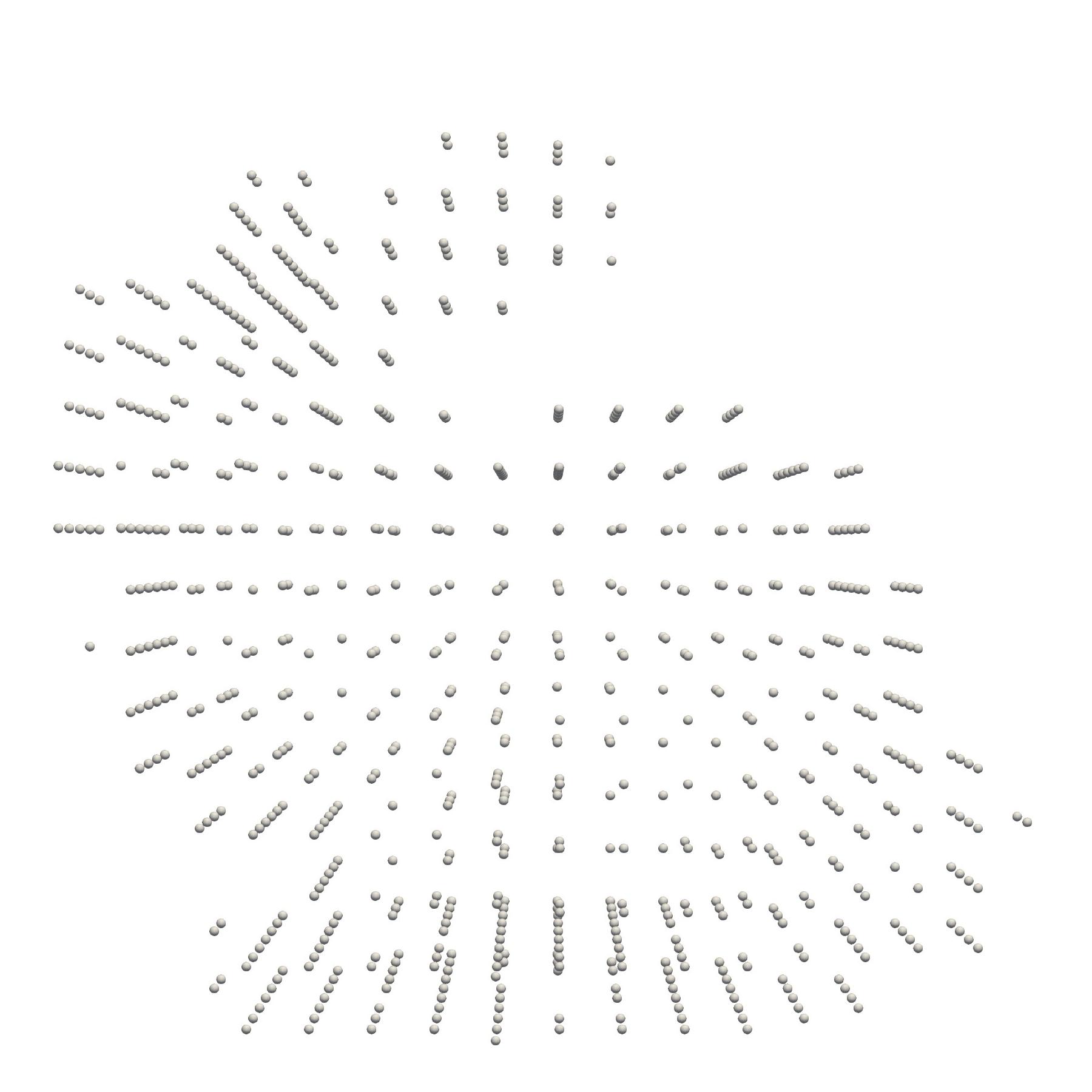}}
}

\vspace{-3.5mm}
\subfigure[{TMC13 (CR=13.93, PSNR=51.12})] 
{
\raisebox{-1cm}{\includegraphics[scale=0.06]{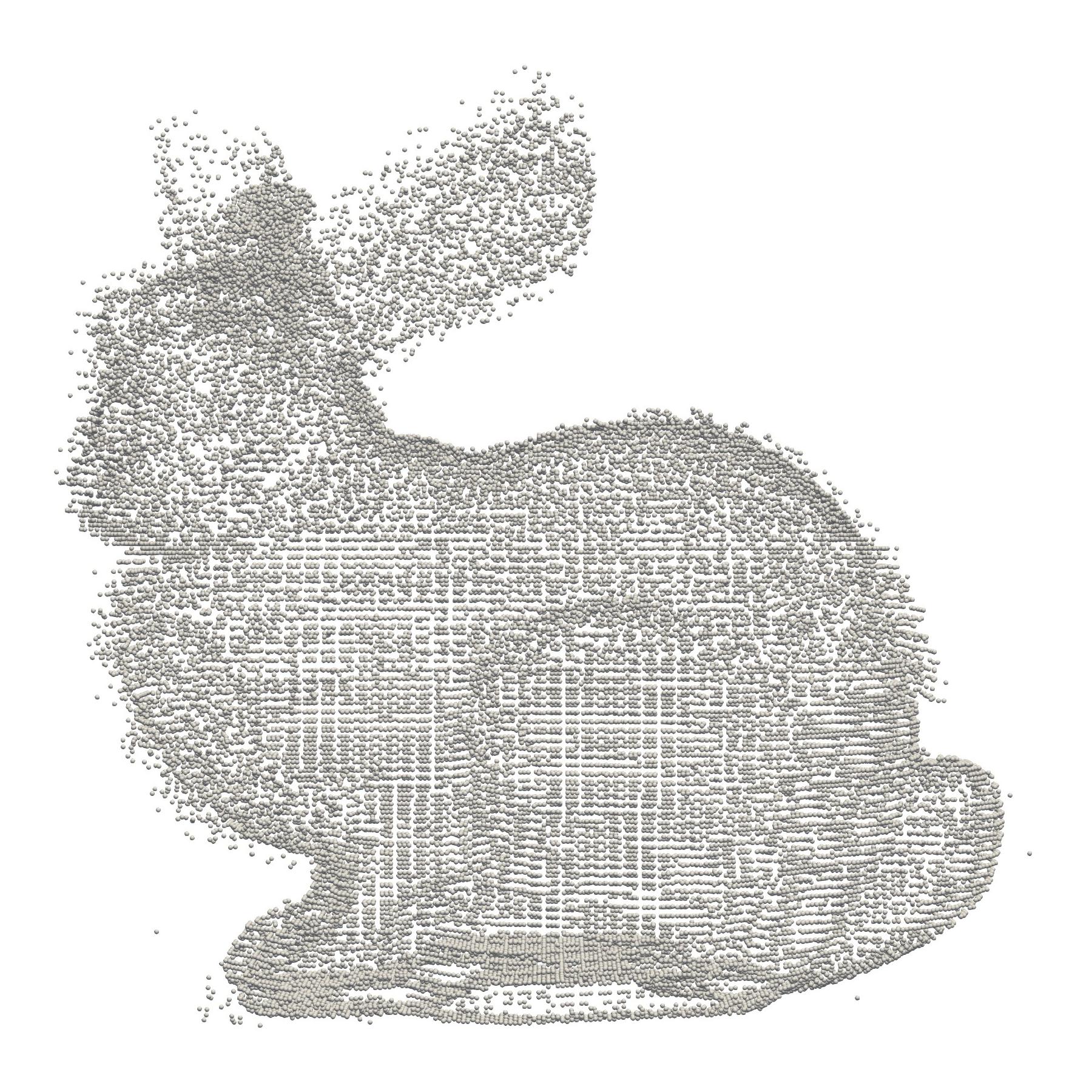}}
}
\subfigure[{DRACO (CR=12.88, PSNR=58.04})] 
{
\raisebox{-1cm}{\includegraphics[scale=0.06]{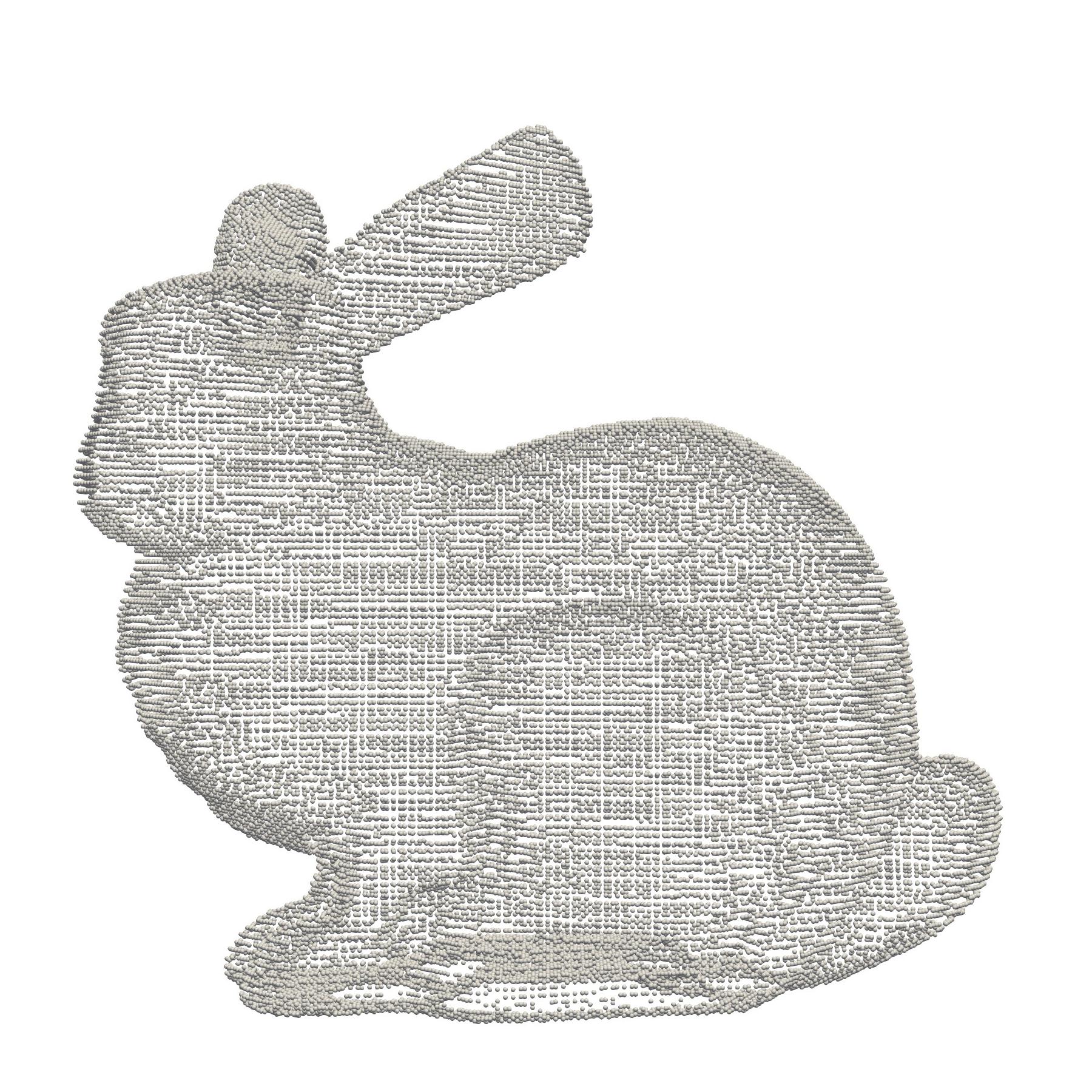}}
}
\subfigure[{SPERR (CR=12.75, PSNR=23.15})] 
{
\raisebox{-1cm}{\includegraphics[scale=0.06]{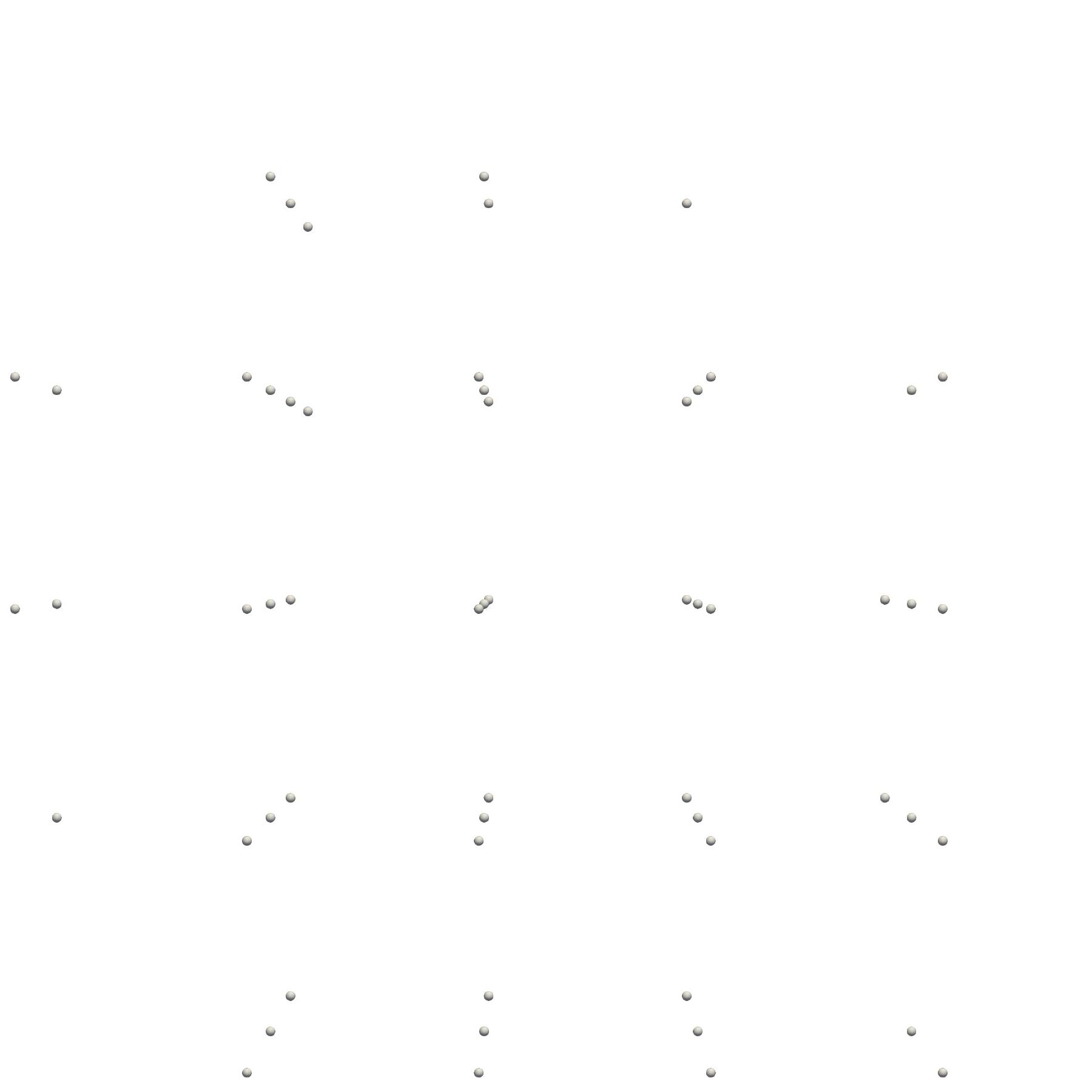}}
}
\subfigure[{LCP (CR=13.99, PSNR=58.17})] 
{
\raisebox{-1cm}{\includegraphics[scale=0.06]{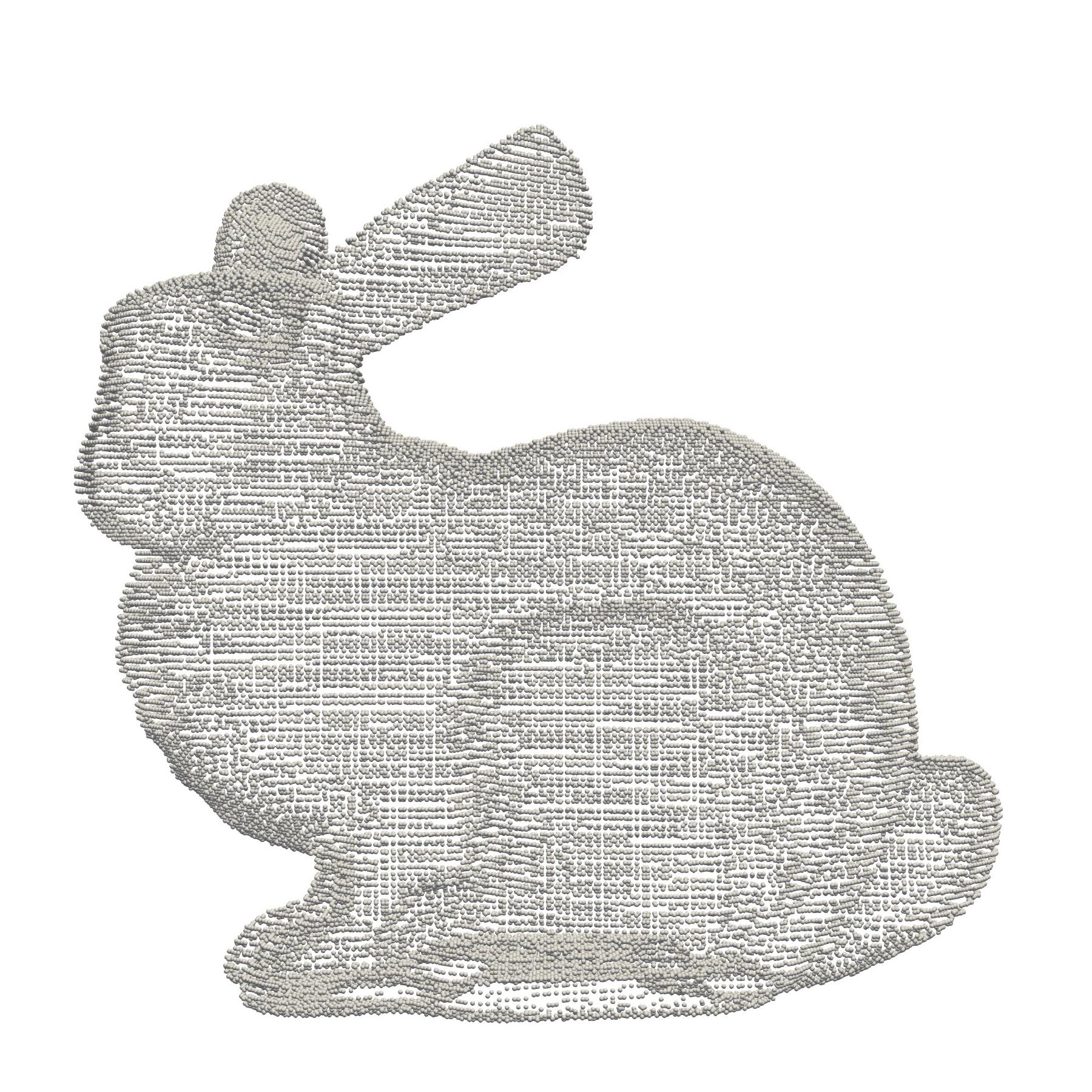}}
}
\vspace{-2mm}

\caption{Visual quality on computer graphics (Bun Zipper) dataset shows the versatility of LCP in non-scientific domains (ZFP is omitted as it has more distortion than SPERR. ZFP's PSNR is 3.64 when CR is 11.58)}
\label{fig:visulization_bun}
\vspace{-2mm}
\end{figure*}




\subsubsection{Ablation study}

We first perform the ablation study to show the essential contributions of LCP's components. LCP-S in~\Cref{fig:version_iteration_comparison} means the LCP-S spatial compressor in ~\Cref{sec: design lcp-s}. The second line in the figure adds the dynamic strategy of finding the best block size (discussed in~\Cref{sec:block_size_in_a_single_frame}) on top of LCP-S, and it proves to be universally beneficial across the datasets. 
The third line further enables the temporal compressor LCP-T and the corresponding hybrid design. (discussed in~\Cref{sec: hybrid lcp s and t}). Considering the temporal domain in compression dramatically increases the compression effectiveness for datasets that are relatively continuous in time. The last line in the figure allows adjusting the error-bound scale for anchor frames to have higher LCP-T effectiveness (discussed in~\Cref{sec:error_bound_in_multiple_frames}). This adjustment works as expected, improving compression quality for Helium (which frequently uses LCP-T) when the bit rate is small (indicating the need for scaling for larger error bounds). 

\subsubsection{Compression error}
We use the absolute error bounds (discussed in~\Cref{eq:abs}) in our experiments for all the compressors except for Draco which only supports setting the targeted bits. To verify the correctness of those compressors, we calculate the actual maximum error in all the cases. Results confirm that all compressors (except for Draco) respect the pre-defined error bound, with~\Cref{fig:error_dist} showing the error distribution of our solution LCP on the Helium dataset. The error bound in~\Cref{fig:error_dist} is 0.1, and it is clear that LCP keeps its maximum error within the bound.

\subsubsection{Compression ratio}
We first reveal the overall ranking of compressors on compression ratio using the Critical Difference (CD) diagram, then present the comparison on four cases representing four datasets. 
CD diagram is a statistical tool used to visually compare the overall rankings of multiple algorithms in various test cases. The CD diagram shown in~\Cref{fig:cd} is generated from the results of all datasets under various settings (different batch sizes and error bounds). The number indicates the rankings of compressors in terms of compression ratio. Compressors connected by a line have no significant difference, and compressors stand alone have a statistically significant difference from others. This figure confirms that our solution LCP has a significantly better compression ratio than all other solutions.


Figure \ref{fig:various_batch} shows the compression ratios of lossy compressors on the multi-frame datasets, under various batch sizes and error bounds. Both error bounds and batch size are user-defined parameters based on their needs. Error bound is adjusted based on applications' tolerance of errors and expectations on compressed sizes (higher error bounds lead to smaller sizes). For batch size, as discussed in~\Cref{sec: background batch compression}, while a large batch size enables higher compression ratios due to the longer temporal domain to use, a smaller batch size makes data retrieval on a single frame faster. Our solution LCP demonstrates the highest compression ratios among all other compressors on all the datasets regardless of error bound and batch size settings. In absolute terms, when batch size is 16, our solution has 78\%, 26\%, 12\%, 104\% compression ratio improvements over the second-best on Helium, Copper, LJ, and YIIP datasets, respectively. This result not only proves the leading position of LCP but also highlights its strong robustness and adaptability in various use cases.

\subsubsection{Rate-distortion}

Rate-distortion graph is one of the most important metrics to evaluate lossy compression quality (the definition of bit rate and PSNR in the graph can be found in~\Cref{sec:problem}). A lower bit rate or higher PSNR indicates better compression quality for such graphs. 
The common practice for interpreting the rate-distortion graph is to compare the compression ratios (which is reverse to bit rate, as defined in~\Cref{sec:problem}) of compressors based on the same distortion, by looking at the x-axis values (bit rate) with a fixed y-axis value (PSNR). 

In this section, we first test all the compressors on single frame inputs (select the middle frame in time from each data) to mimic the use cases when each frame needs to be compressed independently. Then, we extend our evaluation to multi-frame to represent the batch compression scenarios. 
The single-frame result is shown in~\Cref{fig:br_psnr_1} and the multi-frame result is shown in~\Cref{fig:br_psnr_16}. We use batch size 16 for the multi-frame compression. Draco has staircase-style results in the figures because it doesn't support arbitrary bounds and can only give output in limited quality levels, also its results are similar in single-frame and multi-frame as Draco cannot use the temporal domain for compression. In terms of our solution, both figures confirm that LCP has the best compression quality no matter if compressing frames independently or in batch. In absolute terms, LCP reaches 34dB, 35dB higher PSNR over the second-best with the same bit rate for single and multi-frame cases, respectively.





\subsubsection{Visual quality}

Visual quality is another critical metric to evaluate the quality of lossy compression, as data visualization is one of the most widely used post-analysis applications. To have a fair comparison, the visual quality is evaluated based on similar compressed sizes for all the compressors. SPERR exhibit significant data degradation with many particles overlapped or misrepresented. We omit the visualization of ZFP as the reconstructed visualization image has more distortion than SPERR at the demonstrated compression ratio. As shown in~\Cref{fig:visulization_copper}
SZ2, SZ3, and MDZ retain the overall particle structure but introduce noticeable visual distortions. 
DRACO, with a PSNR of 46.61, maintains high visual similarity to the original dataset but artifacts become noticeable when viewed in detail (especially the particles in the bottom and in the right corner). Our solution LCP stands out by almost perfectly preserving the particle structure and distribution (reaching a PSNR of 57.49), thus demonstrating superior visual quality among all baselines.

Although LCP is tailored for scientific particle data, it demonstrates strong versatility on non-scientific data. As shown in~\Cref{fig:visulization_bun}, both LCP and Draco have the highest PSNR at 58, but LCP achieves this with a smaller compressed data size (LCP’s compression ratio is 13.99 and Draco’s is 12.88).


\subsubsection{Compression/decompression Speed}

\begin{figure}[ht] \centering

\hspace{-6mm}
\subfigure[Warpx]
{
\raisebox{-1cm}{\includegraphics[scale=0.25]{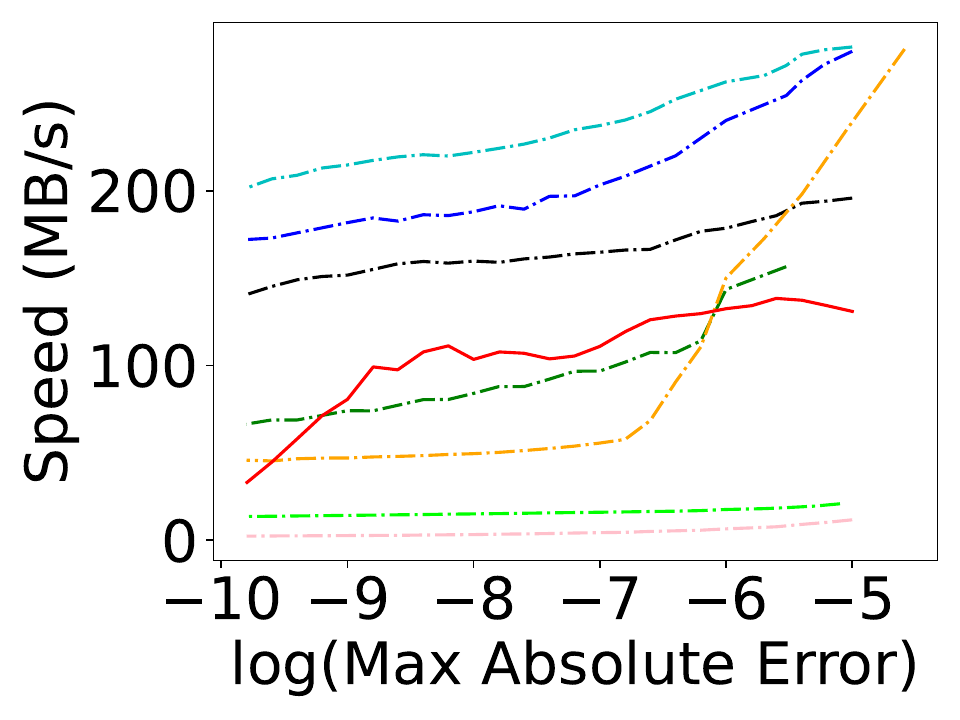}}
}
\hspace{-4mm}
\subfigure[HACC]
{
\raisebox{-1cm}{\includegraphics[scale=0.25]{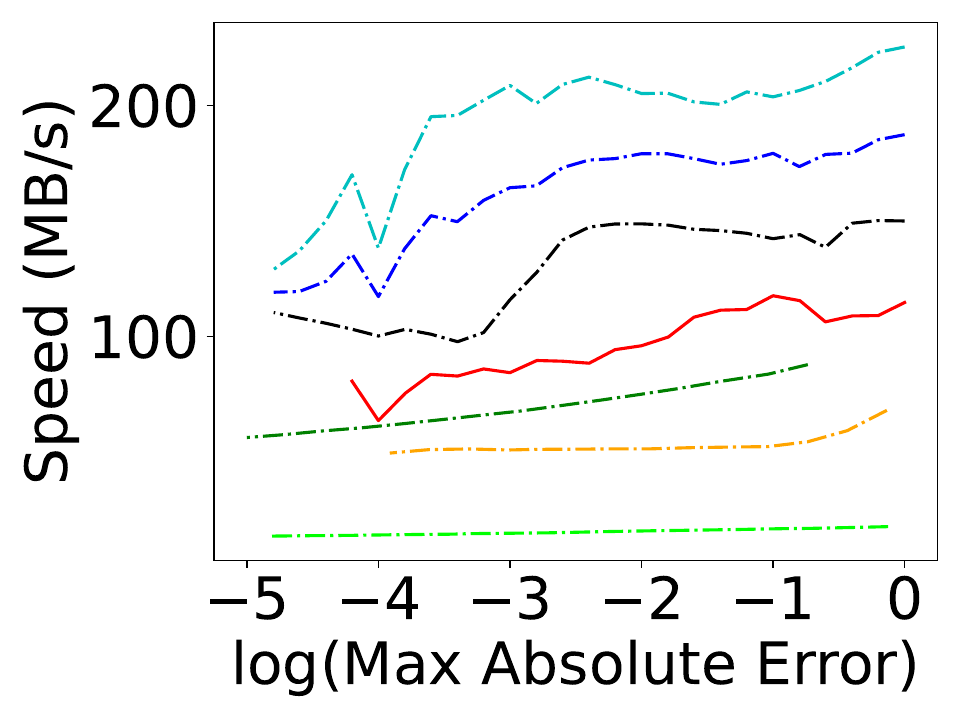}}
}
\hspace{-8mm}

\hspace{-6mm}
\subfigure[Helium]
{
\raisebox{-1cm}{\includegraphics[scale=0.25]{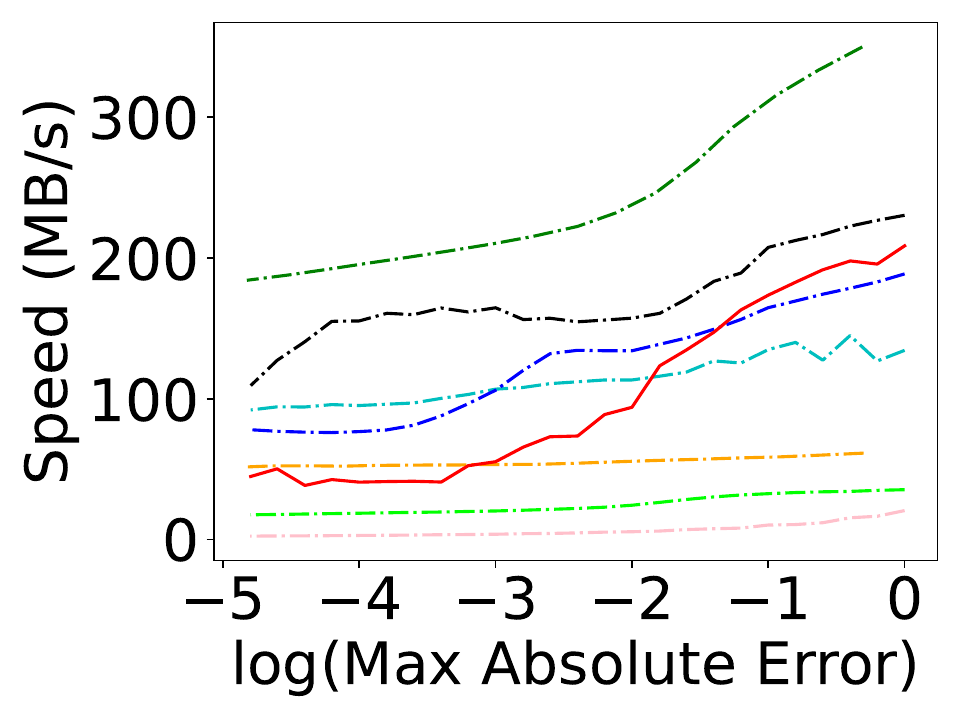}}
}
\hspace{-4mm}
\subfigure[LJ]
{
\raisebox{-1cm}{\includegraphics[scale=0.25]{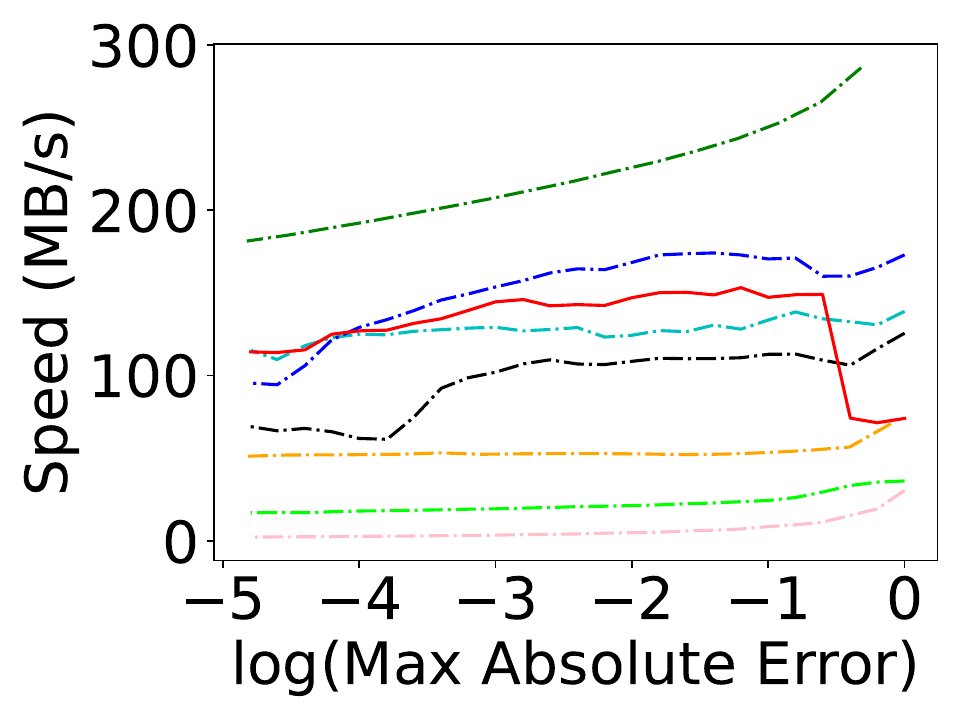}}
}
\hspace{-8mm}

\includegraphics[scale=0.25]{figures/batch1/legend.pdf}

\vspace{-2mm}
\caption{LCP has a mid-tier compression speed among the baselines}
\label{fig:cmpt}
\vspace{-5mm}

\end{figure}








\begin{figure}[ht] \centering

\hspace{-8mm}
\subfigure[HACC]
{
\raisebox{-1cm}{\includegraphics[scale=0.25]{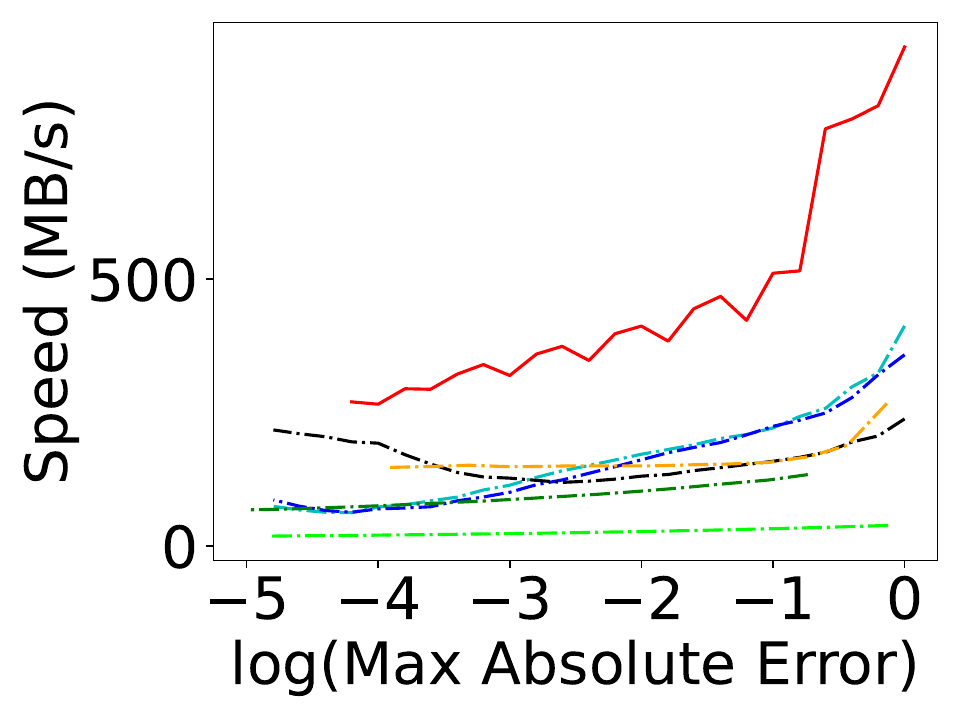}}
}
\hspace{-4mm}
\subfigure[Helium]
{
\raisebox{-1cm}{\includegraphics[scale=0.25]{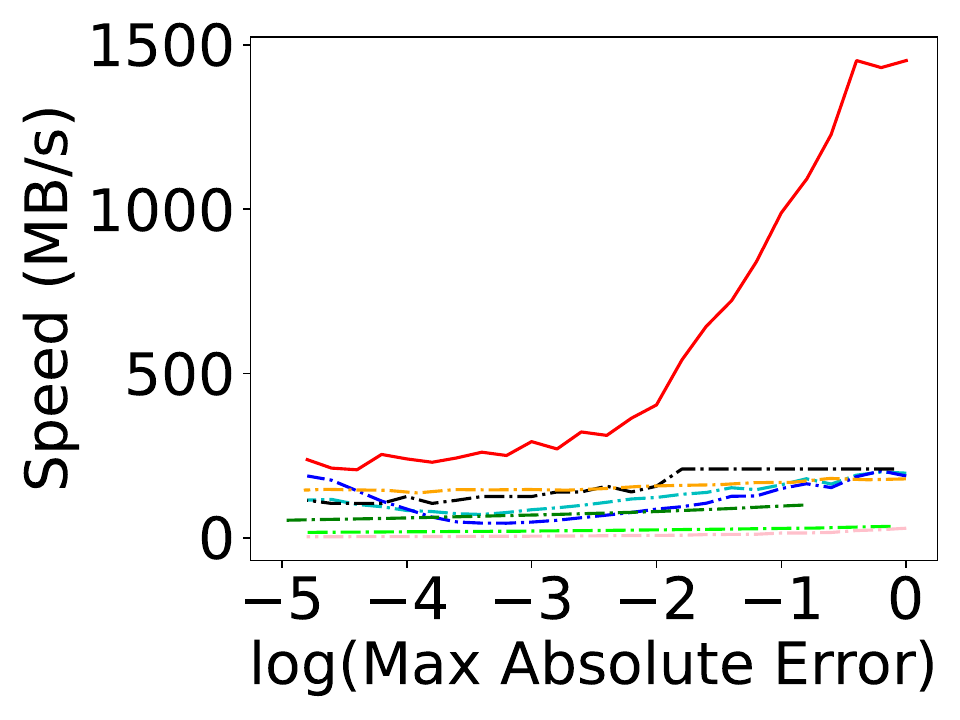}}
}
\hspace{-8mm}
\vspace{-3mm}

\hspace{-8mm}
\subfigure[BUN-ZIPPER]
{
\raisebox{-1cm}{\includegraphics[scale=0.25]{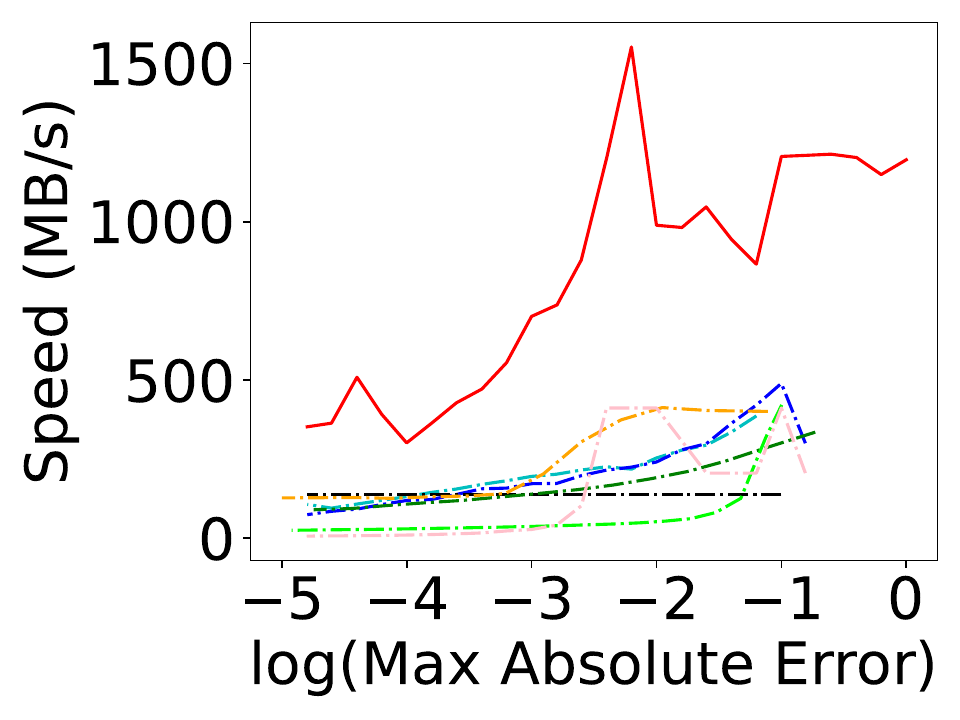}}
}
\hspace{-4mm}
\subfigure[3DEP]
{
\raisebox{-1cm}{\includegraphics[scale=0.25]{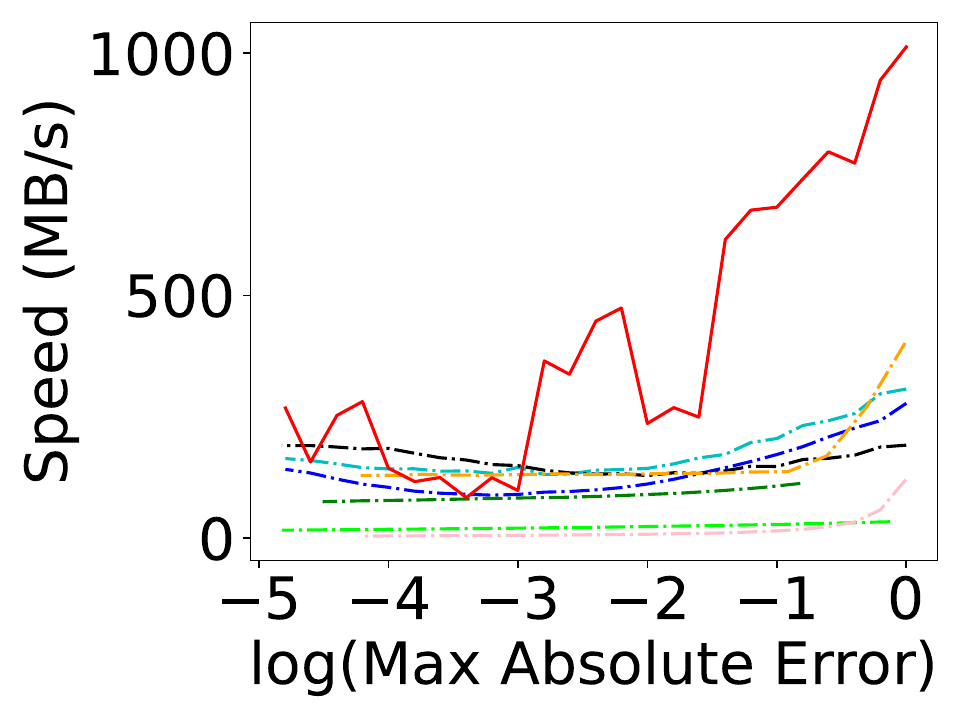}}
}
\hspace{-8mm}

\includegraphics[scale=0.25]{figures/batch1/legend.pdf}

\vspace{-2mm}
\caption{In single frame mode, LCP has the highest data retrieving speed compared to all baselines}
\label{fig:decmpt1}
\vspace{-5mm}

\end{figure}

\begin{figure}[ht] \centering

\hspace{-8mm}
\subfigure[Copper]
{
\raisebox{-1cm}{\includegraphics[scale=0.25]{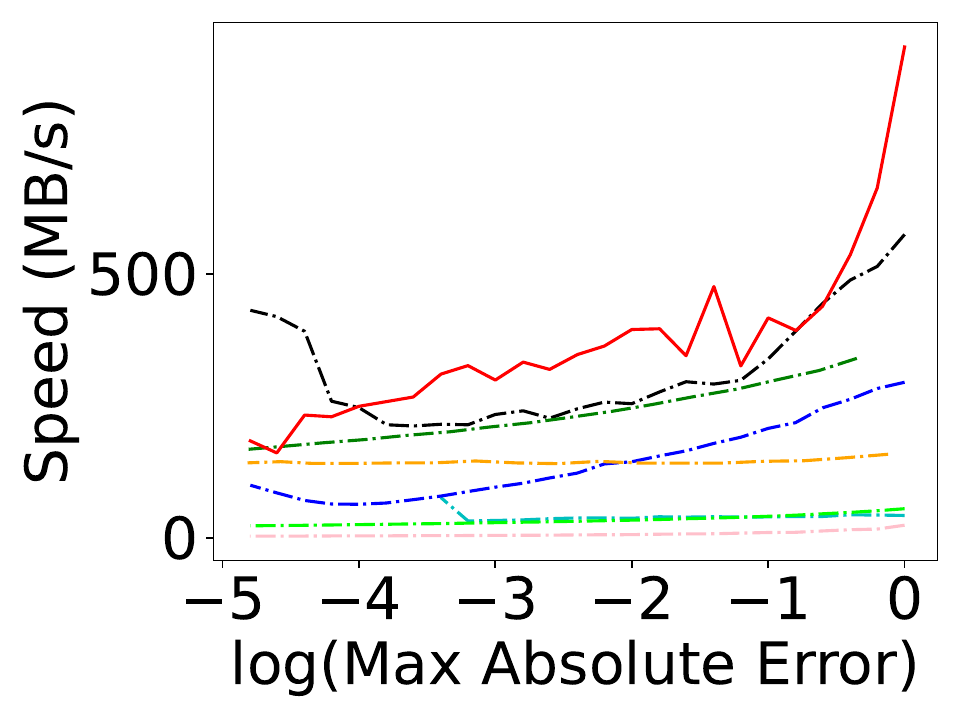}}
}
\hspace{-4mm}
\subfigure[Helium]
{
\raisebox{-1cm}{\includegraphics[scale=0.25]{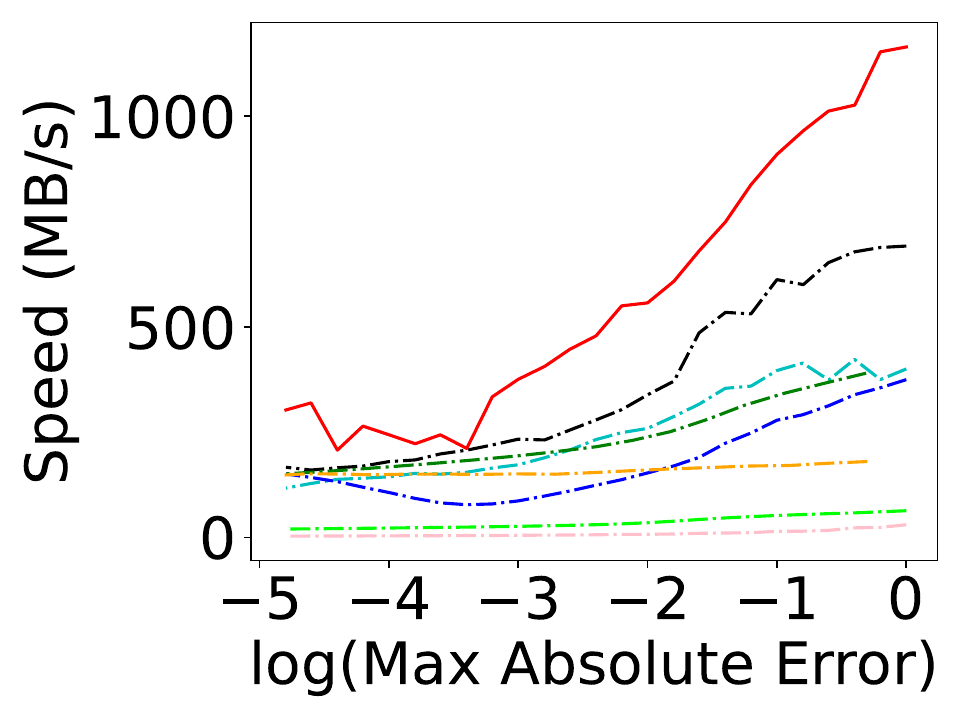}}
}
\hspace{-8mm}
\vspace{-3mm}

\hspace{-8mm}
\subfigure[LJ]
{
\raisebox{-1cm}{\includegraphics[scale=0.25]{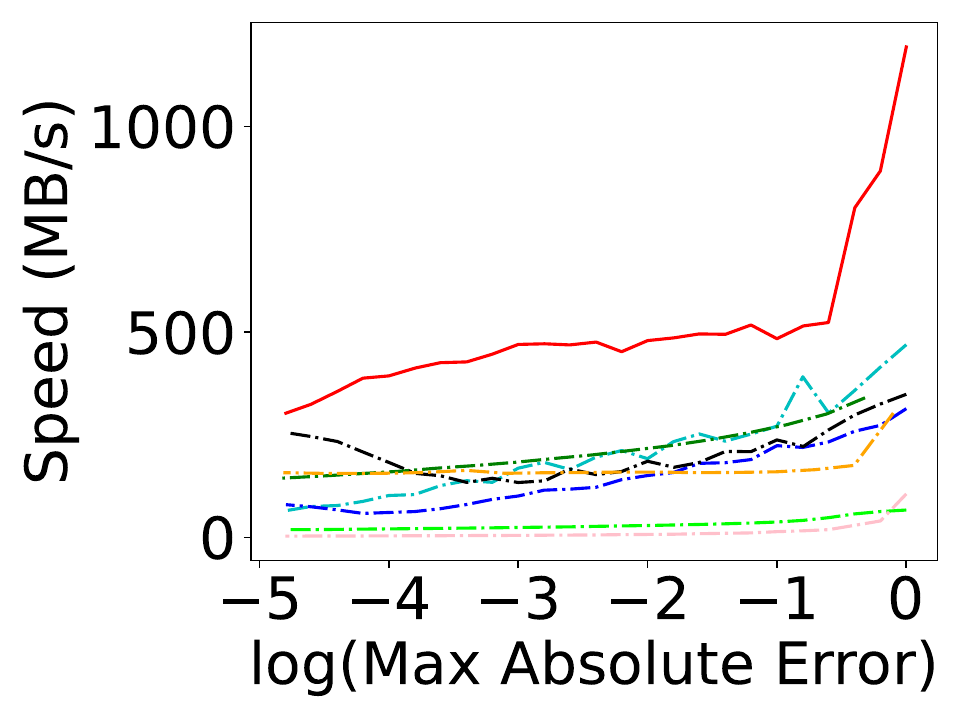}}
}
\hspace{-4mm}
\subfigure[YIIP]
{
\raisebox{-1cm}{\includegraphics[scale=0.25]{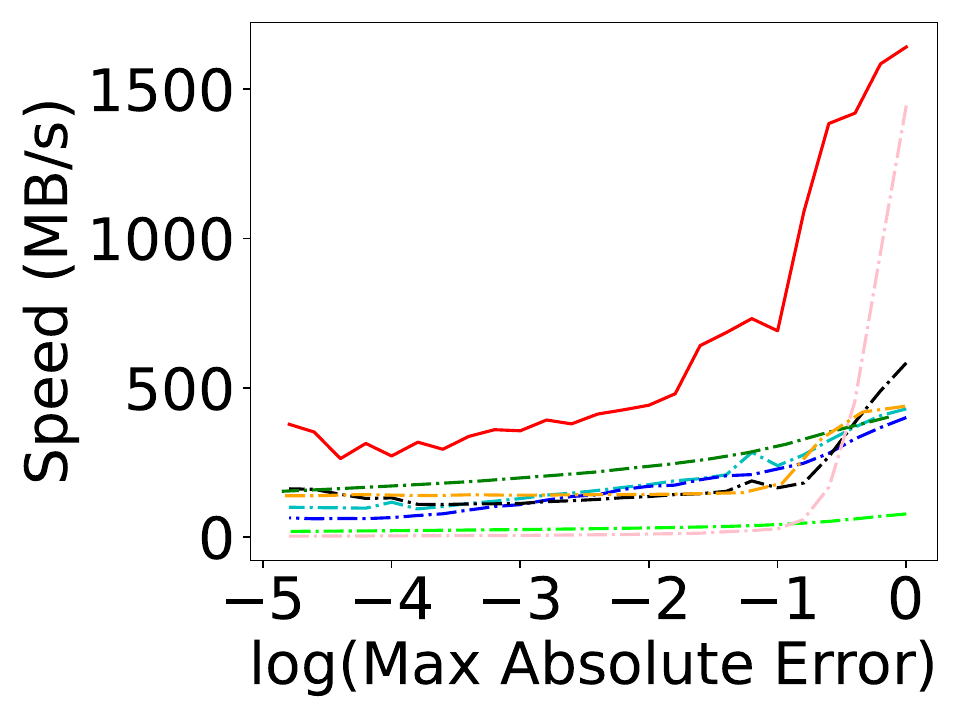}}
}
\hspace{-8mm}

\includegraphics[scale=0.25]{figures/batch16/legend.pdf}

\vspace{-2mm}
\caption{In batch mode, LCP still has the highest retrieving speed (batch Size = 16)}
\label{fig:decmpt16}
\vspace{-4mm}

\end{figure}
We evaluate both the compression and decompression speed of all compressors. 
LCP has a mid-tier compression speed compared to the other baselines, as shown in fig~\ref{fig:cmpt}, that is faster than SPERR, TMC13, and Draco in most cases, but slower than ZFP or SZ3.
Compared to compression, decompression speed is usually much more important for applications because data retrieval directly impacts the efficiency of post-analysis and decision-making processes, as discussed in~\Cref{sec:problem}.  Fig~\ref{fig:decmpt1} shows LCP has the highest speed in decompressing a single frame in most of the cases, with advantages as much as 202\%, 593\%, 397\%, 257\% over the second best on HACC, Helium, BUN-ZIPPER, and 3DEP datasets, respectively. Moreover, because of our efficient hybrid spatial-anchor-frame based design (discussed in~\Cref{sec: hybrid lcp s and t} and~\Cref{sec: compress-in-batch}), LCP is still the fastest compressor in batch compression mode, with up to 64\%, 99\%, 181\%, 318\% higher speed than the second best on the four datasets respectively.

The data retrieval process includes both I/O and decompression. On one hand, the I/O speed is proportional to the size of the data. As shown in~\Cref{fig:br_psnr_1} and~\Cref{fig:br_psnr_16}, under the same data fidelity (PSNR), LCP leads to the smallest data (bit rate) size, such that LCP will result in the shortest I/O time over all other lossy compressors. On the other hand, ~\Cref{fig:decmpt1} and~\Cref{fig:decmpt16} confirm that LCP already has the fastest decompression speed. As a result, LCP is the best compressor for efficient data retrieval as it requires the shortest I/O time and decompression time compared to all other solutions.



\section{Conclusions and Future Work}
\label{sec:conclusion}

In this paper, we propose LCP, an error-bounded lossy compressor designed to enhance the management of scientific particle data. LCP is built upon a hybrid design with block-wise spatial compressor LCP-S and temporal compressor LCP-T, together with strategies including dynamic method selection and parameter optimization, and spatial-anchor-frame-based batch compression. In our thorough evaluation of eight real-world particle datasets from seven distinct domains, compared with eight state-of-the-art lossy compressors, LCP outperforms other compressors in most of the cases, with up to 104\% higher compression ratio and 592\% faster speed over the second best.

In the future, we will improve the compression speed of LCP by designing offline optimizations that can narrow down the candidates for online parameter optimizations. 
We also plan to explore prioritized data retrieval techniques in LCP to expedite the data retrieval and decision-making pipeline. Additionally, we will enable LCP to run on heterogeneous platforms, including GPUs and accelerators, through platform-specific optimizations and hardware co-design. Lastly, we aim to enhance LCP’s data fidelity by exploring machine learning-based error correction solutions. These enhancements will make LCP more versatile, efficient, and capable of meeting diverse user needs.

\section{Acknowledgments}
\footnotesize
This research was supported by the National Science Foundation under Grant OAC-2104023, OAC-2311875, OAC-2311876, OAC-2311878, OAC-2344717. 
This work used the Purdue Anvil CPU cluster through allocation CIS230308 and CIS240192 from the Advanced Cyberinfrastructure Coordination Ecosystem: Services \& Support (ACCESS) program.

\bibliographystyle{ACM-Reference-Format}
\bibliography{citations.bib}

\end{document}